\documentclass[11pt,a4paper]{article}

\usepackage{amsmath,amssymb,relsize}
\usepackage{appendix}
\usepackage{mathtools}
\usepackage{dsfont}
\usepackage{todonotes}
\usepackage{subcaption}
\usepackage{multirow}
\usepackage{array}
\usepackage{mwe}
\usepackage{tikz}
\usepackage{xcolor}
\usepackage[com pat=1.1.0]{tikz-feynman}
\usepackage{textcomp,gensymb}

\usepackage{setspace}

\newcommand{\CW}{c_{\rm\textsc{w}}}
\newcommand{\SW}{s_{\rm\textsc{w}}}
\newcommand{\MW}{m_{\rm\textsc{w}}}
\newcommand{\beq}{\begin{equation}\displaystyle}
\newcommand{\eeq}{\end{equation}}
\newcommand{\ds}{\displaystyle}

\def\lsim{\mathrel{\rlap{\lower3pt\hbox{\hskip0pt$\sim$}}
   \raise1pt\hbox{$<$}}}         
\def\gsim{\mathrel{\rlap{\lower4pt\hbox{\hskip1pt$\sim$}}
   \raise1pt\hbox{$>$}}}         

\newcommand{\tw}{t_{\rm\textsc{w}}}

\title{ \bf Learning from Radiation at a 
\\Very High Energy Lepton Collider}
\author{Siyu Chen$^{a}$, Alfredo Glioti$^{a}$, Riccardo Rattazzi$^{a}$, Lorenzo Ricci$^{a}$, Andrea Wulzer$^{a,b}$ \\ \\
{\small\emph{$^a$ Theoretical Particle Physics Laboratory (LPTP), Institute of Physics,}}\\
{\small\emph{EPFL, Lausanne, Switzerland}}\\
{\small\emph{$^b$ Dipartimento di Fisica e Astronomia, Universit\`a di Padova, Italy}}\\
}

\usepackage[numbers,compress]{natbib}
\usepackage{graphicx}
\date{}

\usepackage[colorlinks = true,
            linkcolor = blue,
            urlcolor  = blue,
            citecolor = blue,
            anchorcolor = blue]{hyperref}

\begin{document}
\baselineskip=13pt

\maketitle

\begin{abstract}
We study the potential of lepton collisions with about $10$~TeV center of mass energy to probe Electroweak, Higgs and Top short-distance physics at the $100$~TeV scale, pointing out the interplay with the long-distance ($100$~GeV) phenomenon of Electroweak radiation. On one hand, we find that sufficiently accurate theoretical predictions require the resummed inclusion of radiation effects, which we perform at the double logarithmic order. On the other hand, we notice that short-distance physics does influence the emission of Electroweak radiation. Therefore the investigation of the radiation pattern can enhance the sensitivity to new short-distance  physical laws. We  illustrate these aspects by studying Effective Field Theory contact interactions in di-fermion and di-boson production, and comparing cross-section measurements that require or that exclude the emission of massive Electroweak bosons. The combination of the two types of measurements is found to enhance the sensitivity to the new interactions. Based on these results, we perform sensitivity projections to Higgs and Top Compositeness and to minimal $Z'$ new physics scenarios at future muon colliders.
\end{abstract}

\newpage

\begingroup
\tableofcontents
\endgroup 

\setcounter{equation}{0}
\setcounter{footnote}{0}
\setcounter{page}{1}

\newpage

\section{Introduction}\label{intro}

The perspective of a future muon collider with high energy and high luminosity~\cite{Delahaye:2019omf}, whose feasibility is  being investigated by the International Muon Collider Collaboration~\cite{muon}, has triggered a growing interest in the physics potential of lepton colliders with a  center of mass energy of $10$~TeV or more~\cite{Buttazzo:2018qqp,Franceschini:2021dxn,Capdevilla:2021fmj,AlAli:2021let,Buttazzo:2020uzc,Han:2020pif,Bartosik:2019dzq,Capdevilla:2020qel,Ruhdorfer:2019utl,Chiesa:2020awd,Costantini:2020stv,Capdevilla:2021rwo,Bartosik:2020xwr,Yin:2020afe,Kalinowski:2020rmb,Huang:2021nkl,Liu:2021jyc,Han:2021udl,Bottaro:2021srh,Li:2021lnz,Asadi:2021gah,Sahin:2021xzt,Chen:2021rid,Haghighat:2021djz,Bottaro:2021snn,Sen:2021fha,Han:2021lnp,Bandyopadhyay:2021pld,Dermisek:2021mhi,Qian:2021ihf,Chiesa:2021qpr,Liu:2021akf,Buttazzo:2021lzi,DiLuzio:2018jwd,Han:2020uak,Chen:2021pqi,Capdevilla:2021kcf,FranceschiniDM}. Such a Very High Energy Lepton  Collider (VHEL) could greatly advance the post-LHC knowledge of fundamental physics~\cite{Delahaye:2019omf} by directly searching for new heavy particles~(see e.g.~\cite{Buttazzo:2018qqp,Franceschini:2021dxn,Capdevilla:2021fmj,AlAli:2021let}), and by precise measurements of ElectroWeak (EW) scale processes  exploiting the high luminosity of virtual vector bosons pairs~(see e.g.~\cite{AlAli:2021let,Buttazzo:2020uzc,Han:2020pif}). By these two search modes, the VHEL reach on new physics is generally comparable to that  of the most ambitious future collider projects, in the corresponding domains of investigation. In particular it is comparable to the combined reach of the -ee and -hh stages of the FCC program.  The sensitivity is slightly weaker or slightly stronger depending on the specific new physics target and, obviously, on the assumed VHEL energy and luminosity. 

At a VHEL, however, there also exists a third strategy of investigation~\cite{Delahaye:2019omf,Buttazzo:2020uzc}, based on hard processes with  energy scale  comparable to the collider energy $E_{\rm{cm}}\sim10$~TeV. As the indirect effects of new heavy particles are enhanced at high energy, these processes are extremely sensitive probes of new physics. With the target integrated luminosity of $10$~ab$^{-1}$, $2\to2$ scattering processes at $E_{\rm{cm}}=10$~TeV can be generically measured with percent or few-percent precision. Such measurements are therefore sensitive to putative new physics at a scale $\Lambda\sim100$~TeV when its effects, relative to the SM  cross-section, scale like $(E_{\rm{cm}}/\Lambda)^2$. In an Effective Field Theory (EFT) description of new physics, this corresponds to an enhanced sensitivity to those dimension-$6$ operators that contribute to the $2\to2$ processes with a quadratically growing-with-energy term. The VHEL sensitivity to new physics through this kind of ``high-energy'' probes vastly and generically exceeds the potential of any other future project that is currently under consideration~\cite{Buttazzo:2020uzc}. In particular it exceeds the sensitivity of precision measurements of EW-scale ($\sim100$~GeV) processes at future Higgs factories, where new physics at $\Lambda\sim100$~TeV produces effects at the unobservable level of one part per million. It also exceeds the potential sensitivity of a $100$~TeV proton collider, because the effective luminosity for partonic collisions above $10$~TeV is significantly  lower than that of the VHEL. The possibility of probing new physics at the $100$~TeV scale, and in particular of probing new physics that is either flavor-universal or endowed with a flavor protection mechanism, is thus a unique feature of the VHEL that deserves an extensive investigation.\footnote{Hard processes are also useful to investigate flavor non-universal effects, as we will see in Section~\ref{Sec:4} for Top Compositeness. See also Ref.~\cite{Buttazzo:2021lzi} for a study of new physics potentially responsible for the $g-2$ muon anomaly.}

The above mentioned high-energy strategy exploits simple $2\to2$ processes with extremely low or negligible background, whose  target accuracy is statistically limited to $1\%$. At a superficial look, it might thus seem that its implementation will not pose any specific challenge, neither as concerns the feasibility of the measurements, nor as concerns the theoretical predictions that are needed for their interpretation. However the situation is slightly more complex, both experimentally and theoretically, owing to the phenomenon of EW radiation~\cite{Ciafaloni:2000df,Ciafaloni:2000rp,Fadin:1999bq,Melles:2000gw,Chiu:2007yn,Chiu:2007dg,PhysRevD.81.014023,Manohar:2014vxa,Manohar:2018kfx,Bauer:2017bnh,Ciafaloni:2001mu,Ciafaloni:2005fm,Fornal:2018znf,Bauer:2017isx,Bauer:2018xag,Han:2020uid,Chen:2016wkt,Dawson:1984gx,Kunszt:1987tk,Borel:2012by,Ruiz:2021tdt}, which becomes prominent at $10$~TeV or above. This happens because of  the existence of a large separation between the hard scale $E$ of the  process and the vector boson mass scale $\MW$, which acts as an IR cutoff. As the hard scale is increased, large logarithms \parbox[c][0pt]{52pt}{$\log E^2/\MW^2$} (squared) enhance both virtual corrections and real emission, up to the point where fixed-order perturbation theory becomes insufficient and resummation is needed. The experimental implications of the copious emission of real EW massive vector bosons should be assessed. It particular it should be studied if and how it impacts the accuracy of the relevant cross-section measurements. In this paper we investigate the implications of EW radiation on the theoretical predictions and, assuming purely statistical experimental errors, on the VHEL sensitivity to new physics.
 
 EW radiation obviously displays some similarities with QCD radiation, but also remarkable differences. First, EW scattering processes violate the KLN theorem assumptions~\cite{Kinoshita:1962ur, PhysRev.133.B1549} because the initial state particles are not EW singlets. Therefore no cancellation takes place between virtual and real contributions, not even in ``fully-inclusive'' cross-sections~\cite{Ciafaloni:2000df,Ciafaloni:2000rp}. Moreover the observables that are fully inclusive in the sense of Ref.~\cite{Ciafaloni:2000df} are insufficient to characterize new physics because they require summing over the ``color'' of the hard final-state particles. In the EW context, color sum means, for instance, including transversely-polarized $W$ and $Z$ bosons and photons (or, longitudinal $W$, $Z$ and Higgs) in the same observable, while we need to keep them separate for a comprehensive exploration of new physics. Unlike QCD (and QED), EW radiation effects cannot be eliminated or systematically mitigated with a judicious choice of the observables. They unavoidably play an important role in the predictions. 

The second peculiarity of EW radiation is that the IR cutoff scale is physical, and the theory is weakly-coupled at the IR scale. It should thus be possible to characterize the radiation fully by first-principle suitably resummed perturbative calculations. Unlike QCD and QED, one can consider observables with an arbitrary degree of radiation exclusiveness, among which ``exclusive'' scattering cross-sections with a fixed number ($2$, in $2\to2$ processes) of hard partons in the final state and no extra massive vector bosons.\footnote{In order to cope with QED and QCD radiation, the observable must still be inclusive over the emission of photons and other light particles. The cross-section we define as ``exclusive''  coincides with  the ``semi-inclusive'' cross-section of Ref.~\cite{Fadin:1999bq}.  Correspondingly,  the ``semi-inclusive'' cross-section we will readily introduce was not considered in Ref.~\cite{Fadin:1999bq}. See Section~\ref{Sec:2} for details.} Fully-inclusive cross-sections can be also considered, however they are not sufficiently informative on new physics as previously mentioned. In this paper we employ ``semi-inclusive'' final states, consisting of   $2$ hard partons with fixed EW color and flavor carrying a large fraction of the total available energy $E_{\rm{cm}}$ and accompanied by an arbitrary number of massive vectors, photons and other light particles. Our resummed predictions for semi-inclusive observables at the double logarithm (DL) accuracy are obtained by  extending the IR Evolution Equation (IREE) treatment of EW radiation developed in Ref.~\cite{Fadin:1999bq}. Similarly, we employ the IREE to compute the more standard exclusive cross-sections. Single-logarithmic terms turn out to be relevant, and they are included at fixed one-loop order in the exclusive cross-sections using the results of Ref.~\cite{Denner:2000jv}. 

Finally, there is an interplay between EW radiation and short-distance physics that has no analog in QED and QCD~\cite{Buttazzo:2020uzc}. Based on a simplistic fixed order intuition, this interplay can be exemplified by noticing that the emission of a soft $W$ from one initial lepton changes the total charge of the initial state of the hard $2\to2$ scattering process. By allowing for the charged $W$ emission one thus obtains a scattering cross-section that is sensitive to charged current new physics interactions, while the exclusive process where no radiation is allowed is only sensitive to neutral currents.~\footnote{More precisely, the charged and neutral current process depend on different linear combinations of the Wilson coefficients of the operators parametrizing new physics.} The combined measurement of the two types of cross-section thus enables a more effective and complete exploration of new physics. In reality the situation is slightly more complex, because the neutral and the charged current hard amplitudes both contribute to the resummed expression of the neutral exclusive and of the charged and neutral semi-inclusive cross-sections. However, since they appear in different combinations in the different cross-sections, the conclusion is unchanged. 

At a more quantitative level, we will see that the VHEL energy sits in an interestingly ``intermediate'' regime for EW radiation. The energy is on one hand large enough for the radiation effects to be important and require resummation. Accurate resummation techniques, more accurate than the one considered in the present paper, will thus be needed to fully exploit the potential of the machine. On the other hand, the energy is not yet in the asymptotic regime where the Sudakov exponentials entail a strong suppression of the non-emission probability. Therefore in this regime the exclusive cross-sections are still large, and comparable with the semi-inclusive (and fully-inclusive) ones. Observables with a different degree of radiation inclusiveness can thus be measured with comparable statistical accuracy and combined, potentially boosting,  as previously explained, the indirect  sensitivity to heavy new physics.

The rest of the paper is organized as follows. We start, in Section~\ref{Sec:2}, by reviewing the IREE approach to the DL resummation of exclusive cross-sections, extending it to semi-inclusive ones. In this section we also present the calculation of the exclusive and semi-inclusive di-fermion and di-boson production processes, to be employed in Section~\ref{Sec:3} to estimate the sensitivity to dimension-$6$ EFT current-current contact interactions of muon colliders with a center of mass energy ranging from $3$ to $30$~TeV. The estimates include experimental reconstruction efficiencies at the level expected for the CLIC detector at $3$~TeV, which we extract from Ref.~\cite{deBlas:2018mhx}. Our results do not depend on the nature of the colliding leptons. In particular they do not include the effect of Beamstrahlung, which is expectedly small at muon colliders but large at $e^+e^-$ colliders of the same energy. Up to this caveat, our results thus also apply to hypothetical linear $e^+e^-$ collider based on plasma wake field acceleration~\cite{ALEGRO:2019alc}.\footnote{Notice that Beamstrahlung potentially entails a strong depletion of the high-energy luminosity peak, which is the part of the luminosity spectrum that is relevant for the high-energy probes.} In Section~\ref{Sec:3} we also study the impact of the EFT sensitivity projections on concrete scenarios for Beyond-the-SM (BSM) physics such as Higgs and Top Compositeness and a minimal $Z^\prime$ extension of the EW sector. Our conclusions and an outlook on future work are reported in Section~\ref{Sec:5}. The first two appendices contain technicalities about the the derivation of the IREE. The third one collects our sensitivity projections on EFT operators relevant for the Top Compositeness scenario, while the fourth appendix extensively summarizes the BSM sensitivity of muon colliders of different center of mass energies. 

\section{Theoretical predictions}\label{Sec:2}

Several approaches have been considered in the literature for the resummation of the effects of EW radiation.  Double logarithm (DL) contributions, of the form $(\alpha\log^2E^2/\MW^2)^n$ with arbitrary $n$, have been resummed in fully-inclusive and exclusive cross-sections, using respectively Asymptotic Dynamics~\cite{Ciafaloni:2000df,Ciafaloni:2000rp} and  IREE~\cite{Fadin:1999bq, Melles:2000gw}. In Soft-Collinear Effective Theory (SCET) the expansion is already organized in exponential form. In that case the resummation of  double logarithms~\cite{Chiu:2007yn} as well as its extension to leading (LL )\footnote {These include but do not coincide with the pure DL, as explained, for instance, in Ref.~\cite{Manohar:2018kfx}.} and next-to-leading (NLL) logarithm \cite{Manohar:2018kfx}  has been studied.  The study of EW parton distribution or fragmentation functions~\cite{Ciafaloni:2005fm,Fornal:2018znf,Bauer:2017isx,Bauer:2018xag,Han:2020uid}  is obviously connected, but not directly relevant for very high energy processes, where the main role is played by the emission of EW radiation that is both collinear and soft. Notice however that in some framework \cite{Ciafaloni:2005fm,Fornal:2018znf,Bauer:2017isx,Bauer:2018xag,Han:2020uid} soft effects are partially or completely included in parton distributions and fragmentation functions. 

In this paper we employ DL predictions based on the robust diagrammatic methodology of the IREE~\cite{Fadin:1999bq}, which we review and further develop in Section~\ref{Sec:IREE}. We also supplement our computations by the available  virtual single logarithms (SL) at 1-loop \cite{Denner:2000jv,Denner:2001gw}.  Based on these results, we present in Sections~\ref{Sec:Diferm} and \ref{Sec:Dibos} our theoretical predictions for the di-fermion and di-boson production processes at the VHEL. While it will emerge that single logarithms  are potentially relevant, a systematic consideration of these effects goes beyond the scope of the present paper. Our finding that electroweak radiation can be used to our own advantage in the exploration of new physics, strongly motivates the systematic inclusion of subleading effects. A first simple step would be to include in our predictions the single logarithms from real emissions at tree level. A complete treatment including resummation would likely best be achieved by using SCET.

\subsection{IR Evolution Equations}\label{Sec:IREE}

The basic idea of the IREE is to introduce an unphysical IR regulator $\lambda$ with dimension $({\mathrm{energy}})^2$  in the calculation of the observables and to write down a differential equation for the evolution with $\lambda$ of the result. Denoting by ``$E^2$'' the hardness of the observable under consideration, the choice $\lambda\sim E^2$  eliminates all the IR enhancements and makes fixed-order perturbation theory well-behaved. For $\lambda\sim E^2$, 
the  Born level computation therefore offers a reliable approximation, which can be used as the initial condition for the evolution equation to lower $\lambda$. The physical predictions are obtained from the solution of the IREE in the limit $\lambda\to0$.

In order to define the IR regulator, consider the $4$-momenta $k_i$ of the external hard particles that participate in the scattering process. They will correspond in our setup to the $4$ legs of a central energetic $2\to2$ processes. With the exception  of the  masses $k_i^2\ll E^2$, all the Lorentz invariants constructed with the $k_i$'s are therefore large and of order $E^2$. Given a pair $ij$ of external hard particles and given  a radiation particle with $4$-momentum $q$ we define  its hardness  relative to the $ij$ pair as
\beq\label{hard}
{\mathfrak{h}}^{ij}(q)\equiv 2\left| \frac{(k_i \cdot q) ( k_j \cdot q)}{(k_i\cdot k_j)} \right|\,.
\eeq
The IR regulator is provided by the lower bound $\lambda$
\beq\label{reg}
{\mathfrak{h}}(q)\equiv\min\limits_{i\neq j}{\mathfrak{h}}^{ij}(q)>\lambda\,.
\eeq
on the ``absolute'' hardness ${\mathfrak{h}}$ of the radiation. Notice that the bound is imposed on the $4$-momentum of each individual radiation particle, either virtual or real. Specifically, eq.~(\ref{reg}) applies to the off-shell loop momenta describing virtual radiation, as well as to the on-shell momenta of real radiation particles in the final state of the process. The specific definition of the radiation hardness in eq.~(\ref{hard}) is convenient because it reflects the structure of the denominators that appear in the calculation of the IR-enhanced diagrams, as we will readily see. At this stage, it is enough to remark that the lower cut on ${\mathfrak{h}}(q)$ in eq.~(\ref{reg}) is a valid IR regulator as it eliminates both the regions where  $q$ is soft and those where it is collinear to one of the hard partons.

The main peculiarity of the IREE formalism applied to EW radiation stems from the presence of the physical scale $\MW\sim100$~GeV associated to the masses of the EW bosons. We will see that $\MW$ acts as a threshold that separates two different regimes, $\lambda\gg\MW^2$ and $\lambda\ll\MW^2$.  In the former regime, the cut on the radiation hardness in eq.~(\ref{reg}) is so strong that the mass of the radiation particles can be safely neglected and  the IREE  computed in the unbroken \mbox{SU$(2)_L\times$U$(1)_Y$} EW gauge theory. Starting from the initial condition at $\lambda\sim E^2$, the evolution is thus performed with the \mbox{SU$(2)_L\times$U$(1)_Y$} evolution kernel down to $\lambda\sim \MW^2$. At $\lambda \ll \MW^2$, the massive vector bosons do not contribute to the evolution and the kernel is purely determined by the unbroken \mbox{U$(1)_Q$} group of electromagnetism.

\subsubsection*{Amplitude evolution}

We start, following Ref.~\cite{Fadin:1999bq}, from the IREE for the scattering amplitude with purely  hard external quanta and with regulator $\lambda$ on the internal lines.  While the discussion holds for an arbitrary number of external legs, we focus for definiteness on $2\to2$ amplitudes, which we indicate by
\beq\label{hardamp}
\mathcal{M}^{\alpha}_\lambda= \mathcal{M}_\lambda\big[{\rm{p}}_1(k_1,\alpha_1)\,{\rm{p}}_2(k_2,\alpha_2)\to {\rm{p}}_3(k_3,\alpha_3)\,{\rm{p}}_4(k_4,\alpha_4) \big]\,,
\eeq
where $\alpha_i$ denotes the gauge group index of the external state ${\rm{p}}_i$, which is taken to transform in an irreducible representation of the group. The amplitude is labeled by the $4$ indices $\alpha=\alpha_1\alpha_2\alpha_3\alpha_4$, and it is IR-regulated according to eq.~(\ref{reg}). Since no real radiation is involved, the cut acts only on the momenta of virtual vector bosons in loop diagrams. We aim at writing down the IREE for $\mathcal{M}^{\alpha}_\lambda$ and to  solve it  given the initial condition
\beq\label{ampBC}
\mathcal{M}^{\alpha}_{E^2}=\mathcal{B}^{\alpha}\equiv {\mathrm{Born\,Amplitude}}\, .
\eeq

As we explained, for $\lambda\gg\MW^2$ the effects of EW symmetry breaking (EWSB) can be ignored, and $\mathcal{M}^{\alpha}_\lambda$ equals the (IR-regulated) amplitude of the unbroken EW gauge theory. More precisely, EWSB gives negligible relative corrections of order $\MW/\sqrt{\lambda}$ (or powers thereof) to all those amplitudes that are not forbidden by the \mbox{SU$(2)_L\times$U$(1)_Y$}  exact symmetry of the unbroken theory. The other amplitudes are sensitive to EWSB at the leading order and therefore they cannot be studied in the unbroken theory.\footnote{For instance the amplitude with $3$ transversely- and one longitudinally-polarized $W$ bosons is suppressed by $\MW/E$ already at the Born level, owing to the fact that it is impossible to form an \mbox{SU$(2)_L$} singlet with one doublet (i.e., the representation of longitudinal $W$'s owing to the Equivalence Theorem) and three triplets.} However their contribution to the physical scattering process is negligible and they can be safely excluded from the discussion. Similarly, for the allowed processes, up to negligible power corrections of order $\MW/E$, the amplitude  $\mathcal{M}^{\alpha}_\lambda$ is an \mbox{SU$(2)_L\times$U$(1)_Y$} invariant tensor satisfying  the charge conservation equation 
\beq\label{chco}
(G_{1^{c}}^A)^{\alpha}_{\;\beta}\mathcal{M}^{\beta}_\lambda+
(G_{2^{c}}^A)^{\alpha}_{\;\beta}\mathcal{M}^{\beta}_\lambda+
(G_3^A)^{\alpha}_{\;\beta}\mathcal{M}^{\beta}_\lambda+
(G_4^A)^{\alpha}_{\;\beta}\mathcal{M}^{\beta}_\lambda\,
\overset{\; {\lambda\gg\MW^2}}{{=}}\;0\,,\;\;\;\;\;\forall\,A,\,\alpha\,.
\eeq
In the equation, $G_i^A$ denotes the generators associated with the representation of each hard particle ``$i$'' under the EW group, acting only on the corresponding index ``$\alpha_i$'' of the amplitude tensor. For instance
\beq\label{capG}
(G_3^A)^{\alpha}_{\;\beta}=\delta^{\alpha_1}_{\beta_1}\delta^{\alpha_2}_{\beta_2} (G_3^A)^{\alpha_3}_{\;\beta_3} \delta^{\alpha_4}_{\beta_4}\,.
\eeq
Notice that, in our notation, $(\alpha_3,\alpha_4)$ run in the representations of the outgoing states, while $(\alpha_1,\alpha_2)$ run in the conjugate representation of the incoming particles. Consequently in eq.~(\ref{chco}),  $G_{1^{c}}=-G_1^*$ and $G_{2^{c}}=-G_2^*$.

\begin{figure}
	\centering
	\begin{tikzpicture}[scale=1.6]
	\begin{feynman}[every blob={/tikz/fill=black,/tikz/minimum size=7pt}, every dot={/tikz/minimum size=2.6pt}]
	\vertex[blob] (m) at (0, 0){};
	\vertex (a) at (-1.25,0.833){$1$};
	\vertex[dot] (a1) at (-0.825,0.55){};
	\vertex[dot] (a2) at (-0.6, 0.4){};
	\vertex[dot] (a3) at (-0.225, 0.15){};
	\vertex (b) at ( -1.25,-0.833){$2$};
	\vertex[dot] (b1) at (-0.825,-0.55){};
	\vertex[dot] (b2) at (-0.6, -0.4){};
	\vertex[dot] (b3) at (-0.225, -0.15){};
	\vertex (c1) at (1.25, 0.833){$3$};
	\vertex (c2) at (1.25, -0.833){$4$};
	\vertex at (-0.22, -0.32){$q_1$};
	\vertex at (-0.45, -0.6){$q_{n-1}$};
	\vertex at (-0.4, 0){${\small{\cdots}}$};
	\diagram* {
		(a) -- [momentum=$k_1$] (a1)-- (m),
		(b) -- [momentum'=$k_2$] (b1)--(m),
		(m) -- [momentum=$k_3$](c1),
		(m) -- [momentum'=$k_4$](c2),
		(a1) -- [boson, edge label'= $q_n$](b1),
		(a2) -- [boson](b2),
		(a3) -- [boson](b3),
	};
	\end{feynman}
	\end{tikzpicture}	
	\hspace{20pt}
	\begin{tikzpicture}[scale=1.6]
	\begin{feynman}[every blob={/tikz/fill=gray!30,/tikz/inner sep=2pt}, every dot={/tikz/minimum size=2.6pt}]
	\vertex[blob] (m) at (0, 0){};
	\vertex (a) at (-1.25,0.833){$1$};
	\vertex[dot] (a1) at (-0.75,0.5){};
	\vertex (b) at ( -1.25,-0.833){$2$};
	\vertex[dot] (b1) at (-0.75,-0.5){};
	\vertex (c1) at (1.25, 0.833){$3$};
	\vertex (c2) at (1.25, -0.833){$4$};
	\diagram* {
		(a) -- [momentum=$k_1$] (a1)-- (m),
		(b) -- [momentum'=$k_2$] (b1)--(m),
		(m) -- [momentum=$k_3$](c1),
		(m) -- [momentum'=$k_4$](c2),
		(a1) -- [boson,line width=1.5pt, momentum'=$q$](b1),
		(a1) -- [boson,line width=0.8pt,white](b1),
	};
	\vertex[dot] (q1) at (-0.75,-0.5){};
	\end{feynman}
	\end{tikzpicture}
	\caption{Left panel: the leading Sudakov diagrams topology. Right panel: a diagrammatic representation of the contributions to the amplitude variation that are logarithmically enhanced. The momentum $q$ is integrated over the infinitesimal strip $\delta\sigma$~(\ref{strip}). \label{Fig:AmpIREE}}
\end{figure}
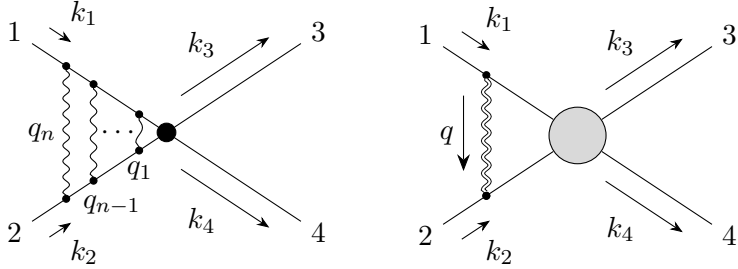

The IREE is obtained by computing the variation of the amplitude under an infinitesimal variation $\lambda\to\lambda+\delta\lambda$ of the IR cutoff in eq.~(\ref{reg}). This computation dramatically simplifies in the leading DL approximation as one can infer by inspecting  diagrams involving a number $n$ of soft/collinear virtual vector bosons. Indeed the maximal logarithm power arises from the region where  momenta are hierarchically separated $E^2\gg {\mathfrak{h}}(q_1)\gg {\mathfrak{h}}(q_2)\gg\dots\gg{\mathfrak{h}}(q_n)$ with the softer legs dressing the subdiagrams involving the harder legs, as shown in the left panel of Figure~\ref{Fig:AmpIREE}. In this configuration  only the outermost virtual vector can reach a virtuality ${\mathfrak{h}}(q_n)\sim \lambda$, the inner ones being much harder in the dominant region of integration. The effect on ${\cal M}^\alpha_\lambda$ of the variation of $\lambda$  is then computed by  considering the variation of the endpoint of the integral over the momentum of such outermost vector. More precisely we have that $-\delta {\cal M}_\lambda^\alpha\equiv {\cal M}_\lambda^\alpha-{\cal M}_{\lambda+\delta\lambda}^\alpha$ equals the integral over the outermost loop momentum in the infinitesimal strip
\beq\label{strip}
\delta\sigma=\{q:\,{\mathfrak{h}}(q)\in[\lambda,\,\lambda+\delta\lambda]\}\,.
\eeq
The contribution to the variation from the vector that connects a given pair of hard external legs can be depicted like on the right panel of Figure~\ref{Fig:AmpIREE}. The vector boson is represented with a double line to indicate that its momentum $q$ must be integrated only over the strip $\delta\sigma$. 

As we already said the leading contribution comes from the integration region where $q$ is soft (and also collinear), in which the vector boson emission is described by the eikonal formula
\begin{align}
\begin{aligned}
\begin{tikzpicture}[scale=1]
  \begin{feynman}[every blob={/tikz/fill=gray!30,/tikz/inner sep=2pt}, every dot={/tikz/minimum size=2.5pt}]
      \vertex[dot] (m) at (0.4, 0){};
      \vertex (a) at (1.5,0){$\alpha_i$};
      \vertex[blob] (c) at (-0.7, 0) {$\mathcal{M}$};
      \vertex (d) at (0.4, -0.6) [label={[xshift=13pt, yshift=-5pt] $A,\mu$} ];
      \diagram* {
        (m) -- [momentum=$k_i$] (a),
        (m) -- [boson, momentum'=$q$](d),
        (c) -- (m)
      };
    \end{feynman}
    \end{tikzpicture}
    \end{aligned}\ds
   \simeq \frac{k_i^{\mu}}{k_i\cdot q} (G_i^A)^{\alpha}_{\;\beta} \mathcal{M}^{\beta}\,,\;\;\;\;\;\;\;\;\;\;
\begin{aligned}
\begin{tikzpicture}[scale=1]
  \begin{feynman}[every blob={/tikz/fill=gray!30,/tikz/inner sep=2pt}, every dot={/tikz/minimum size=2.5pt}]
      \vertex[dot] (m) at (-0.2, 0){};
      \vertex (a) at (-1.5,0){$\alpha_i$};
      \vertex[blob] (c) at (+0.6, 0) {$\mathcal{M}$};
      \vertex (d) at (-0.2, -0.6) [label={[xshift=13pt, yshift=-5pt] $A,\mu$} ];
      \diagram* {
        (a) -- [momentum=$k_i$] (m),
        (m) -- [boson, momentum'=$q$](d),
        (c) -- (m)
      };
    \end{feynman}
    \end{tikzpicture}
    \end{aligned}\ds
   \simeq \frac{k_i^{\mu}}{k_i\cdot q} (G_{i^c}^A)^{\alpha}_{\;\beta} \mathcal{M}^{\beta}\,,   
\label{Eq:EikAppr}
\end{align}
with $G_i$ the group generator acting  on particle ``$i$'' as in eq.~(\ref{capG}). In line with our conventions, as explained above, the generators of the charge-conjugate representation $G_{i^c}$ appear in the eikonal formula for vector boson emission from an incoming particle. For brevity, we have included the gauge coupling constants in the definition of the generators $G_i$. In terms of the canonical \mbox{SU$(2)_L\times$U$(1)_Y$} generators we then have
\beq\label{cang}
G_i^{1,2,3}=g\,T_i^{1,2,3}\,,\;\;\;\;\;G_i^{Y}=g^\prime\,Y_i\,.
\eeq

The integration over the soft $q$ momentum factorizes with respect to the integral over the harder lines, represented as a blob in Figure~\ref{Fig:AmpIREE}. Indeed in the eikonal ($q\to0$) limit the virtual vector boson momentum can be neglected and the sub-amplitude blob evaluated on the momenta $k_i$ of the external legs before the virtual vector boson emission/absorption. Therefore the blob gives us back the original amplitude, with one less loop but this is immaterial as $\mathcal{M}^{\alpha}_\lambda$ is the all-loops amplitude. We can thus express the amplitude variation in terms of the amplitude itself, eventually obtaining an evolution equation. In covariant gauges, the leading DL contributions only arise from virtual vectors connecting two distinct external lines. Therefore, we have 
\begin{align}\ds
    \delta \mathcal{M}^{\alpha}_\lambda=\frac{-i}{(2\pi)^4}\sum_{j\,<\,i}\; \, \int_{\ds\delta\sigma}
    \hspace{-2pt} d^4 q\,
    \frac{1}{q^2 + i \epsilon} \frac{k_i \cdot k_j }{(q \cdot k_i)(q \cdot k_j)}\big[\sum_A G^{A}_i \cdot G^{A}_j \big]^{\alpha}_{\;\;\beta} \, \mathcal{M}_{\lambda}^{\beta}\,,
    \label{Eq:VarAmp}
\end{align}
where the sum extends over the unordered $ij$ pairs of distinct external legs and it is understood that the conjugate generators must be employed for the incoming legs $i,j=1,2$, due to eq.~(\ref{Eq:EikAppr}). 

The evaluation of the integral in eq.~(\ref{Eq:VarAmp}) is quite straightforward, and it is reported in Appendix~\ref{AppDL} for completeness. This gives
\begin{align}\ds
    \delta \mathcal{M}^{\alpha}_\lambda=-\frac{1}{8\pi^2}\frac{\delta\lambda}\lambda\,\log\frac{E^2}\lambda\,\frac12\sum_{A,\,i} (G^{A}_i)^{\alpha}_{\;\;\beta} \sum_{j\neq i}  (G^{A}_j)^{\beta}_{\;\;\gamma} \, \mathcal{M}_{\lambda}^{\gamma}\,,
    \label{Eq:VarAmp1}
\end{align}
up to non logarithmically enhanced terms. Notice that in the equation we traded the sum over unordered $ij$ pairs for an ordered sum times $1/2$. The sum over $j\neq i$ can be performed using charge conservation according to eq.~(\ref{chco}), giving
\begin{align}\ds
    \delta \mathcal{M}^{\alpha}_\lambda&=\frac{1}{16\pi^2}\frac{\delta\lambda}\lambda\,\log\frac{E^2}\lambda\,\sum_{i} \big[\sum_A G^{A}_i G^{A}_i\big]^{\alpha}_{\;\;\beta} \, \mathcal{M}_{\lambda}^{\beta}\nonumber\\
    &=
    \ds
    \frac{1}{16\pi^2}\frac{\delta\lambda}\lambda\,\log\frac{E^2}\lambda\,\sum_{i} \left[g^2 c_i+g^{\prime\,2}y_i^2\right] \, \mathcal{M}_{\lambda}^{\alpha}
    \,,
    \label{Eq:VarAmp2}
\end{align}
where  for any given external particle with weak isospin spin $t_i$ and hypercharge $y_i$, the coefficients $c_i=t_i(t_i+1)$ and $y_i^2$ are nothing but the Casimirs of respectively SU$(2)_L$ and U$(1)_Y$. We thus recovered the familiar result that, in  DL accuracy, IR effects are universal for each individual external particle  and purely determined by the Casimir of the corresponding gauge group representation. 

We finally obtain an IREE
\beq\label{ampIREEhigh}
\frac{d \mathcal{M}^{\alpha}_\lambda}{d\log^2({E^2}/\lambda)}=-\frac12{\cal{K}} \, \mathcal{M}_{\lambda}^{\alpha}\,,\;\;{\rm{where}}\;\;{\cal{K}} \overset{\; {\lambda\gg\MW^2}}{{=}}\;\frac{1}{16\pi^2}\sum_{i} \left[g^2 c_i+g^{\prime\,2}y_i^2\right] 
\,,
\eeq
with, since the Casimir operators are proportional to the identity, an evolution kernel ${\cal{K}}$ that is a mere multiplicative constant. Solving eq.~(\ref{ampIREEhigh}) starting from the initial condition~(\ref{ampBC}) gives the amplitude evaluated with an IR cutoff scale $\lambda=\MW^2$
\beq\label{ampBC1}
 \mathcal{M}^{\alpha}_{\MW^2}=\exp\left[
 -\sum_{i} \frac{g^2 c_i+g^{\prime\,2}y_i^2}{32\pi^2}\, \log^2 (E^2/\MW^2) \right]\mathcal{B}^{\alpha}\,.
\eeq

In order to continue the amplitude evolution to lower $\lambda$, we should now consider the regime $\lambda\ll\MW^2$, write the corresponding IREE and solve them using eq.~(\ref{ampBC1}) as initial condition. This is straightforward, because we have seen that all that matters for the derivation of the IREE are the loop integrals in a strip where the virtual radiation hardness is infinitesimally close to the cutoff $\lambda$ as in eq.~(\ref{strip}). In this region, a logarithmic enhancement of the amplitude variation only originates from photon exchange diagrams.\footnote{The calculation of the loop integral in Appendix~\ref{AppDL} shows explicitly that no enhancement emerges from the exchange of vectors with mass $m_V$ much larger than $\lambda$.} The IREE evolution kernel is thus immediately obtained by specifying the previous formulae to the \mbox{U$(1)_Q$} gauge group of QED
\beq\label{ampIREElow}
{\cal{K}} \overset{\; {\lambda\ll\MW^2}}{{=}}\;\frac{1}{16\pi^2}\sum_{i} \left[e^{2}q_i^2\right]\,.
\eeq
Notice that in order to derive the IREE in this regime, only  conservation of  electric charge must  be employed. The  conservation of the full \mbox{SU$(2)_L\times$U$(1)_Y$} charges of eq.~(\ref{chco}) is not valid for $\lambda\ll\MW^2$, where the effects of electroweak symmetry breaking are important.

Solving eq.~(\ref{ampIREElow}) produces the regular QED Sudakov factors, which go to zero in the physical limit $\lambda\to0$ where the IR regulator is removed. Therefore the amplitude $\mathcal{M}^{\alpha}_{0}$ vanishes, and so  does the cross-section of the corresponding fully-exclusive scattering process, in which no extra radiation is present in the final state. More inclusive observables need to be considered for a non-vanishing result. One possibility is to allow for the presence of real photon radiation up to an upper threshold of order $\MW^2$ on the hardness ${\mathfrak{h}}$. This defines a  cross-section that we denote as {\it{exclusive}} because it indeed excludes the radiation of massive EW bosons. In fact, it is easy to check that ${\mathfrak{h}}(q)>m^2$ for the emission of a real radiation quantum  with $q^2=m^2$. An upper cut ${\mathfrak{h}}(q)<\MW^2$ then excludes  the presence of massive EW bosons in the final state, but  allows for (sufficiently) soft photons. Ref.~\cite{Fadin:1999bq} considered this same observable (but calling it ``semi-inclusive'') showing that it stops evolving with $\lambda$ below $\MW^2$, due to the cancellation of real and virtual IR effects in QED. Cross-sections that are exclusive according to our definition can thus be computed at the DL accuracy by just squaring the $\lambda=\MW^2$ amplitude~(\ref{ampBC1}). At the end of the next  section we will re-derive the result of Ref.~\cite{Fadin:1999bq} for exclusive cross-sections by a slightly different methodology, which is also suited for the calculation of the other type of cross-sections we are interested in.

\subsubsection*{Density matrix evolution}

It is possible to extend the IREE methodology to more complex quantities than the hard Feynman amplitude. Specifically, we consider the hard ``density matrix''~\footnote{The same object was dubbed ``overlap matrix'' in Ref.~\cite{Ciafaloni:2000rp}.}
\beq\label{Eq:DensityMatrixDefi}
{\cal{D}}^{\alpha \bar{\alpha}}_\lambda \equiv \mathcal{M}^{\alpha}_\lambda(\mathcal{M}^{\bar{\alpha}}_\lambda)^* +\sum\limits_{N=1}^{\infty} \int {\rm{dPh}}_{N,\lambda}^{\cal{H}} \sum\limits_{\rho_1\ldots \rho_N} \mathcal{M}^{\alpha;\rho}_\lambda(\mathcal{M}^{\bar{\alpha};\rho}_\lambda)^*\,,
\eeq
which incorporates the emission of an arbitrary number $N$ of radiation particles, with gauge group indices denoted as $\rho=\rho_1\ldots\rho_N$. In the equation, $\mathcal{M}^{\alpha}_\lambda$ is the hard amplitude with no extra emissions as in the previous section, while  $\mathcal{M}^{\alpha;\rho}_\lambda$ is the amplitude for the production of the $2$ hard particles plus the radiation. The virtual radiation particles exchanged in the Feynman diagrams for the amplitude are subject to the IR hardness cutoff $\lambda$ as in eq.~(\ref{reg}). The phase-space volume element ${\rm{dPh}}_{N,\lambda}^{\cal{H}}=\prod_{k=1}^{N}{\rm{dPh}}_{k,\lambda}^{\cal{H}}$ for the emission of real radiation  is also constrained by eq.~(\ref{reg}). The~ \parbox[c][0pt]{10pt}{$^{\cal{H}}$} superscript refers to the possible presence of an upper cutoff on the radiation hardness ${\mathfrak{h}}(q)<{\cal{H}}$. In what follows we will first consider processes we define as  {\it {semi-inclusive}}, for which  ${\cal{H}}\sim E^2$.  For these processes the upper radiation cut is effectively absent, and plays no role in the discussion. The exclusive processes defined in the previous section instead simply correspond to ${\cal{H}}=\MW^2$.

It should be noted that eq.~(\ref{Eq:DensityMatrixDefi}) formally violates the conservation of the total energy and momentum, because in the radiation terms we are employing the same hard $4$-momenta that obey energy and momentum conservation in the absence of radiation. It is understood that this makes sense only in the presence of an upper cutoff on the total energy and momentum of the radiation, say a one tenth of $E$. In this way, the radiation plays a minor role in the total balance of energy and momentum conservation or, equivalently, the hard $4$-momenta can be readjusted to balance the radiation emission up to small corrections in the corresponding Feynman amplitudes. In practice, the cutoff allows us to factorize the total phase-space into that for radiation, on one hand, and that for the hard  $2\to2$ process on the other, with the latter also including  the delta function of $4$-momentum conservation. The density matrix~(\ref{Eq:DensityMatrixDefi}) can thus be related to the physical scattering cross-section.

An upper cut $E_{rad}< E/10$ on the total radiation energy and momentum does not affect the predictions at the double logarithm accuracy. Indeed a simple modification of the real radiation integral (see the discussion around eq.~(\ref{Eq:VarDM}) computation in Appendix~\ref{AppDL} shows that the effect of this cut on the $q$ momentum of the radiated particle merely entails reduction of the double logarithm from $\log^2E^2/\lambda$ to  $\log^2E_{rad}^2/\lambda$. The difference is then of order $\log E^2/\lambda \times \log E^2/E_{rad}^2$ and falls into the same class as single logarithms as long as $E/E_{rad}$ is not too small, with $1/10$ qualifying.

The hard density matrix~(\ref{Eq:DensityMatrixDefi}) is a simple generalization of the scattering cross-section in which the conjugated amplitude indices $\bar\alpha$ are not equal to the indices $\alpha$ of the non-conjugated amplitude. It is a useful generalization because it obeys charge conservation equations similar to eq.~(\ref{chco}). Namely, in the regime $\lambda\gg\MW^2$, we have
\beq\label{chcoDM}
\sum\limits_{i=1^{c},2^{c},3,4}\big[
(G_i^A)^{\alpha}_{\;\beta}{\cal{D}}^{\beta \bar{\alpha}}_\lambda +
(G_{i^c}^A)^{\bar\alpha}_{\;\bar\beta}{\cal{D}}^{\alpha \bar{\beta}}_\lambda\big] \,
\overset{\; {\lambda\gg\MW^2}}{{=}}\;0\,,\;\;\;\;\;\forall\,A,\,\alpha,\,\bar\alpha\,,
\eeq
where the obvious relations  $[1^{c}]^c\equiv 1$ $[2^{c}]^c\equiv 2$ should be understood. That way
 the generators acting on the indices $\bar \beta$ of the complex conjugated amplitude are those of the corresponding charge conjugated representation.  Eq.~(\ref{chcoDM}) holds only for $\lambda\gg\MW^2$, because in this regime both the virtual and the real emissions are nearly insensitive to EWSB effects as previously explained. For $\lambda\ll\MW^2$, only the electric charge generator is conserved.

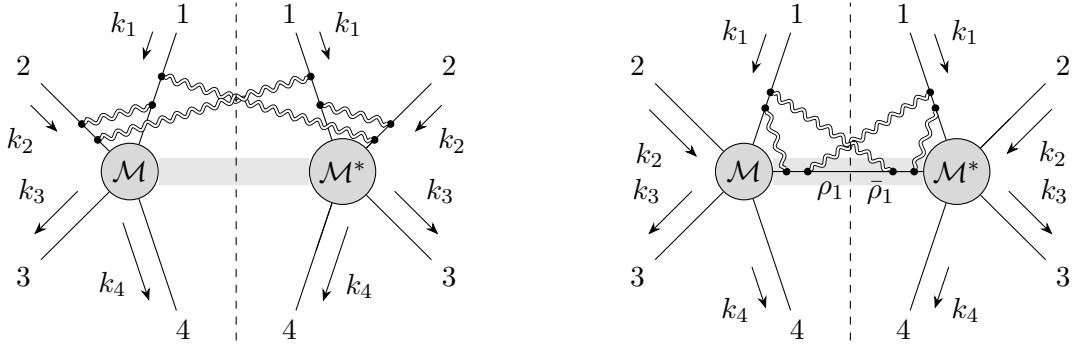
\begin{figure}
\begin{minipage}{0.5\textwidth}
\centering
\begin{tikzpicture}[scale=1.4]
  \begin{feynman}[every blob={/tikz/fill=gray!30,/tikz/inner sep=2pt}, every dot={/tikz/minimum size=2.5pt}]
      \vertex (a) at (-2,1){$2$};
      \vertex[dot] (aa) at (-1.45,0.45){};
      \vertex[dot] (aabis) at (-1.3,0.3){};
      \vertex (aaa) at (-1.52,0.52);
      \vertex[blob] (m) at (-1,0){$\mathcal{M}$};
      \vertex (a1) at (-0.5,1.5){$1$};
      \vertex[dot] (a1a1) at (-0.79,0.63){};
      \vertex[dot] (a1a1R) at (0.79,0.63){};
       \vertex[dot] (a1a1BIS) at (-0.7,0.9){};
       \vertex[dot] (A1A1BIS) at (0.7,0.9){};
      \vertex (b) at ( -2,-1){$3$};
      \vertex (bb) at (-1.25,-0.25);
      \vertex (b1) at (-0.5,-1.5){$4$};
      \vertex (b1b1) at (-0.75,-0.75);
       \vertex (A) at (2,1){$2$};
       \vertex[dot] (AA) at (1.3,0.3){};
      \vertex[dot] (AAbis) at (1.45,0.45){};
      \vertex (AAA) at (1.50,0.53);
      \vertex[blob] (M) at (1,0){$\mathcal{M}^*$};
      \vertex (A1) at (0.5,1.5){$1$};
      \vertex (A1A1) at (0.75,0.75);
      \vertex (B) at ( 2,-1){$3$};
      \vertex (BB) at (1.25,-0.25);
      \vertex (B1) at (0.5,-1.5){$4$};
      \vertex (B1B1) at (0.75,-0.75);
      \diagram* {
        (a) -- [momentum'=$k_2$](aa)--(m),
        (m) --(a1a1),
        (a1)-- [momentum'=$k_1$](a1a1BIS)--(a1a1),
        (bb) -- [momentum'=$k_3$](b),
        (bb)--(m),
        (m) --[momentum'=$k_4$] (b1),
        (M) -- (AA),
        (A)--[momentum=$k_2$](AAbis)--(M),
        (A1) --[momentum=$k_1$](A1A1BIS)-- (M),
        (M) -- (BB)--[momentum=$k_3$](B),
        (M)--[momentum=$k_4$](B1),
        (B1B1)--(M),
        (a1a1BIS)--[boson,line width=1.5pt](AA),
        (a1a1BIS)--[boson,line width=0.8pt,white](AA),
        (A1A1BIS)--[boson,line width=1.5pt](aabis),
        (A1A1BIS)--[boson,line width=0.8pt,white](aabis),
        (a1a1)--[boson,line width=1.5pt](aa),
        (a1a1)--[boson,line width=0.8pt,white](aa),
        (a1a1R)--[boson,line width=1.5pt](AAbis),
        (a1a1R)--[boson,line width=0.8pt,white](AAbis),
      };
    \end{feynman}
      \draw[line width=10.25pt,gray!20] (m) -- (M);
\draw[dashed] (0,-1.6) -- (0,1.6);
\end{tikzpicture}
\end{minipage}
\begin{minipage}{0.5\textwidth}
\centering
\begin{tikzpicture}[scale=1.4]
  \begin{feynman}[every blob={/tikz/fill=gray!30,/tikz/inner sep=2pt}, every dot={/tikz/minimum size=2.5pt}]
      \vertex (a) at (-2,1){$2$};
      \vertex(aa) at (-1.5,0.5);
      \vertex (aaa) at (-1.52,0.52);
      \vertex[blob] (m) at (-1,0){$\mathcal{M}$};
      \vertex (a1) at (-0.5,1.5){$1$};
      \vertex[dot] (a1a1) at (-0.8,0.6){};
      \vertex[dot] (A1A1BIS) at (0.8,0.6){};
       \vertex[dot] (a1a1BIS) at (-0.75,0.75){};
      \vertex (a1a1a1) at (-0.77,0.77);
      \vertex (b) at ( -2,-1){$3$};
      \vertex (bb) at (-1.25,-0.25);
      \vertex (b1) at (-0.5,-1.5){$4$};
      \vertex (b1b1) at (-0.75,-0.75);
       \vertex (A) at (2,1){$2$};
      \vertex (AA) at (1.5,0.5);
      \vertex (AAA) at (1.50,0.53);
      \vertex[blob] (M) at (1,0){$\mathcal{M}^*$};
      \vertex (A1) at (0.5,1.5){$1$};
      \vertex (B) at ( 2,-1){$3$};
      \vertex (BB) at (1.25,-0.25);
      \vertex (B1) at (0.5,-1.5){$4$};
      \vertex (B1B1) at (0.75,-0.75);
       \vertex (MMM) at (0,0);
             \draw[line width=10.25pt,gray!20] (m) -- (M);
         \vertex[dot] (c) at (-0.4,0){};
         \vertex[dot] (C) at (-0.6,0){};
          \vertex[dot] (cc) at (0.4,0){};
         \vertex[dot] (CC) at (0.6,0){};
           \vertex[dot] (A1A1) at (0.75,0.75){};
      \diagram* {
        (a) -- [momentum'=$k_2$](m),
        (m) --(a1a1),
        (a1)-- [momentum'=$k_1$](a1a1BIS)--(a1a1),
        (bb) -- [momentum'=$k_3$](b),
        (bb)--(m),
        (m) --(b1b1)--[momentum'=$k_4$] (b1),
        (M) -- (AA),
        (A)--[momentum=$k_2$](M),
        (A1) --[momentum=$k_1$](A1A1)-- (M),
        (M) -- (BB)--[momentum=$k_3$](B),
        (B1B1)--[momentum=$k_4$](B1),
        (B1B1)--(M),
         (m)--(c)--[edge label'=$\rho_1$](MMM),
         (a1a1)--[boson,line width=1.5pt](C);
         (c)--[boson,line width=1.5pt](A1A1);
         (a1a1)--[boson,line width=0.8pt,white](C);
         (c)--[boson,line width=0.8pt,white](A1A1);
         (M)--(MMM),
         (CC)--[edge label=$\bar{\rho}_1$](MMM),
         (CC)--[boson,line width=1.5pt](A1A1BIS),
         (cc)--[boson,line width=1.5pt](a1a1BIS),
         (CC)--[boson,line width=0.8pt,white](A1A1BIS),
         (cc)--[boson,line width=0.8pt,white](a1a1BIS),
      };
    \end{feynman}
\draw[dashed] (0,-1.6) -- (0,1.6);
\end{tikzpicture}
\end{minipage}
\caption{Diagrammatic representation of the contributions to the density matrix variation from primary (left panel) and secondary (right panel) radiation. The vector bosons are represented as double lines to indicate that their momenta have to be integrated over the infinitesimal strip~(\ref{strip}).\label{Fig:DMEvDiag}}
\end{figure}

The IREE can be obtained like in the previous section by computing the variation of ${\cal{D}}_\lambda$ under $\lambda\to\lambda+\delta\lambda$, taking now also  into account also the effect of
the IR cutoff on  real emission. The contribution of virtual loop momentum integrals is thus accompanied by that of integrals over the momentum of real radiation.  All integrals have to be performed over the infinitesimal strip $\delta\sigma$ defined in eq.~(\ref{strip}). Logarithmically enhanced terms only arise from the exchange of
 virtual or real gauge bosons between different external legs ($i\not = j$), like in Figure~\ref{Fig:DMEvDiag}.
 The effects and the corresponding diagrams can be divided into two classes. The first, in the left panel of Figure~\ref{Fig:DMEvDiag}, is given by
 {\it{primary}} radiation diagrams where vector bosons are exchanged between  the hard legs. 
 The second, in the right panel, is given by  {\it{secondary}} radiation diagrams where vector bosons connect to at least one real radiation leg.
 
 We will first consider  the effects of primary radiation. The virtual radiation integral gives the result already mentioned in eq.~(\ref{Eq:VarAmp1}), and, as we show in  Appendix~\ref{AppDL},  the result is exactly the same for the real radiation integral. The total variation from primary radiation is then
\begin{align}
\ds
\delta {\cal{D}}^{\alpha \bar{\alpha}}_\lambda=-\frac{1}{16\pi^2}\frac{\delta\lambda}\lambda\,\log\frac{E^2}\lambda
\sum\limits_{i=1^{c},2^{c},3,4}
\sum_{A} \bigg[
&(G^{A}_i)^{\alpha}_{\;\;\beta} \sum_{j\neq i} \big[ (G^{A}_j)^{\beta}_{\;\;\gamma}{\cal{D}}^{\gamma \bar{\alpha}}_\lambda 
+
(G_{j^c}^A)^{\bar\alpha}_{\;\;\bar\beta}{\cal{D}}^{\beta \bar{\beta}}_\lambda\big]\nonumber\\\ds
+&(G^{A}_{i^c})^{\bar\alpha}_{\;\;\bar\beta} \sum_{j\neq i} \big[ (G^{A}_j)^{\alpha}_{\;\;\beta}{\cal{D}}^{\beta \bar{\beta}}_\lambda 
+
(G_{j^c}^A)^{\bar\beta}_{\;\;\bar\gamma}{\cal{D}}^{\alpha \bar{\gamma}}_\lambda\big]
\bigg]\,.
    \label{Eq:VarDM}
\end{align}
The argument of the first sum, over the four external legs, collects the contributions of all the radiation emitted from the leg ``$i$'' of the amplitude and of the conjugated amplitude. A factor $1/2$ is included to avoid double-counting. Notice that both virtual and real radiation connecting one leg with itself is excluded from the sum, because, as we already mentioned, no enhancement arises from those diagrams.

We can now proceed as in the previous section, and use the charge conservation in eq.~(\ref{chcoDM}) to perform the sum over $j$ in eq.~(\ref{Eq:VarDM}). We find the IREE
\beq\label{dmIREEhigh}
\frac{d {\cal{D}}^{\alpha \bar{\alpha}}_\lambda}{d\log^2({E^2}/\lambda)}=-
{\cal{K}}^{\alpha \bar{\alpha}}_{\beta \bar{\beta}} \, {\cal{D}}^{\beta \bar{\beta}}_\lambda\,,
\eeq
with an evolution kernel that is the direct sum of universal terms for each external leg
\begin{eqnarray}\ds\label{KernelDMHIGH}
&&{\cal{K}}^{\alpha \bar{\alpha}}_{\beta \bar{\beta}} \overset{\; {\lambda\gg\MW^2}}{{=}}\;
\frac{1}{32\pi^2} 
\sum\limits_{i}\left[
\big[\sum_A G^{A}_i G^{A}_i\big]^{\alpha}_{\;\;\beta}\delta^{\bar\alpha}_{\bar\beta}
+
\delta^{\alpha}_{\beta}\big[\sum_A G^{A}_{i^c} G^{A}_{i^c}\big]^{\bar\alpha}_{\;\;\bar\beta}
+ 
2\,\sum_A (G^{A}_i)^{\alpha}_{\;\;\beta} (G^{A}_{i^c})^{\bar\alpha}_{\;\;\bar\beta}
\right]\nonumber\\
&&\hspace{29pt}\ds
\,=\;\;\;\,
\frac{g^2}{16\pi^2} 
\sum\limits_{i}\big[
c_i\,\delta^{\alpha_i}_{\beta_i}\delta^{\bar\alpha_i}_{\bar\beta_i}+\sum_{A=1,2,3} (T_i^A)^{\alpha_i}_{\;\;\beta_i} (T_{i^c}^A)^{\bar\alpha_i}_{\;\;\bar\beta_i}
\big]\big[\prod_{j\neq i}\delta^{\alpha_j}_{\beta_j}\delta^{\bar\alpha_j}_{\bar\beta_j}\big]\nonumber\\
&&\hspace{29pt}\ds
\,=\;\;\;\,\frac{g^2}{16\pi^2} 
\sum\limits_{i}\big[{{K}}_i\big]^{\alpha_i\bar\alpha_i}_{\beta_i\bar\beta_i}\big[\prod_{j\neq i}\delta^{\alpha_j}_{\beta_j}\delta^{\bar\alpha_j}_{\bar\beta_j}\big]
\,.
\end{eqnarray}
The kernel contains one term, provided by the \mbox{SU$(2)_L$} Casimir $c_i=t_i(t_i+1)$, which is proportional to the identity in the color indices of the density matrix tensor, plus a non-diagonal term constructed with the  \mbox{SU$(2)_L$} group generators matrices $T^A_i$ of the external legs. Notice that the contribution of the \mbox{U$(1)_Y$} hypercharge generator cancels. 

There is one peculiarity of eq.~(\ref{KernelDMHIGH}) that is worth emphasizing. The semi-inclusive cross-sections we are interested in are the diagonal entries ($\alpha=\bar\alpha$) of the density matrix, with no sum performed over the gauge group index $\alpha$ of the scattering particles.\footnote{This is true only in a basis where the gauge indices $\alpha_i$ label the on-shell SM particles, while for the calculation of di-boson cross-sections we work in a different basis. See Section~\ref{Sec:Dibos} and Appendix~\ref{AppEWBasis} for details.} However one can also consider {\it{inclusive}} cross-sections, where the sum over the gauge index $\alpha_i$ is performed for one or several external legs. By setting $\bar\alpha_i=\alpha_i$ and summing over $\alpha_i$, the \mbox{SU$(2)_L$} generators in eq.~(\ref{KernelDMHIGH}) recombine to form the Casimir operator, and the contribution to the evolution kernel from leg ``$i$'' cancels. We thus find that, at DL accuracy, the cancellation between real and virtual IR effects in inclusive cross-sections occurs  on a leg-by-leg basis. Namely, the effects of soft/collinear emissions associated to each individual leg cancel in the cross-section (and in the entire density matrix) for processes that are inclusive over the color of the corresponding particle. This result is stronger than the KLN theorem, which foresees a cancellation only when summing over the color of all legs. The reason for the added strength is that we are here considering radiation that is both soft and collinear. Notice  however that  fully inclusive observables of practical relevance can only involve summation on the color of the final state particles. This retains the IR effects associated with  the colliding particles in the initial state (e.g., two left-handed leptons $\ell^+_L\ell^-_L$) which are not \mbox{SU$(2)_L$} singlets.  The resulting non-cancellation of IR effects in ``fully-inclusive'' cross-sections, coincides with the result  of Ref.~\cite{Ciafaloni:2000rp}.

So far we have ignored the secondary radiation diagrams, depicted in the right panel of Figure~\ref{Fig:DMEvDiag}. We show now that their contribution vanishes, giving full justification to eq.~(\ref{KernelDMHIGH}). Secondary radiation diagrams correspond to the effect of the $\lambda$ cutoff variation on virtual or real vector bosons attached to one of the intermediate ``$\rho$'' particles in the definition of the density matrix~(\ref{Eq:DensityMatrixDefi}). Clearly these effects are potentially enhanced only if the intermediate particle is relatively hard, such that a significant separation is present between the IR cutoff $\lambda$ and the scalar product between the intermediate particle and the external leg momenta. We thus start considering vector bosons attached to the hardest intermediate particle, with gauge index ``$\rho_1$'' as in the figure. The density matrix is inclusive over the color of the intermediate particle. However we can momentarily define an ``extended'' density matrix ${\cal{D}}^{\alpha\,;\rho_1\, \bar{\alpha}\,;\bar\rho_1}_\lambda$ with labels $\rho_1$ and $\bar\rho_1$ for the gauge indices of the amplitude and of the conjugate amplitude, as in the figure. The actual density matrix is eventually obtained by first setting $\rho_1=\bar\rho_1$ and then  summing. The effect on the extended density matrix variation of all the radiation emitted from $\rho_1$ and $\bar\rho_1$ can be written in a form similar to eq.~(\ref{Eq:VarDM}) and then simplified  using the analog of eq.~(\ref{chcoDM}) for the extended density matrix. The resulting  contribution to the evolution kernel from the intermediate $\rho_1$ leg  is the analog of that from the hard external legs in eq.~(\ref{KernelDMHIGH}). But this contribution cancels out in the evolution of the actual density matrix, which is inclusive over the $\rho_1$ leg, because of the previously explained  leg-by-leg cancellation mechanism. The argument can of course be repeated for the diagrams involving the second hardest intermediate particle, showing,  as anticipated, that all the secondary radiation diagrams can be ignored in the calculation of the evolution kernel.

It is straightforward to adapt the previous results to the regime $\lambda\ll\MW^2$, in which only the exchange of photons contributes to the evolution, as discussed in the previous section. By specifying eq.~(\ref{KernelDMHIGH}) to the Abelian \mbox{U$(1)_Q$} group we immediately find that the kernel vanishes, owing to the well-known cancellation between real and virtual IR effects in QED. For the calculation of the physical ($\lambda\to0$) density matrix, and in turn of the semi-inclusive cross-section, we thus only need to solve the IREE with the $\lambda\gg\MW^2$ kernel (\ref{KernelDMHIGH}), down to $\lambda=\MW^2$. 

For $\lambda=E^2$ the hard density matrix~(\ref{Eq:DensityMatrixDefi}) is well-approximated by its tree-level expression, which serves as the initial condition for the evolution
\beq\label{dminc}
{\cal{D}}^{\alpha \bar{\alpha}}_{E^2} = \mathcal{B}^{\alpha}(\mathcal{B}^{\bar{\alpha}})^*\,.
\eeq
The kernel is the direct sum of tensors, denoted as $K_i$ in eq.~(\ref{KernelDMHIGH}), each acting on the pair $\alpha_i, \bar\alpha_i$ associated to the $i$-th external particle. Therefore the solution of the IREE reads
\beq\label{sidm}
{\cal{D}}^{\alpha \bar{\alpha}}_{\rm{si}}\equiv {\cal{D}}^{\alpha \bar{\alpha}}_{\MW^2}=\bigg\{\prod_i
\bigg[\exp\big[
{-\frac{g^2}{16\pi^2} K_i\log^2(E^2/\MW^2)}
\big]\bigg]^{\alpha_i\bar\alpha_i}_{\beta_i\bar\beta_i}\bigg\}
\mathcal{B}^{\beta}(\mathcal{B}^{\bar{\beta}})^*\,,
\eeq
where the ``$_{\rm{si}}$'' subscript denotes the density matrix of the semi-inclusive process, with no upper cut on the real radiation hardness. The explicit form of the $K_i$ exponentials in the above equation is reported in eq.s~(\ref{dfact}) and~(\ref{tfact}) for external legs in the doublet and triplet \mbox{SU$(2)_L$} representations. Applications of eq.~(\ref{sidm}) to specific processes are shown in Sections~\ref{Sec:Diferm} and~\ref{Sec:Dibos}.

We have defined the density matrix~(\ref{Eq:DensityMatrixDefi}) allowing for the presence of an upper cutoff ${\cal{H}}$ on the real radiation, but this played no role in the previous discussion because this cutoff is effectively absent (${\cal{H}}\sim E^2$) in our definition of semi-inclusive processes. In exclusive processes  we instead set ${\cal{H}}=\MW^2$, namely we veto real radiation particles with hardness above $\MW^2$. 
Obviously, for $\lambda\gg\MW^2$ this upper cut is in contradiction with the IR cutoff in eq.~(\ref{reg}) on the radiation phase-space. Therefore in the density matrix for the exclusive process no real radiation is present and  in the $\lambda\gg\MW^2$ regime the result simply equals the square of the hard amplitude in eq.~(\ref{hardamp}). The evolution up to $\lambda=\MW^2$ can thus be obtained from the hard amplitude evolution~(\ref{ampBC1}) we obtained in the previous section, or easily re-derived by dropping the terms in eq.~(\ref{Eq:VarDM}) (namely, the second and the third) that are due to real radiation. The contribution of real radiation is instead restored for $\lambda\ll\MW^2$ and the evolution stops due to the cancellation between virtual and real QED radiation as previously explained. The physical ($\lambda\to0$) density matrix for exclusive processes can thus be written in a simple closed form as
\beq\label{exdm}
{\cal{D}}^{\alpha \bar{\alpha}}_{\rm{ex}}=\exp\left[
 -\sum_{i} \frac{g^2 c_i+g^{\prime\,2}y_i^2}{16\pi^2}\, \log^2 (E^2/\MW^2) \right] \mathcal{B}^{\alpha}(\mathcal{B}^{\bar{\alpha}})^*\,.
\eeq
In Sections~\ref{Sec:Diferm} and~\ref{Sec:Dibos} we employ this formula to compute exclusive di-fermion and di-boson production cross-sections, and discuss the need of supplementing it with fixed-order single-logarithmic terms, from Ref.~\cite{Denner:2000jv,Denner:2001gw}.
 
Before concluding this section it is worth commenting on the experimental definition of the semi-inclusive and exclusive processes, and on the perspectives for their actual experimental detectability. The semi-inclusive process is characterized by two central (specifically, emitted from $30$ to $150$ degrees from the beam line) energetic particles of specific EW color and flavor. In particular we will require them to carry a total center of mass energy above $85\%$ of the VHEL $E_{\rm{cm}}$, enforcing in this way the upper cut on the total radiation $4$-momentum required for the definition of the hard density matrix as discussed below eq.~(\ref{Eq:DensityMatrixDefi}). The two particles can be accompanied by the radiation of EW bosons, photons, or any other soft particle. 

Notice that in our calculation at the DL order we could ignore all the effects of collinear (rather than soft-collinear) radiation, which emerge at the single logarithm. On the other hand, the single logarithms associated with low-virtuality (below $\MW$) photon splittings are much larger than $\log E^2/\MW^2$. In particular, the emission of real photons that are energetic but collinear to a light charged hard particle (e.g., an electron or a muon) with mass $m_\ell$ produces terms proportional to $\log E^2/m_\ell^2$. By the KLN theorem these terms will be canceled by the corresponding virtual contributions, but only in suitably-defined observables that recombine the emitted photons in the experimental definition of the hard particle $4$-momentum. With a lower threshold of order $\MW$ on the energy of the photons to be recombined, the net effect on our prediction should be of the order of a single EW logarithm $\log E^2/\MW^2$. A more detailed assessment of this aspect, and of the possible interplay between the QED and the EW bosons collinear emissions, requires the inclusion of single logarithms and goes beyond the scope of the present paper. Similar considerations hold for the collinear emission of QCD gluons to be collected into jets, in the case of colored final states. 

Up to the caveats outlined above, there are good perspectives for the actual direct experimental detectability of semi-inclusive cross-sections. The situation is arguably more problematic for the exclusive cross-section. In exclusive final states, we require the presence of the two hard particles defined as above, plus the absence of any massive vector boson (since ${\mathfrak{h}}(q)>q^2=m^2$, as discussed at the end of the previous section), or photons above the hardness upper threshold $\MW^2$. However, it is experimentally impossible to impose this radiation veto strictly because the limited coverage of the detector in the forward and backward regions will not allow to tag EW bosons or photons that are collinear to the beam. Furthermore our definition of the exclusive cross-section is problematic in the case of QCD-colored final states. Indeed if the upper cut ${\mathfrak{h}}(q)<\MW^2$ had to be imposed also on gluon radiation, QCD effects should be included in the exclusive density matrix evolution (but not in the semi-inclusive one, where they cancel because of color inclusivity), resulting in a large QCD Sudakov suppression factor in eq.~(\ref{exdm}). This factor is as small as $\exp[-\alpha_s/(4\pi)(8/3) \log^2E_{\rm{cm}}^2/\MW^2]\sim 0.03$ for di-quark final states at the highest VHEL energy $E_{\rm{cm}}=30$~TeV, entailing a strong suppression of the cross-section. Avoiding this suppression requires a definition of the exclusive cross-section with a higher threshold on the QCD radiation. We will further comment in the Conclusions on the limitations of the exclusive cross-section definition employed in this paper.

\subsection{Di-fermion production}\label{Sec:Diferm}
The first process we investigate is the production of a highly energetic pair of fermions
\beq\label{Eq:DifermionGen}
    \ell^+(k_1)\,\ell^-(k_2) \rightarrow \bar{f}(k_3) \, g (k_4) +X \,,
\eeq
where $f$ and $g$ can be one of the six quarks, a lepton $\ell^\prime\neq\ell$ or a neutrino $\nu_{\ell^\prime}$. We do not discuss explicitly the final states with the same leptonic flavor as the initial state, $\ell^\prime=\ell$, but these processes will be employed for the muon collider sensitivity projections in Section~\ref{Sec:3}. As previously discussed, the final state is characterized (both for exclusive and semi-inclusive processes) by an invariant mass for the $(\bar{f},g)$ pair that is almost equal to the center of mass energy $E_{\rm{cm}}$ of the colliding leptons and by central scattering angle $\theta_*\in[30^{\circ},150^{\circ}]$. Here $\theta_*$ is the angle between the incoming $\ell^+$ and the final anti-fermion $\bar{f}$ in the lab frame. Notice that $\theta_*$ almost coincides with the scattering angle in the center of mass frame of the hard process, because of the tight cut on the invariant mass of the $(\bar{f},g)$ pair.

In order to resum the DL it is convenient to organize the calculation of the cross-section in terms of amplitudes and density matrices whose external legs are canonical irreducible representations of the EW group. This is trivial to achieve for the di-fermion process because the helicity eigenstates of quarks and leptons in the massless limit do indeed transform as canonical representations (doublets and singlets, with specific hypercharge), reported for completeness in Appendix~\ref{AppEWBasis}. Furthermore, since we restrict our attention to inelastic processes $\ell^\prime\neq\ell$, the only sizable helicity amplitudes are those with the same chirality $\chi_I$ ($\chi_O$) for the two incoming (outgoing) fermions, corresponding to helicities ${\bar\psi}_{+1/2}{\psi}_{-1/2}$ for $\chi=L$ and ${\bar\psi}_{-1/2}{\psi}_{+1/2}$ for $\chi=R$. The dominance of such amplitudes holds in the SM because of the vector-like structure of gauge interaction, and it will be preserved by the $4$-fermions new physics contact interaction operators we will study in Section~\ref{Sec:3}. We thus have to deal with four polarized cross-sections for each di-fermion production process, labeled by $\chi_I\chi_O=LL,LR,RL,RR$. Each such cross-section will be obtained from the diagonal $\alpha=\bar\alpha$ entries of the density matrices of Section~\ref{Sec:IREE}, times the appropriate phase-space factors.

\subsubsection*{Exclusive processes}

Exclusive cross-sections are readily obtained from eq.~(\ref{exdm}), and take the form
\beq\label{excpar}
\frac{d\sigma_{\rm{ex}}}{d\cos\theta_*}=e^{\rm{DL}}\frac{d\sigma_B}{d\cos\theta_*}\,,
\eeq
in terms of the corresponding Born-level differential cross-sections. The Double Log exponent DL is of order $g^2/16\pi^2\log^2(E_{\rm{cm}}^2/\MW^2)$, which ranges from $0.14$ at $E_{\rm{cm}}=3$~TeV up to $0.25$ ($0.38$) for $E_{\rm{cm}}=10(30)$~TeV, times the sum of the four SU$(2)$ Casimir of the external legs. For $LL$ chirality processes this factor is as large as $4\times 1/2(1/2+1)=3$, showing that DL resummation is mandatory at VHEL energies $E_{\rm{cm}}\geq10$~TeV, at least for this chirality. Double logs are still considerable for $LR$ and $RL$ chirality, while they get smaller in the $RR$ configuration because $g^{\prime\,2}\sim g^2/4$. Resummation might instead not be necessary for $E_{\rm{cm}}=3$~TeV. However it will still be needed to include the effects of radiation at fixed order since we aim, eventually, at theoretical predictions with percent-level accuracy.

\begin{table}
\begin{center}
 \begin{tabular}{||c|| c | c | c || c | c | c || c | c | c ||} 
 \cline{2-10}
 \multicolumn{1}{c|}{}&\multicolumn{3}{|c||}{3 TeV}&\multicolumn{3}{|c||}{10 TeV}&\multicolumn{3}{c|}{30 TeV}\\
 \cline{2-10}
 \multicolumn{1}{c|}{}&\rule[-6pt]{0pt}{18pt} DL& $e^{\rm{DL}}\hspace{-3pt}-\hspace{-2pt}1$& SL$(\frac\pi{2})$ &DL& $e^{\rm{DL}}\hspace{-3pt}-\hspace{-2pt}1$& SL$(\frac\pi{2})$&DL& $e^{\rm{DL}}\hspace{-3pt}-\hspace{-2pt}1$& SL$(\frac\pi{2})$ \\ \hline 
 $\ell_L\rightarrow \ell^\prime_L $&-0.46&-0.37&0.25& -0.82&	-0.56&	0.33&	-1.23&	-0.71&	0.41 \\ \hline 
  $ \ell_L\rightarrow q_L$&-0.44&-0.36&0.25& -0.78&	-0.54&	0.34&	-1.18&	-0.69&	0.42\\ \hline $\ell_L\rightarrow e_R$&-0.32&-0.27&0.13& -0.56&	-0.43&	0.17&	-0.85&	-0.57&	0.21 \\\hline 
  $\ell_L\rightarrow u_R $& -0.27&-0.24&0.11&  -0.48	&-0.38&	0.15&	-0.72&	-0.51&	0.18 \\\hline $\ell_L\rightarrow d_R $&-0.24&-0.21&0.10& -0.43&	-0.35&	0.13&	-0.64&	-0.47&	0.16 \\\hline 
  $\ell_R\rightarrow \ell^\prime_L $&-0.32&-0.27&0.13 &-0.56&	-0.43&	0.17&	-0.85&	-0.57&	0.21  \\\hline $ \ell_R\rightarrow q_L $&-0.30&-0.26&0.12&-0.53	&-0.41	&0.16&	-0.79&	-0.55	&0.21\\\hline 
  $\ell_R\rightarrow \ell^\prime_R$& -0.17&-0.16&0.07 &-0.30	&-0.26&	0.09&	-0.46	&-0.37&	0.12 \\\hline
 $\ell_R\rightarrow u_R$&-0.12&-0.12&0.05& -0.22&	-0.20&	0.07&	-0.33&	-0.28&	0.08\\\hline
$ \ell_R\rightarrow d_R$&-0.09&-0.09&0.04& -0.17&-0.16&	0.05&	-0.25&	-0.22&	0.06\\\hline
\end{tabular}
\caption{Double and single logarithmic corrections to the exclusive processes $\ell^+ \ell^- \rightarrow \bar{f} f$. The single-logarithmic corrections are evaluated at $\theta_*=\pi/2$. \label{Tab:DL4Fermions}}
\end{center}
\end{table}

The DL Sudakov exponents in eq.~(\ref{excpar}) are listed in Table~\ref{Tab:DL4Fermions}. The processes are labeled taking into account that electric charge conservation enforces $g=f$ in eq.~(\ref{Eq:DifermionGen}), since a charge mismatch cannot be compensated by the emission of charged $W$ bosons, which is forbidden in exclusive processes. The table also reports single logarithm (SL) contributions computed at the fixed one loop order, which we extract from Ref.s~\cite{Denner:2000jv}.\footnote{Two loops NLL results for four-fermion processes are also available in \cite{Denner:2006jr,Kuhn:2001hz}.} Specifically, we employ the general formulae of Ref.s~\cite{Denner:2000jv} to compute the 1-loop log-enhanced cross-section, we subtract the corresponding DL and normalize to the Born cross-section. We also subtract the single logarithms from the Renormalization Group evolution, because we decided to compute the Born amplitude with the EW couplings at the hard scale $E_{\rm{cm}}$.\footnote{The calculation is similar to the one performed by two of us in Ref.~\cite{Torre:2020aiz}. We refer the reader to Section~2.3 of~\cite{Torre:2020aiz} for additional details, concerning in particular the inclusion of non-log-enhanced angular-dependent terms.} Notice that the threshold for photon recombination into the hard final state particles matters at the single-logarithmic order. Here we assume a scale of recombination of order $\MW$, for which the SL terms can be easily obtained by adding a fictitious photon mass $m_{\gamma}=\MW$ to the calculations of Ref.s~\cite{Denner:2000jv,Denner:2001gw}. The SL terms obtained in this way can be used for ``improved'' theoretical predictions
\beq\label{excparplus}
\frac{d\sigma_{\rm{ex}}^{{\rm{SL}}_1}}{d\cos\theta_*}=e^{\rm{DL}}(1+{\rm{SL}(\theta_*)})\frac{d\sigma_B}{d\cos\theta_*}\,,
\eeq
that include single logarithms at fixed  $1$-loop order. We see in Table~\ref{Tab:DL4Fermions} that the SL contributions are relatively large. It is unclear whether they require resummation or if including them at fixed order (definitely higher than 1-loop, if we target $1\%$ accuracy) is sufficient.

Notice that, unlike double logarithms, the single logarithm contributions  are not proportional to the Born-level amplitude of the same scattering process. Namely the amplitudes of the neutral-current processes in Table~\ref{Tab:DL4Fermions} receive SL corrections that are proportional to Born charged-current amplitudes. Therefore it should be kept in mind the SL terms in eq.~(\ref{excparplus}), which we normalized to the Born cross-section of the process, depend on the ratio between charged and neutral current Born amplitudes. We evaluated the amplitude ratio within the SM to produce the results in Table~\ref{Tab:DL4Fermions}. However the amplitude ratio depends on the new physics contact interactions we consider in Section~\ref{Sec:3}, entailing a dependence of the SL terms on the new physics parameters. This is not the case for the double logarithms, which are completely universal and insensitive to short-distance physics. The single logarithms also carry a non-trivial dependence on the scattering angle $\theta_*$, as explicitly indicated in eq.~(\ref{excparplus}). In Table~\ref{Tab:DL4Fermions} they are evaluated at central angle $\theta_*=\pi/2$, where they are always positive. They can become negative, and typically increase in magnitude, in the forward and backward scattering regions, which we however exclude with the central cut $\theta_*\in[30^{\circ},150^{\circ}]$. Finally, notice that the SL terms are affected by the sizable mass of the top quark, which we do include in the $t\bar{t}$ production process. 

\begin{figure}
	\centering
	\includegraphics[scale=.5]{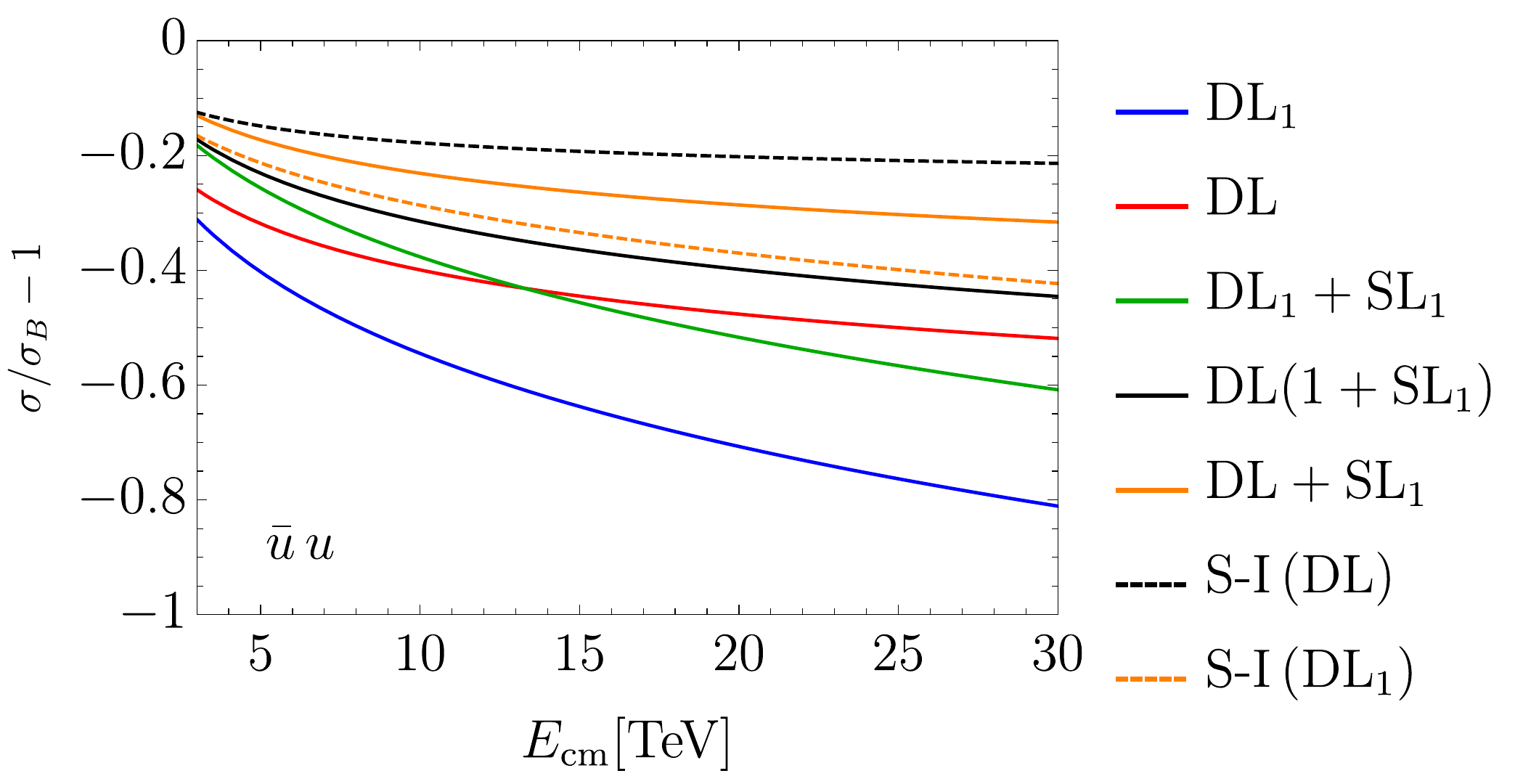}
	\caption{Impact of radiative corrections on the production of two up quarks at the VHEL. The solid lines represent different predictions for the exclusive cross-section. The dashed lines are double-logarithm semi-inclusive cross-sections resummed (in black) or at one loop (in orange). 
	\label{Fig:XSDifermion}}
\end{figure}

The impact of EW radiation effects on the total (unpolarized) cross-section in the central region, relative to the Born, is displayed in Figure~\ref{Fig:XSDifermion} as a function of $E_{\rm{cm}}$. The production of two light up-type quarks is considered for illustration, but the results for the other final states are similar. The blue line is the one-loop DL prediction without exponentiation, while in red we report the resummed DL prediction in eq.~(\ref{excpar}). The green line (labeled DL$_1+$SL$_1$) represents the fixed-order one loop DL plus SL, while in black we report the SL-improved prediction in eq.~(\ref{excparplus}). The dashed lines are semi-inclusive cross-sections computed below. We notice a significant cancellation between double and single logarithmic terms. However this cancellation is not expected to be structural and to survive at higher orders in perturbation theory. 

We do not try to assign theoretical uncertainties to our predictions. However an upper bound can be obtained by considering the orange line in the figure, in which the resummed DL are combined additively with the SL (i.e., as $e^{\rm{DL}}+{\rm{SL}}$), rather than multiplicatively. An alternative estimate of the uncertainties could be obtained by varying the scale of the EW couplings employed for the evaluation of the radiation terms DL and SL. Varying this scale from $\MW$ (which we employ for our predictions) to $E_{\rm{cm}}$, the relative change of the radiation effects is rather small, typically at the $10\%$ level or less.

\subsubsection*{Semi-inclusive processes}

The semi-inclusive cross-sections are the diagonal $\alpha=\bar\alpha$ entries of the semi-inclusive density matrix in eq.~(\ref{sidm}), with the appropriate $K_i$ exponential factors for each external leg. The factors only depend on the \mbox{SU$(2)_L$} quantum numbers of the legs and not of their hypercharge (and QCD color). They are provided by eq.~(\ref{dfact}) for $L$-chirality external external legs (which transform as doublets or conjugate-doublets) and they are trivial for the $R$-handed singlets. Notice that eq.~(\ref{dfact}) (and the same is true for the triplet exponential factor~(\ref{tfact})) does not mix diagonal with off-diagonal entries of the density matrix. Namely if we set $\alpha=\bar\alpha$ we obtain a tensor that is diagonal in $\beta$ and $\bar\beta$. Therefore the DL resummed cross-sections, collected in a vector $d\vec\sigma_{\rm{si}}$, are linear combinations of the Born cross-sections $d\vec\sigma_B$. We express this relation as 
\beq\label{sieqdif}
\frac{d\vec\sigma_{\rm{si}}}{d\cos\theta_*}=e^{\rm{DL}}\cdot\frac{d\vec\sigma_B}{d\cos\theta_*}\,,
\eeq
where the Double-Logarithm terms ``DL'' are now matrices connecting the Born cross-sections of different processes unlike for exclusive processes~(\ref{excpar}).

For an explicit illustration of the semi-inclusive cross-section calculation, and of the main features of the result, we consider the $RL$-chirality production processes. In this case, we have
\beq\label{RLxs}
d \vec{\sigma}_{\rm{si}} \hspace{-2pt}=\hspace{-2pt} \begin{pmatrix} d\sigma_{\rm{si}} (\ell^+_{R} \ell^-_{R}\hspace{-3pt}\rightarrow\hspace{-2pt} \bar{u}_L u_L) \\
	d\sigma_{\rm{si}} (\ell^+_{R} \ell^-_{R}\hspace{-3pt}\rightarrow\hspace{-2pt} \bar{u}_L d_L) \\
	d\sigma_{\rm{si}} (\ell^+_{R} \ell^-_{R}\hspace{-3pt}\rightarrow\hspace{-2pt} \bar{d}_L u_L)\\
	d\sigma_{\rm{si}} (\ell^+_{R} \ell^-_{R}\hspace{-3pt}\rightarrow\hspace{-2pt}  \bar{d}_L d_L) \end{pmatrix},
\;\;\;
d \vec{\sigma}_{B} \hspace{-2pt}=\hspace{-2pt} \begin{pmatrix} d\sigma_{B} (\ell^+_{R} \ell^-_{R}\hspace{-3pt}\rightarrow\hspace{-2pt} \bar{u}_L u_L) \\
	0\\
	0\\
	d\sigma_{B} (\ell^+_{R} \ell^-_{R}\hspace{-3pt}\rightarrow\hspace{-2pt}  \bar{d}_L d_L) \end{pmatrix}
\hspace{-2pt}=\hspace{-2pt}	d\sigma_{B} (\ell^+_{R} \ell^-_{R}\hspace{-3pt}\rightarrow\hspace{-2pt} \bar{q}_L q_L) \hspace{-2pt}\begin{pmatrix} 1 \\
	0\\
	0\\
	1 \end{pmatrix}
	,
\eeq
where ``$u$'' and ``$d$'' denote here the up and down components of a $L$-handed fermion doublet. The exponentiated DL matrix reads
\beq\label{DLM}
e^{\rm{DL}}= \frac14\,e^{-\mathcal{L}}
\begin{pmatrix}
4\cosh^2 (\mathcal{L}/2) & 2\sinh (\mathcal{L}) & 2\sinh (\mathcal{L}) & 4\sinh^2(\mathcal{L}/2) \\
2\sinh(\mathcal{L}) & 4\cosh^2(\mathcal{L}/2)  & 4\sinh^2(\mathcal{L}/2) & 2\sinh (\mathcal{L}) \\
2\sinh (\mathcal{L}) & 4\sinh^2(\mathcal{L}/2)  & 4\cosh^2(\mathcal{L}/2)  & 2\sinh(\mathcal{L}) \\
4\sinh^2 (\mathcal{L}/2) & 2 \sinh (\mathcal{L}) & 2\sinh (\mathcal{L}) & 4\cosh^2(\mathcal{L}/2)  \\
\end{pmatrix}\,,
\eeq
where $\mathcal{L}={g^2}/16\pi^2\log^2 (E^2_{cm}/\MW^2)$. 

We see that DL effects induce a non-vanishing cross-section for charged processes with $g\neq{f}$ in eq.~(\ref{Eq:DifermionGen}), such as $\bar{u}_L d_L$ and $\bar{d}_L u_L$ production. Clearly this stems from the emission of real soft $W$-bosons, which is allowed in the semi-inclusive final state. Such charged cross-sections are proportional to the Born cross-section for the corresponding neutral ($\bar{u}_L u_L$ or $\bar{d}_L d_L$) processes, and they are not drastically smaller than those because the double-logarithm is sizable at VHEL energies. Therefore they can be measured bringing additional sensitivity to the charge-preserving Born amplitudes and to the corresponding short-distance new physics effects. The interplay with short-distance physics is even more interesting for the $LL$-chirality process. In that case, $\vec{\sigma}_{\rm{si}}$ is a $16$-dimensional vector that contains $4$ observable ($\ell^+\ell^-$-initiated) processes with final states $\bar{u}_L u_L$, $\bar{u}_L d_L$, $\bar{d}_L u_L$ and $\bar{d}_L d_L$. DL is a $16\times16$ matrix that relates the observable processes to $16$ Born amplitudes, among which those (like, e.g., ${\bar\nu}_\ell \ell^-\to \bar{u}_L d_L$) that are sensitive to new charged current interactions. We can thus probe the latter interactions even with the neutral $\ell^+\ell^-$ VHEL collisions.

The black dashed lines in Figure~\ref{Fig:XSDifermion} quantify the impact of the EW radiation effects on the neutral semi-inclusive cross-sections relative to the Born predictions. The effects are smaller than for exclusive cross-sections, as qualitatively expected owing to the partial cancellation between virtual and real radiation. While this suggests that resummation might play a less relevant role in semi-inclusive predictions, we point out that one-loop double logarithms are insufficient for accurate predictions. This is shown in the purple dashed line in the figure, which is obtained by truncating at the one-loop order the exponentiated DL matrix. It would be interesting to study the impact of single logarithms on the predictions. This could be achieved by combining the single radiative logs from Ref.s~\cite{Denner:2000jv,Denner:2001gw} with the factorized formulas for real emissions in Ref.s~\cite{Chen:2016wkt,Cuomo:2019siu} (which however would have to be extended to include also the soft radiation region), but is left to future work.

As a final technical note, we remark that the DL matrix is negative semi-defined with a single vanishing eigenvector that corresponds to the ``fully-inclusive'' cross-section, further averaged over the \mbox{SU$(2)_L$} color of the initial states. Specifically the vanishing eigenvector of eq.~(\ref{DLM}) is $(1,1,1,1)^t$, which corresponds to the sum of the cross-sections over the \mbox{SU$(2)_L$} gauge indices of the final states. Therefore in this case the double logarithmic effects cancel on the ``fully-inclusive'' cross-section, in accordance with the KLN theorem since the right-handed initial leptons are \mbox{SU$(2)_L$} singlets. Clearly this does not happen for the $LL$-chirality processes (nor for $LR$-chirality) and the average over leptons and neutrinos in the initial states would be necessary for the cancellation. The vanishing eigenvalue controls the behavior of the DL exponential at asymptotically high energies. In the case of eq.~(\ref{DLM}), we have
\beq
e^{\rm{DL}}\;
\overset{E_{\rm{cm}}\to\infty}{\longrightarrow}\frac14\; \begin{pmatrix}
1 & 1 & 1 & 1 \\
1 &1 & 1 & 1 \\
1 & 1 & 1 & 1 \\
1 & 1 & 1 & 1 \\
\end{pmatrix}\,,
\eeq
and all the semi-inclusive cross-section listed in eq~(\ref{RLxs}) become equal. Notice however this only holds at asymptotic energies, way above the VHEL energies. Cross-sections equality becomes a reasonable (better than order-one) approximation only for if $g^2/16\pi^2\log^2(E_{\rm{cm}}^2/\MW^2)$ is as large as $\sim1.5$, i.e. $E_{\rm{cm}}\gtrsim 10000$~TeV.

\subsection{Di-boson production}\label{Sec:Dibos}

We now turn to the production of two energetic vector or Higgs bosons. We are interested in reactions that are not power-like suppressed at high energy, therefore we restrict our attention to  ``longitudinal'' processes entailing the production of zero-helicity $W$ and $Z$ bosons and Higgs, and to ``transverse'' di-boson processes where the $W$ and the $Z$ (or, the photon) have $\pm1$ helicities. Indeed the ``mixed'' longitudinal/transverse production processes are suppressed by $\MW/E_{\rm{cm}}$ at the amplitude level, as readily understood (see e.g.~\cite{Borel:2012by,Cuomo:2019siu}) by combining the Goldstone Boson Equivalence Theorem with the selection rules associated with the \mbox{SU$(2)_L\times$U$(1)_Y$} SM group.

The new physics interactions we consider in Section~\ref{Sec:3} only affect longitudinal di-boson production cross-sections, which thus play the role of the signal in our analysis. We nevertheless also need the transverse cross-sections for an estimate of the background. We discuss the calculation of the (exclusive and semi-inclusive) cross-sections for the two type of processes in turn.

\subsubsection*{Longitudinal di-boson}

We consider the production, out of $\ell^+\ell^-$, of one of the following hard final states
\beq\label{2bproc}
W_0^+W_0^-\,,\;\;\;\;\;Z_0 h\,,\;\;\;\;\;W_0^\pm Z_0\,,\;\;\;\;\;W_0^\pm h\,,
\eeq
where the subscript ``$_0$'' refers to the helicity of the massive vectors, and ``$h$'' denotes the physical Higgs particle. Obviously only the first  two  final states can be produced in an exclusive process, while the latter ones require the emission of at least one charged $W$ and therefore they only occur at the semi-inclusive level. Notice that the ones listed above are the only hard final states with longitudinal bosons and Higgs that can be produced by soft EW bosons radiation out of sizable Born-level $2\to2$ cross-sections. Therefore they are the only longitudinal di-boson processes that can be considered for precise VHEL measurements in the high-energy regime.

At energies much above $\MW$, the adequate description of longitudinally-polarized massive vectors is provided by the charged and neutral Goldstone boson scalars $\pi^\pm$ and $\pi_0$ (see Appendix~\ref{AppEWBasis}). Together with the Higgs, they form a canonical  \mbox{SU$(2)_L\times$U$(1)_Y$} doublet $H$ with $1/2$ hypercharge, reported in eq.~(\ref{Lgauge}). We thus need to consider amplitudes and density matrices associated with the hard processes
\begin{eqnarray}\ds
&\ell^+_{-1/2}(k_1)\ell^-_{+1/2}(k_1) \to {\bar{H}}(k_3,\,\alpha_3^{\bar{\rm{d}}})\, H(k_4,\,\alpha_4^{\rm{d}}) \,,\nonumber\\
&\ds{{\bar{\ell}}}_{+1/2}(k_1,\,\alpha_1^{\bar{\rm{d}}})\ell_{-1/2}(k_2,\,\alpha_2^{\rm{d}}) 
\to {\bar{H}}(k_3,\,\alpha_3^{\bar{\rm{d}}})\, H(k_4,\,\alpha_4^{\rm{d}}) \,,\label{procdbl}
\end{eqnarray}
for, respectively, $L$-handed and $R$-handed production.\footnote{The production from opposite-chirality leptons is negligible, both in the SM and in the presence of the new contact interactions we investigate in the following section.} For the gauge group indices we employ the same notation as in eq.~(\ref{hardamp}), supplemented by the superscripts $^{{\rm{d}}}$ ($^{\bar{\rm{d}}}$) to indicate that the indices belong to the doublet (conjugate-doublet) representation. With a slight abuse of notation we are denoting as $\ell_{-1/2}=(\nu_{\ell,-1/2},{\ell^-}_{\hspace{-7pt}-1/2})^t$ the lepton doublet with $-1/2$ helicity and with $\ell_{+1/2}$ the conjugate-doublet with helicity $+1/2$. Notice that final states with two $H$ or two $\bar{H}$ need not to be included because they are power-like suppressed at high energy by hypercharge conservation.

The relevant density matrices are obtained as a straightforward application of the results in Section~\ref{Sec:IREE}. The need for employing $H$ and $\bar{H}$ as external states does not pose any additional difficulty (relative to the di-fermion processes) in the evaluation of exclusive cross-sections. That is  because the double logs are mere multiplicative factors in front of the Born-level density matrix~(\ref{exdm}). Therefore the exclusive cross-sections still take the form of eq.~(\ref{excpar}) and  are proportional to the corresponding Born-level predictions. For the semi-inclusive cross-section, we can proceed as for di-fermions in the determination of the $K_i$ exponential factors, using in particular eq.~(\ref{dfact}) which also holds in the present case because $H$ and $\bar{H}$ are doublets. However in order to apply eq.~(\ref{sidm}) we must first express the ${\cal{D}}^{\alpha \bar{\alpha}}_{\rm{si}}$ density matrix, which is written in the isospin basis ($H$ and $\bar{H}$), in the physical basis of the charge and CP eigenstates
$h$, $Z_0=\pi_0$ and $W_0^\pm=\pi^\pm$. This is achieved by simply inverting eq.~(\ref{Lgauge}). The final result can again be expressed in terms of the Born-level cross-sections in the form of  eq.~(\ref{sieqdif}). 

\begin{figure}
\centering
	\includegraphics[scale=.45]{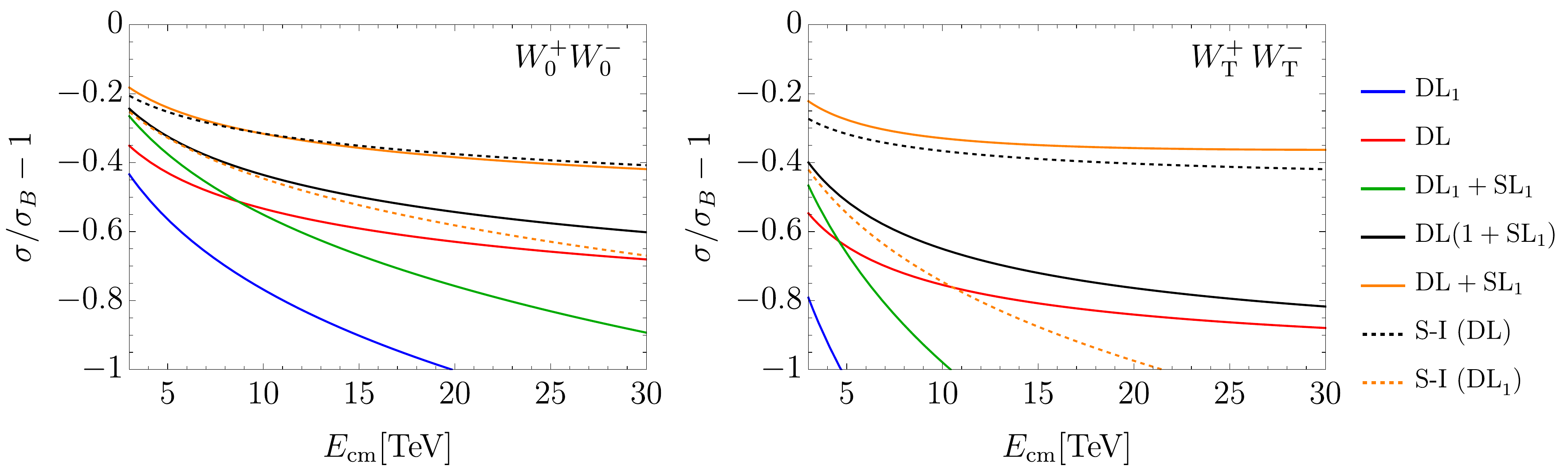}
	\caption{Same as Figure~\ref{Fig:XSDifermion}, but for di-boson production. As explained in Section~\ref{Sec:3.2}, the cross-sections for $W^+ W^-$ production are integrated in the angular region $\theta_*\in[67^\circ,150^{\circ}]$.}
	\label{Fig:XSDiboson}
\end{figure}

The results display the same qualitative features as di-fermions. In particular we observe the same interplay between short-distance physics affecting the neutral- and the charged-current Born amplitudes, which we investigate in Section~\ref{Sec:3} in details. Also at the quantitative level, the relative impact of radiation radiation is similar, as expected because \mbox{SU$(2)_L$} doublets are involved also in these processes. This is shown in the left panel of Figure~\ref{Fig:XSDiboson}, where we show the exclusive and semi-inclusive cross-section predictions for $W^+_0W^-_0$. The different predictions are obtained as explained in the previous section for the di-fermion processes. Notice in particular the exclusive predictions that include one-loop single logarithms as in eq.~(\ref{excparplus}). We employ these predictions for exclusive cross-section in the phenomenological studies of Section~\ref{Sec:IREE}.

\subsubsection*{Transverse di-boson}

Vector bosons ($W$, $Z$, or $\gamma$) with transverse helicity ${\rm{T}}=\pm1$ have zero hypercharge and they decompose as a real triplet plus a singlet under the SM \mbox{SU$(2)_L$}, as in eq.~(\ref{Tgauge}). Therefore three non-power-suppressed hard processes have to be considered for $L$-handed production
\begin{eqnarray}\label{tdbp}
&\ds
{{\bar{\mu}}}_{+1/2}(k_1,\,\alpha_1^{\bar{\rm{d}}})\ell_{-1/2}(k_2,\,\alpha_2^{\rm{d}}) \to B(k_3)\, B(k_4) 
\,,
\nonumber\\
&\ds
{{\bar{\mu}}}_{+1/2}(k_1,\,\alpha_1^{\bar{\rm{d}}})\ell_{-1/2}(k_2,\,\alpha_2^{\rm{d}})  \to W(k_3,\,\alpha_3^{{\rm{t}}})\, B(k_4) 
\,,\\
&\ds
{{\bar{\mu}}}_{+1/2}(k_1,\,\alpha_1^{\bar{\rm{d}}})\ell_{-1/2}(k_2,\,\alpha_2^{\rm{d}})  \to W(k_3,\,\alpha_3^{{\rm{t}}})\, W(k_4,\,\alpha_4^{\rm{t}}) \,, \nonumber
\end{eqnarray}
while only one is relevant for the production initiated by $R$-handed leptons~\footnote{The Born process $\ell^+_{-1/2}(k_1)\ell^-_{+1/2}(k_2) \to W(k_3,\,\alpha_3^{{\rm{t}}})\, W(k_4,\,\alpha_4^{{\rm{t}}})$ is power-suppressed in the SM.}
\beq
\ell^+_{-1/2}(k_1)\ell^-_{+1/2}(k_2)
\to B(k_3)\, B(k_4) \,,
\eeq
The ``$^{\rm{t}}$'' superscript in eq.~(\ref{tdbp}) refers to the triplet nature of the $W$ indices. 

Unlike for di-fermion and longitudinal di-boson, the transverse di-boson cross-sections for $L$-handed initial leptons are linear combinations of several distinct density matrices with different \mbox{SU$(2)_L$} quantum numbers. Therefore the exclusive cross-sections are not proportional, unlike in eq.~(\ref{excpar}), to the corresponding Born cross-sections in general. For instance in the $\gamma\gamma$ cross-section the contribution from the $WW$ final state experiences a stronger Sudakov suppression~(\ref{exdm}) than one from the $BB$ (or $WB$) final state, owing to the higher \mbox{SU$(2)_L$} Casimir of the final states. 

The evaluation of the semi-inclusive cross-sections proceeds as for the longitudinal di-bosons. Namely we derive the cross-sections for the physical states by inverting eq.~(\ref{Tgauge}) and we compute the double-logarithm exponentials using eq.~(\ref{tfact}) on the $SU(2)$ triplet subspace. Clearly the need of combining different density matrices complicates the calculation, but it does not introduce any novel conceptual aspect. At the quantitative level instead, the situation is significantly different than for di-fermions and longitudinal di-bosons. As shown on the right panel of Figure~\ref{Fig:XSDiboson}, EW radiation effects are much larger due to the larger Casimir $c_{\rm{t}}=2$ of the triplet representation. A sufficiently accurate modeling of these effects will probably require resummation even at the lowest VHEL energy $E_{\rm{cm}}=3$~TeV.

The figure reports the cross-section of the $W^+_{\rm{T}}W^-_{\rm{T}}$ final state. This final state, together with $W^+_{\rm{T}}Z_{\rm{T}}$, is  the only transverse di-boson process  we will consider in Section~\ref{Sec:3} (as a background to the corresponding longitudinal processes). Notice however that there are many other transverse di-boson processes (namely $ZZ$, $Z\gamma$, $\gamma\gamma$, and $W\gamma$) that can be measured at the VHEL. These processes probe heavy new physics in the EW sector. In particular,  as shown in Refs.~\cite{DiLuzio:2018jwd, Han:2020uak, FranceschiniDM}, they are sensitive (together with di-fermions) to  minimal Dark Matter in large-multiplets. The large effects of EW radiation might have a strong impact on these studies.

\section{Sensitivity projections}\label{Sec:3}

As described in the Introduction, we target effects from short-distance new physics that grow quadratically with the collision energy, to be probed in $\ell^+ \ell^-$ collisions at the highest available energy $E=E_{\rm{cm}}$. In this section we consider the dimension-$6$ EFT operators listed in Table~\ref{Tab:OpWarsaw}, and we estimate the sensitivity of muon colliders of energies $E_{\rm{cm}}=3,\,10,\,14$ or $30$~TeV to their Wilson coefficients. We assume a baseline integrated luminosity~\cite{Delahaye:2019omf}
\beq\label{intum}
\widehat{\mathcal{L}}=10\text{ ab}^{-1}\left(\frac{E_{\text{CM}}}{10\text{ TeV}}\right)^2 \,.
\eeq
Semi-quantitative comments on the impact of a reduced luminosity target are postponed to the Conclusions. We base our projections on statistically-dominated measurements of exclusive and semi-inclusive cross-sections for the processes listed in Table~\ref{Tab:Processes}. In the table, for each process we label with a check mark the operators that produce a quadratically growing-with-energy correction to the SM cross-section.

The target EFT operators are selected to represent generic manifestations, at energies much below the new physics scale, of the BSM scenarios we investigate in Section~\ref{Sec:4}. These are Composite Higgs, Composite Top and a minimal $Z'$ model, which we select as concrete examples of new physics in the Higgs, Top and EW-gauge sectors. Among the many operators that emerge in these scenarios, we focused our attention on those that display energy growth in $2\to2$ scattering processes at the muon collider. We will see in Section~\ref{Sec:4} that other operators offer a weaker sensitivity to the same BSM scenarios.

The phenomenological analysis of the various processes listed in Table~\ref{Tab:Processes} is described in Sections~\ref{Sec:3.1} and~\ref{Sec:3.2}, focusing respectively on the effects of  the ``W\&{Y}'' and  of  the ``Di-boson'' operators of Table~\ref{Tab:OpWarsaw}. In an attempt to mimic realistic experimental results, we include reconstruction (and, in some case, mistag) efficiencies at a level that is comparable with the CLIC detector performances, which  we extract, whenever possible, from Refs.~\cite{deBlas:2018mhx,Abramowicz:2018rjq}. Table~\ref{Tab:Processes} displays surprisingly low efficiencies for certain processes (e.g., $t\bar{t}$), entailing a considerable degradation of the measurement uncertainty. In Sections~\ref{Sec:3.1} and~\ref{Sec:3.2} we also present our results for the sensitivity of muon colliders to the corresponding set of operators, with the main aim of outlining the impact of the EW radiation effects on the analysis. The operators in the last class, dubbed ``$3^{\text{rd}}$ family'' in Table~\ref{Tab:OpWarsaw}, are not discussed explicitly but the sensitivity projection results are reported in Appendix~\ref{3rd}. The relevant final states, $tt$, $bb$ and $tb$ are discussed in Section~\ref{Sec:3.1}.

\begin{table}
	\begin{center}
		\begingroup
		\renewcommand*{\arraystretch}{1.7}
		\begin{tabular}{|c|| c | c |} 
			\cline{2-3}
			\multicolumn{1}{c|}{}&SILH basis& Warsaw-like basis \\
			\hline \multirow{2}{*}{W\&{Y}}& $O_{2W}=\left( D_{\mu} W^{\mu \nu,a} \right)^2$ &$O_{2W}'= J^{a,\mu}_L J_{L,\mu}^a \qquad J^{a,\mu}_L =\frac{1}{2} \sum_{f} \bar{f} \gamma^{\mu} \sigma^a f$
			\\ 
			&$O_{2B}=\left( \partial_{\mu} B^{\mu \nu} \right)^2$&
			$O_{2B}'= J^{\mu}_Y J_{Y,\mu} \,\, \qquad J^{\mu}_Y = \sum_{f} Y_f \bar{f} \gamma^{\mu}  f$
			\\[0.5ex]
			\hline
			\multirow{3}{*}{Di-boson}& $O_W = \dfrac{ig}{2} (H^\dagger \sigma^a \overleftrightarrow{D}_\mu H) D^\nu W^a_{\mu\nu}$ & $O_{W}'=\dfrac{g^2}{4}(H^\dagger i\hspace{-2pt} \overleftrightarrow{D}_{\hspace{-3pt}\mu} \sigma^a H)(\bar{L}_L \gamma^\mu \sigma^a L_L)$\ \\
			&$O_B = \dfrac{ig'}{2} (H^\dagger \overleftrightarrow{D}_{\hspace{-2pt}\mu} H) \partial^\nu B_{\mu\nu}$& 
			$O_{B}'\hspace{-2pt}=\hspace{-2pt}-\dfrac{g^{\prime2}}{4}(\hspace{-2pt}H^\dagger i \hspace{-2pt}\overleftrightarrow{D}_{\hspace{-3pt}\mu} H\hspace{-2pt})(\hspace{-2pt}\bar{L}_L \hspace{-1pt}\gamma^\mu L\hspace{-1pt}_L\hspace{-2pt})\hspace{-2pt}$\\
			& &$\hspace{40pt}-\dfrac{g^{\prime2}}{2}(\hspace{-2pt}H^\dagger i\hspace{-2pt} \overleftrightarrow{D}_{\hspace{-3pt}\mu} H\hspace{-2pt})(\bar{l}_R \gamma^\mu l_R)$ \\[0.5ex] \hline 
			&$O^{(3)}_{qD} = \left( \bar{q} \gamma^{\mu} \sigma^a q \right) \left( D^{\nu} W_{\mu \nu}^a \right)$&$O^{\prime(3)}_{qD} = \left( \bar{q} \gamma^{\mu} \sigma^a q \right) J^{a}_{L,\mu}$\\
			\multirow{1}{*}{\shortstack{$3^{\text{rd}}$ family}}&$O^{(1)}_{qD} = \left( \bar{q} \gamma^{\mu} q \right) \left( \partial^{\nu} B_{\mu \nu} \right)$&$O^{\prime (1)}_{qD} = \left( \bar{q} \gamma^{\mu} \sigma^a q \right) J_{Y,\mu}$\\
			&$O_{tD} = \left( \bar{t} \gamma^{\mu} t \right) \left( \partial^{\nu} B_{\mu \nu} \right)$&$O^{\prime}_{tD} = \left( \bar{t} \gamma^{\mu} \sigma^a t \right) J_{Y,\mu}$\\[0.5ex] \hline
		\end{tabular}
		\endgroup
		\caption{The operators under consideration in their ``SILH''~\cite{Giudice:2007fh} form and, after using the equations of motion, expressed as a linear combination of Warsaw~\cite{Grzadkowski:2010es} operators. $Y_f$ is the hypercharge of the fermionic field $f$. In the operators involving the $3^{\text{rd}}$ family the fields $t$ and $q$ denote respectively the right-handed and left-handed top quark. \label{Tab:OpWarsaw}}
		\vspace{20pt}
		\begingroup
		\renewcommand*{\arraystretch}{1.2}
		\begin{tabular}{| c | c | c | c | c | c | c | c | c | c | c |} 
			\hline
			Process&$N$ (Ex) & $N$ (S-I)&Eff.&$O_{2W}'$ & $O_{2B}'$ & $O_{W}'$ & $O_{B}'$ & $O_{qD}^{\prime (3)}$ & $O_{qD}^{\prime (1)}$ & $O_{uD}^{\prime}$\\ \hline
			$e^+ \, e^-$ &6794&9088&$100\%$&\checkmark&\checkmark&&&&& \\ \hline
			$e \nu_{e}$ &---&2305&$100\%$&\checkmark&\checkmark&&&&& \\ \hline
			$\mu^+ \, \mu^-$ &206402&254388&$100\%$&\checkmark&\checkmark&&&&& \\ \hline
			$\mu \,\nu_{\mu}$ &---&93010&$100\%$&\checkmark&\checkmark&&&&& \\ \hline
			$\tau^+\, \tau^-$ &6794&9088&$25\%$  &\checkmark&\checkmark&&&&& \\\hline
			$\tau \nu_{\tau}$ &---&2305&$50\%$&\checkmark&\checkmark&&&&& \\ \hline
			$jj$ (Nt) &19205&25725&$100\%$&\checkmark&\checkmark&&&&&\\ \hline
			$jj$ (Ch)&---&5653&$100\%$& \checkmark&\checkmark&&&&& \\ \hline
			$c \, \bar{c}$ &9656&12775&$25\%$& \checkmark&\checkmark&&&&& \\ \hline
			$cj$ &---&5653&$50\%$ &\checkmark&\checkmark&&&&& \\ \hline 
			$b \, \bar{b}$ &4573&6273&$64\%$& \checkmark&\checkmark&&&\checkmark&\checkmark& \\ \hline
			$t \, \bar{t}$ & 9771&11891&$5\%$& \checkmark&\checkmark&&&\checkmark&\checkmark&\checkmark \\ \hline
			$b\,{t}$ &---&5713&$57\%$& \checkmark&\checkmark&&&\checkmark&\checkmark&\checkmark \\ \hline
			$Z_0h$ & $680$ & $858$ & $26\%$ & & & \checkmark & \checkmark & & & \\ \hline
			$W^+_0W^-_0$ & $1200$ & $1456$ & $44\%$ & & & \checkmark & \checkmark & & & \\ \hline
			$W^+_{\rm{T}}W^-_{\rm{T}}$ & $2775$ & $5027$ & $44\%$ & & & & & & & \\ \hline
			$W^\pm h$ & --- & $506$ & $19\%$ & & & \checkmark & \checkmark & & & \\ \hline
			$W^\pm_0 Z_0$ & --- & $399$ & $23\%$ & & & \checkmark & \checkmark & & & \\ \hline
			$W^\pm_{\rm{T}} Z_{\rm{T}}$ & --- & $2345$ & $23\%$ & & & & & & & \\ \hline
		\end{tabular}
		\endgroup
		\caption{The exclusive and semi-inclusive processes employed for the sensitivity projections. The operators that give a growing-with-energy contribution to each operator are labeled with a check mark. The expected number of events (before efficiencies) is for $E_{\rm{cm}}=10$~TeV with the integrated luminosity~(\ref{intum}). \label{Tab:Processes}}
	\end{center}
\end{table}

\subsection[W\&{Y} operators]{W$\boldsymbol{\&}${Y} operators}\label{Sec:3.1}

The first two operators we consider are those associated with the W and Y parameters of LEP EW precision tests~\cite{Barbieri:2004qk}, namely $O_{2W}$ and $O_{2B}$ defined as in Table~\ref{Tab:OpWarsaw}. These operators arise in the so-called {\it{universal}} scenarios ~\cite{Barbieri:2004qk, Wells:2015uba}, that is  new physics that couples dominantly to the bosonic sector of the SM. 
 Employing $O_{2W}$ and $O_{2B}$ is convenient in the low-energy context of the LEP experiment, however for our purpose it is better to trade them for the current-current operators $O_{2W}'$ and $O_{2B}'$ (see again Table~\ref{Tab:OpWarsaw}), using the SM equations of motion. In doing so, we neglect the contribution to the $O_{W}'$ and $O_{B}'$ operators, which are expected to have no impact on the sensitivity. In what follows we parameterize the $O_{2W}'$ and $O_{2B}'$ operator coefficients
 \beq\label{WYW}
G_{2W}' = -\frac{g^2 {\rm{W}}}{2 \MW^2}\,,\;\;\;\;\; G_{2B}' = -\frac{g'^2 {\rm{Y}}}{2 \MW^2}\,,
\eeq
in terms of the dimensionless parameters W and Y.

The relevant scattering processes, listed in Table~\ref{Tab:Processes}, are the production of two energetic fermions in the central region of the detector. Specifically, as explained at the end of Section~\ref{Sec:2}, we have in mind the two hard particles whose invariant mass is higher than around $85\%$ of the total collider $E_{\rm{cm}}$, and a scattering angle $\theta_* \in [30^\circ, 150^\circ]$. We assume perfect detector sensitivity to massive $W$ and $Z$ bosons of arbitrary low $3$-momentum, enabling the measurement of exclusive scattering cross-sections where the emission of massive vectors (and of photons with hardness above $\MW^2$) is vetoed. The exclusive cross-section measurements are combined with the semi-inclusive cross-sections, where the emission of an arbitrary number (including zero) of massive vectors or hard photons is allowed.

For each inclusive and semi-inclusive final state, we employ cross-section measurements in $10$ equally-spaced  bins of $\cos\theta_*$ in the range $[ -\sqrt{3}/2,  \sqrt{3}/2]$. In processes (e.g., $jj$, or $b{\bar{b}}$) where the two final states are effectively indistinguishable, $\cos\theta_*$ is defined to be positive and $5$ bins are employed. We assume cross-section measurements with purely statistical uncertainties, which we estimate based on the number of events that are expected in the SM. 

Of course in order to combine the exclusive and semi-inclusive cross-sections for the same (neutral) hard final state we must take into account that the exclusive events are also counted in the measurement of the semi-inclusive cross-section. It is thus convenient to consider a  cross-section {\it with radiation}, defined as the difference between the semi-inclusive and the exclusive cross-sections
\begin{equation}
\sigma_{\rm{rad}}\equiv\sigma_{\rm{si}} - \sigma_{\rm{ex}} \,.
\end{equation}
The measurement of $\sigma_{\rm{rad}}$ can be combined with the one of $\sigma_{\rm{ex}}$ since they are statistically independent. For charged hard final states there is instead only one type of cross-section, which necessarily involves EW radiation emission by charge conservation. We will refer to the charged cross-section as ``semi-inclusive'' or ``with radiation'' interchangeably.\\[5pt]
\indent{}We now discuss the di-fermion processes individually. \\[-16pt]
\begin{itemize}
	\setlength\itemsep{+1pt}
	\item ${\boldsymbol{e}}^{\boldsymbol{+}}{\boldsymbol{e}}^{\boldsymbol{-}}$,~${\boldsymbol{\mu}}^{\boldsymbol{+}}{\boldsymbol{\mu}}^{\boldsymbol{-}}$~{\bf{and}}~${\boldsymbol{\tau}}^{\boldsymbol{+}}{\boldsymbol{\tau}}^{\boldsymbol{-}}${\bf{:}} 
	We assume $100{\%}$ reconstruction efficiency for muon and electrons, and an efficiency of $50{\%}$~\cite{deBlas:2018mhx} for each $\tau$ lepton. Notice that the cross-section for muons is around $30$ times larger than for the other leptons. This is mostly due to the $t$-channel enhancement of the elastic $\mu^+\mu^-$ scattering. 
	\item ${\boldsymbol{c}}{\boldsymbol{\overline{c}}}$~{\bf{and}}~${\boldsymbol{b}}{\boldsymbol{\overline{b}}}$\,{\bf{:}} We assume $50{\%}$ and $80{\%}$ efficiency for tagging respectively charm and bottom quark jets~\cite{deBlas:2018mhx}. We ignore the mis-tag of light jets, as well as $c$/$b$ misidentification. No information on the charge of the tagged quark is employed.
	\item ${\boldsymbol{j}}{\boldsymbol{{j}}}$\,{\bf{:}} We consider the production of two light quarks $u$, $d$ or $s$, which we suppose to be reconstructed as jets with $100{\%}$ efficiency. In Table~\ref{Tab:Processes} we report separately the production of a neutral (Nt) and of a charged (Ch) quark/anti-quark pair, but the two processes are collected into a single $2$-jets final state. We also include the contribution from mistagged $b$ and $c$ quarks.
	\item ${\boldsymbol{t}}{\boldsymbol{\overline{t}}}$\,{\bf{:}} Based on Ref.s~\cite{Abramowicz:2018rjq,Durieux:2018ekg}, we estimate as $5{\%}$ the total efficiency for the reconstruction of the $t \bar{t}$ pair. This (somewhat low) efficiency estimate only includes the semi-leptonic $t \bar{t}$ final states, in which the charge of the tagged top quarks can be measured.
	\item ${\boldsymbol{t}}{\boldsymbol{{b}}}$~{\bf{and}}~${\boldsymbol{c}}{\boldsymbol{j}}$\,{\bf{:}} We use $50\%$ and $80\%$ tag efficiency for the charm and the bottom, respectively, and $\sqrt{0.05}=20\%$ efficiency for the top. The charge of the top quark is assumed to be reconstructed.
	\item ${\boldsymbol{e}}{\boldsymbol{\nu_e}}$,~${\boldsymbol{\mu}}{\boldsymbol{\nu_\mu}}$~{\bf{and}}~${\boldsymbol{\tau}}{\boldsymbol{\nu_\tau}}${\bf{:}} The efficiency is $100\%$ for muons and electrons, and $50{\%}$ for the $\tau$. It should be noted that, because of the invisible neutrino, the hard scattering region of this final state can not be selected with a cut on the invariant mass of the two particles. The selection will  have instead to be performed on the energy and the transverse momentum of the observed charged lepton.
\end{itemize}

\begin{figure}[t]
	\begin{minipage}{.5\linewidth}
		\centering
		\includegraphics[scale=.52]{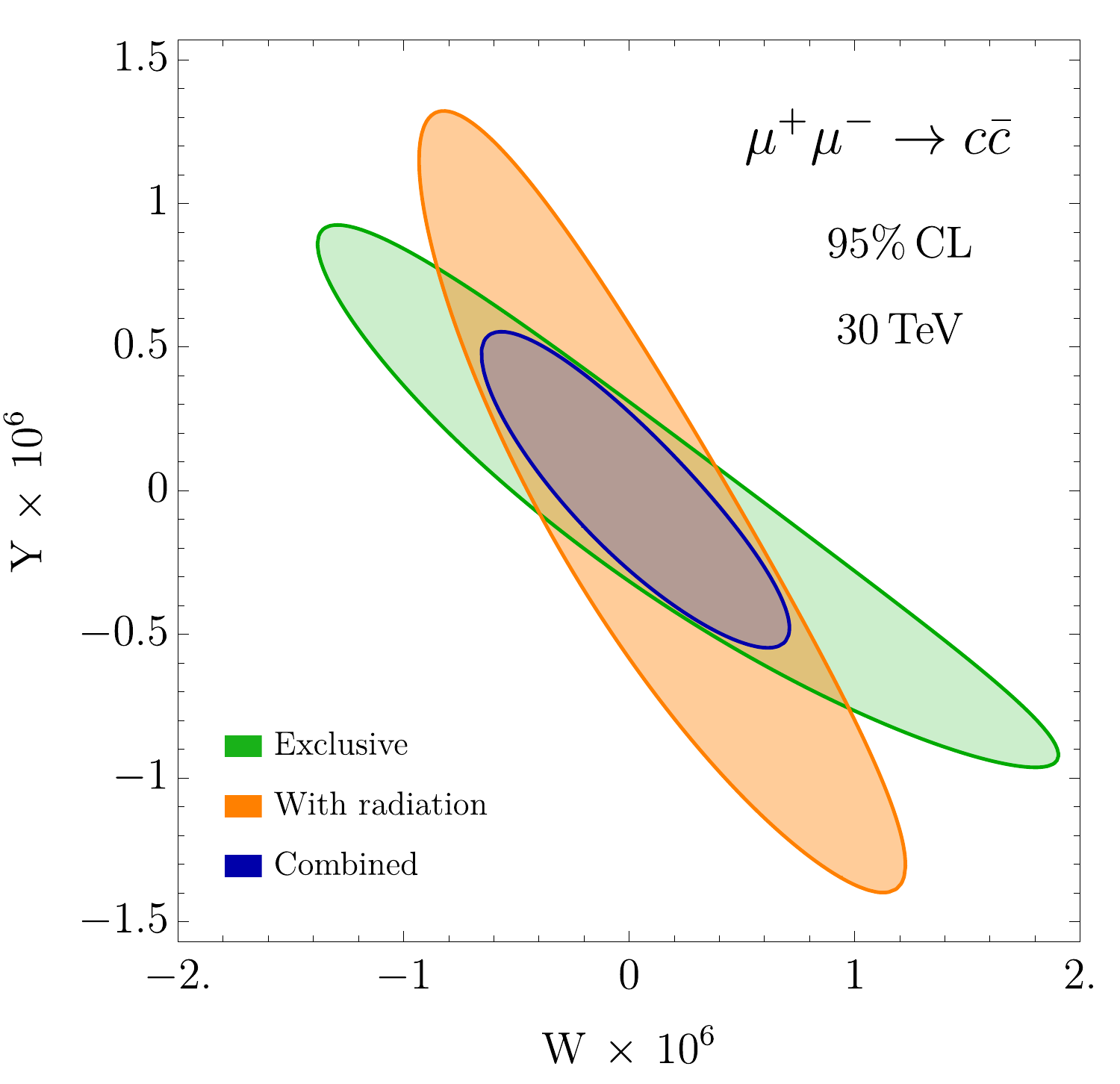}
	\end{minipage}%
	\begin{minipage}{.5\linewidth}
		\centering
		\includegraphics[scale=.52]{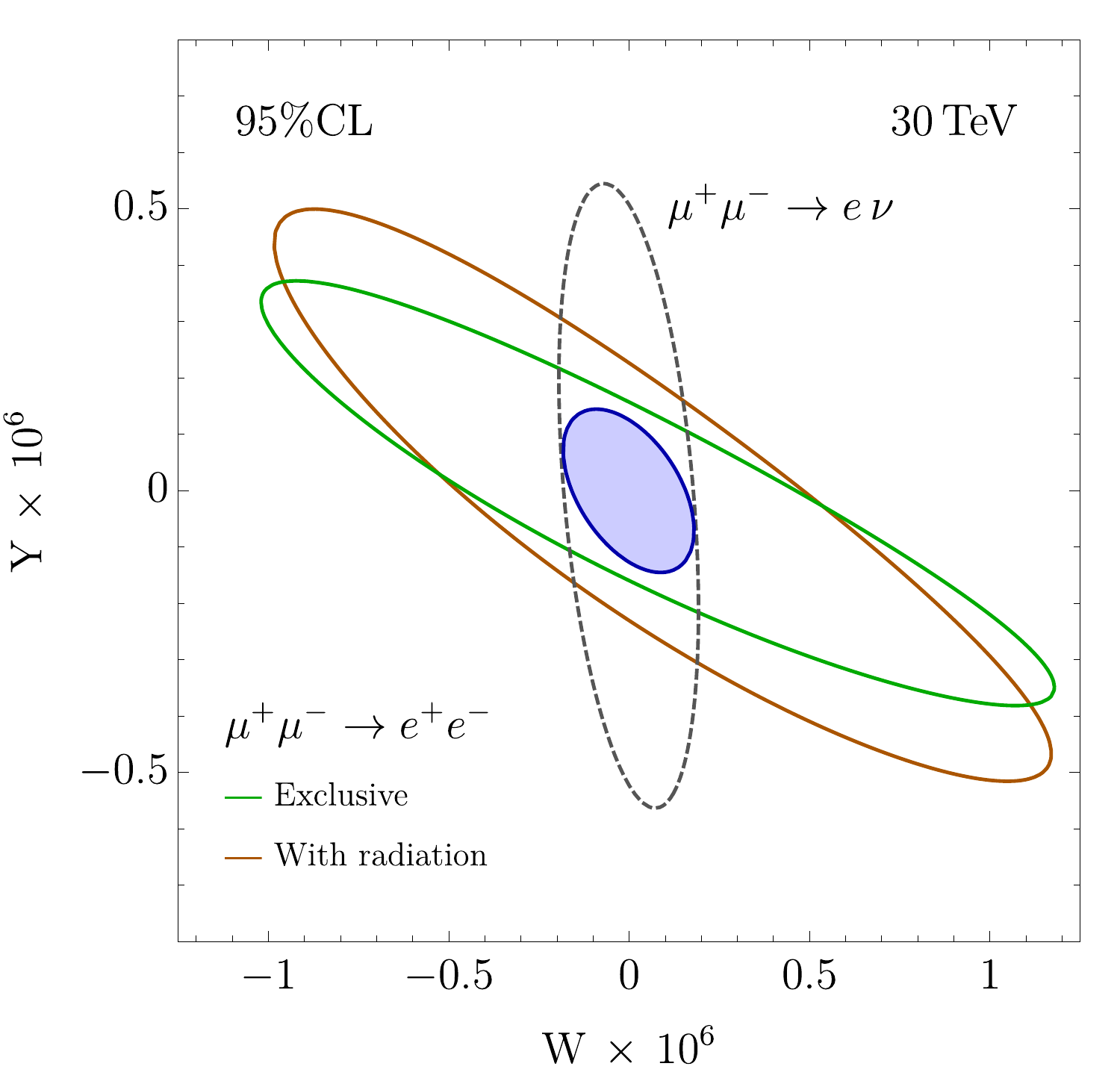}
	\end{minipage}
	\caption{$95\%$~CL sensitivities to the W and Y parameters of the $30$~TeV muon collider. Exclusive and ``with radiation'' (i.e., semi-inclusive minus exclusive) cross-section measurements of the $c{\overline{c}}$ process are considered in the left panel. The right panel shows the impact of $e^+e^-$ (exclusive and ``with radiation'') and $e\nu$ (that only exists at the semi-inclusive level) final states.
		\label{Fig:ExclusionReachWY0}}
\end{figure}

The different dependence on W and Y of the neutral- and charged-current Born amplitudes entails (see Section~\ref{Sec:Diferm}) a different dependence on these parameters of the exclusive and semi-inclusive cross-sections. The statistical combination of the two types of cross-sections can thus increase  the sensitivity, as illustrated in Figure~\ref{Fig:ExclusionReachWY0}. The left panel displays the $95\%$~CL sensitivity of $c{\overline{c}}$ production to W and Y, comparing the impact of the exclusive cross-section (in green) to that (in orange) of the  cross-section with radiation. The two measurements probe different regions of the W and Y parameter space, and a significant sensitivity gain is observed in their combination (in blue). The green and blue lines on the right panel of Figure~\ref{Fig:ExclusionReachWY0} display a similar complementarity pattern for the $e^+e^-$ final state. There also appears an even stronger complementarity with the measurement of the $e\nu$ cross-section, reported as a gray dashed line. The emergence of the $e\nu$ process, as well as the other charged final states in Table~\ref{Tab:Processes}, is entirely due to EW radiation. Nevertheless its (semi-inclusive) cross-section is large, because EW radiation is indeed a prominent phenomenon at $E_{\rm{cm}}\simeq10$~TeV. Furthermore the cross-section displays a peculiar dependence on new physics, producing a sensitivity contour that is different from that  of the $e^+e^-$ measurements. The statistical combination of the three measurements (in blue) improves the sensitivity significantly.

\begin{figure}[t]
	\centering
	\includegraphics[scale=.5]{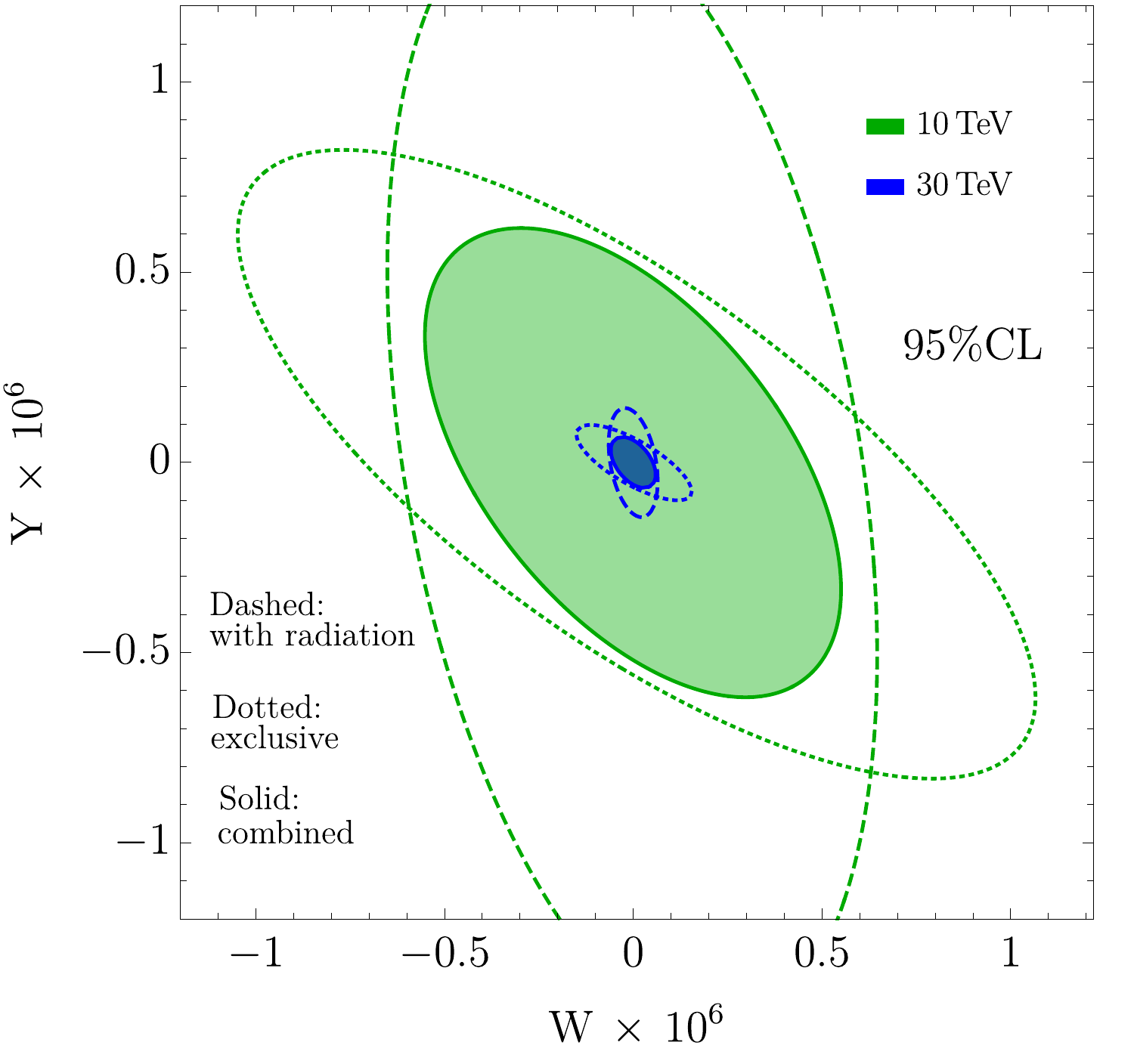}
	\caption{$95\%$~CL sensitivities to the W and Y at the $10$ and $30$~TeV muon collider.}
	\label{Fig:ExclusionReachWY}
\end{figure}

The final results of our analysis including all channels are summarized in Figure~\ref{Fig:ExclusionReachWY} and in Table~\ref{Tab:SORDifermion}. The figure displays the sensitivity contours of exclusive measurements as dotted lines, and the combined impact of charged and of neutral ``with radiation'' cross-sections, in dashed. The combination of all measurements is also shown. The table reports the results for $3$, $10$, $14$ and $30$~TeV, comparing the sensitivity of exclusive cross-sections alone with the total combination. 

At the High-Luminosity LHC (HL-LHC), it will be possible to probe the W and Y parameters at the level of $4\cdot10^{-5}$ and $8\cdot10^{-5}$, respectively, at $95\%$~CL~\cite{Torre:2020aiz,Farina:2016rws,Panico:2021vav}. Table~\ref{Tab:SORDifermion} shows that the $3$~TeV muon collider would improve by one order of magnitude or more, and the sensitivity improves quadratically with the muon collider energy. Among the other future collider projects~\cite{deBlas:2019rxi}, CLIC at $3$~TeV has the best sensitivity, of around $4\cdot10^{-6}$ for both parameters~\cite{deBlas:2018mhx}. This is of course comparable with the $3$~TeV muon collider sensitivity, and a factor $10$ worst than that of the muon collider at $10$~TeV. The comparison with FCC-hh projections is even more favorable to the muon collider.

\subsection{Diboson operators}\label{Sec:3.2}

\begin{table}
	\begin{center}
		\begin{tabular}{||c|| c | c | c || c | c | c ||} 
			\cline{2-7}
			\multicolumn{1}{c|}{}&\multicolumn{3}{|c||}{Exclusive-only [$95\%$~CL]}&\multicolumn{3}{c|}{Combined [$95\%$~CL]}\\
			\cline{2-7}
			\multicolumn{1}{c|}{}&W$\times10^7$&Y$\times10^7$&$\rho_{\text{W},\text{Y}}$&W$\times10^7$&Y$\times10^7$&$\rho_{\text{W},\text{Y}}$\\ \hline 3 TeV& $[-53, 53]$&$[-48, 48]$&	-0.72&	$[-41,41]$&	$[-46,46]$&	-0.60 \\ \hline 10 TeV& $[-5.71, 5.71]$&$[-4.47, 4.47]$&	-0.74&	$[-3.71, 3.71]$&	$[-4.16, 4.16]$&	-0.54 \\ \hline 14 TeV& $[-3.11, 3.11]$&$[-2.31,2.31]$&	-0.74&	$[-1.90, 1.90]$&	$[-2.13,2.13]$&	-0.52 \\ \hline 30 TeV& $[-0.80, 0.80]$&$[-0.52,0.52]$&	-0.75&	$[-0.42, 0.42]$&	$[-0.47,0.47]$&	-0.48\\ \hline
		\end{tabular}
		\caption{Single-operator $95\%$~CL reach and correlation for the W$\&$Y parameters at different muon collider energies including only exclusive cross-sections and combining all measurements. Since the likelihood is dominated by the linear terms in the new physics parameters, the single parameter reach plus the correlation characterizes our results completely. \label{Tab:SORDifermion}}
	\end{center}
\end{table}

The setup for this analysis is similar to that of Ref.~\cite{Buttazzo:2020uzc}. Namely we consider the SILH operators $O_W$ and $O_B$, we convert them into the current-current interactions $O_W'$ and $O_B'$ as in Table~\ref{Tab:OpWarsaw}, and we study their effect on the production of high-energy vector bosons and Higgs. Notice that, by the equivalence theorem, $O_W'$ and $O_B'$ only significantly affect the production of longitudinally polarized vector bosons. We are therefore here studying the production of high-energy longitudinally vector bosons and Higgs, with the production of transversely polarized vector bosons playing merely the role of background.
Since the effects are quadratically enhanced by the energy, such high-energy di-boson processes are by far the best probe of these operators at the muon collider~\cite{Buttazzo:2020uzc}.\\[5pt]
\indent{}We thus consider, among those in Table~\ref{Tab:Processes}, the following final states \\[-16pt]
\begin{itemize}
	\setlength\itemsep{+1pt}
	\item ${\boldsymbol{Z}}{\boldsymbol{h}}$\,{\bf{:}} Following Ref.~\cite{Buttazzo:2020uzc}, we consider an efficiency of $26\%$ for tagging the two hard and central final state particles, with a selection that reduces the background to a manageable level. Notice that this final state is dominated by the longitudinal helicity channel $Z_0h$.
	\item ${\boldsymbol{W}}^{\boldsymbol{+}}{\boldsymbol{W}}^{\boldsymbol{-}}${\bf{:}} Again like in~\cite{Buttazzo:2020uzc}, we assume a $44\%$ efficiency for the detection of the two $W$ bosons in the semi-leptonic decay channel, where the charge of the $W$'s can be reconstructed. Transverse $WW$ production plays here the role of background.
	\item ${\boldsymbol{W}}{\boldsymbol{h}}$\,{\bf{:}} We consider an efficiency of $19\%$, having in mind the leptonic $W$ decay, and $h\to{b}\overline{b}$. Like for $Zh$, there is no relevant background from transverse production.
	\item ${\boldsymbol{W}}{\boldsymbol{Z}}$\,{\bf{:}} We apply an efficiency of $23\%$, which corresponds to the leptonic $W$ and the hadronic $Z$ decay. The background from transverse $WZ$ production is considerable, and is taken into account. 
\end{itemize}

In our analysis we do not consider the possibility of employing the decay angles of the bosons to extract information on their polarization. Therefore the transverse di-bosons processes $W^+_{\rm{T}}W^-_{\rm{T}}$ and $W_{\rm{T}}Z_{\rm{T}}$ are effectively irreducible backgrounds to the corresponding longitudinal processes, and the scattering angle $\theta_*$ is the only discriminating variable. An increased lower cut on $\theta_*$ benefits the sensitivity, as it suppresses the $t$-channel enhancement of the transverse background processes. After optimization we find, like in Ref.~\cite{Buttazzo:2020uzc}, that a good signal sensitivity is obtained by the measurement of fiducial $WW$ and $WZ$ cross-sections in the range
\beq
\theta_*\in [ 67^\circ,150^\circ]\,.
\eeq
The possibility of binning $\theta_*$ has been considered, but found not to improve the sensitivity. Our analysis will thus be only based on the measurement of the fiducial $WW$ and $WZ$ cross-sections in the above region, and of the $Zh$ and $Wh$ cross-sections for $\theta_*\in [30^\circ,150^\circ]$. As in the previous section, both exclusive and semi-inclusive cross-sections will be employed for the neutral processes $WW$ and $Zh$, plus the semi-inclusive charged cross-sections for $Wh$ and $WZ$. 

\begin{figure}
	\begin{minipage}{.43\linewidth}
		\centering
		\includegraphics[scale=.43]{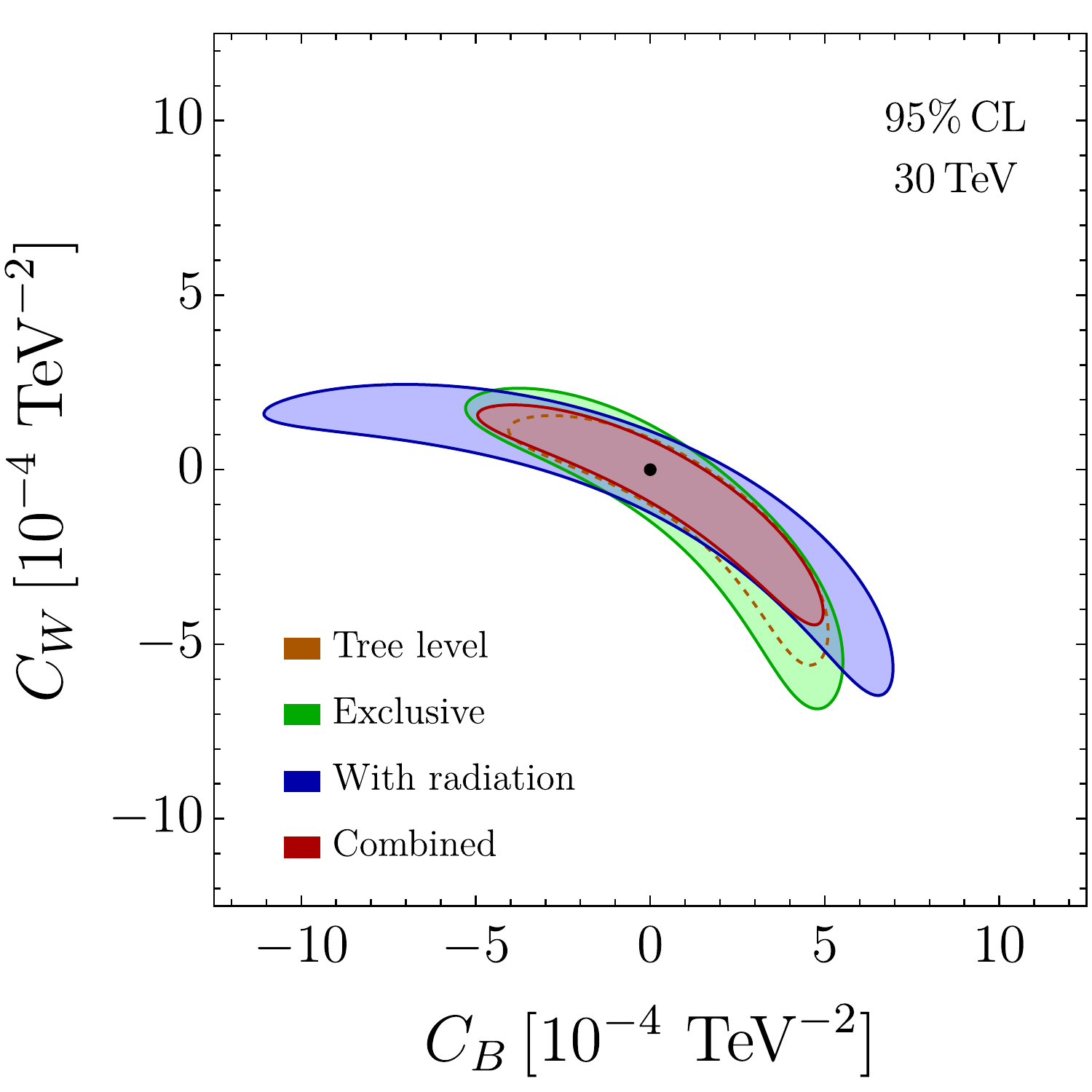}
	\end{minipage}%
	\hspace{20pt}
	\begin{minipage}{.43\linewidth}
		\hfill
		\includegraphics[scale=.43]{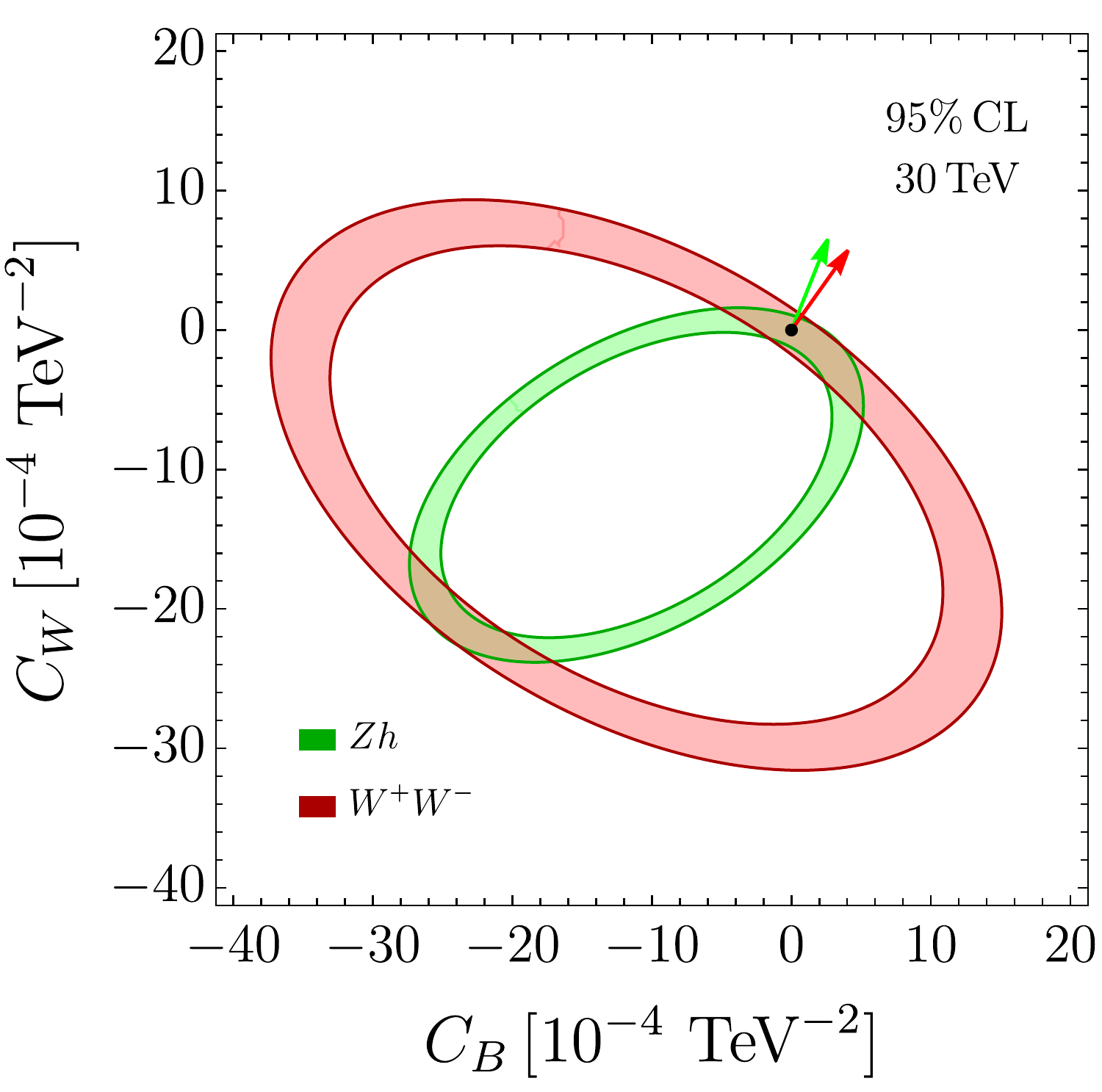}
	\end{minipage}
	\caption{Left: $95\%$ sensitivity contours in the $(C_B,C_W)$ plane at the $30$~TeV muon collider. A second allowed region, not shown in the figure, can be eliminated by other measurements~\cite{Buttazzo:2020uzc}. Right: $Zh$ and $WW$ likelihood contours at tree-level. Notice that the ellipses for $WW$ and $Zh$ are tangent in two points, one being the SM, the other being the point where the amplitudes have the same magnitude as in the SM but opposite sign.}
	\label{Fig:ExclusionReachDiboson}
\end{figure}

\begin{table}
	\begin{center}
		\begin{tabular}{||c|| c | c || c |  c ||} 
			\cline{2-5}
			\multicolumn{1}{c|}{}&\multicolumn{4}{|c|}{Single Operator (Exclusive-only) [$95\%$~CL]}\\
			\cline{2-5}
			\multicolumn{1}{c|}{}&\multicolumn{2}{|c||}{$C_B[ 10^{-4}\text{ TeV}^{-2}]$}&\multicolumn{2}{|c||}{$C_W [ 10^{-4}\text{ TeV}^{-2}]$} \\ 
			\cline{2-5}
			\multicolumn{1}{c|}{}& Linear & Quadratic & Linear & Quadratic\\
			\hline 
			3 TeV& $[-170, 170]$&$[-189, 157]$&$[-77.4, 77.4]$&$[-81, 74.4]$ \\
			\hline 
			10 TeV& $[-15.3, 15.3]$&$[-17, 14.2]$&$[-8.18, 8.18]$&$[-8.62, 7.82]$ \\
			\hline
			14 TeV& $[-7.86, 7.86]$&$[-8.69, 7.25]$&$[-4.40, 4.40]$&$[-4.65, 4.20]$ \\
			\hline 
			30 TeV& $[-1.73, 1.73]$&$[-1.92, 1.6]$&$[-1.1, 1.1]$&$[-1.16, 1.04]$ \\
			\hline
		\end{tabular}
		
		\vspace{0.2cm}
		
		\begin{tabular}{||c|| c | c || c |  c ||} 
			\cline{2-5}
			\multicolumn{1}{c|}{}&\multicolumn{4}{|c|}{Single Operator (Combined)  [$95\%$~CL]}\\
			\cline{2-5}
			\multicolumn{1}{c|}{}&\multicolumn{2}{|c||}{$C_B[ 10^{-4}\text{ TeV}^{-2}]$}&\multicolumn{2}{|c||}{$C_W [ 10^{-4}\text{ TeV}^{-2}]$} \\ 
			\cline{2-5}
			\multicolumn{1}{c|}{}& Linear & Quadratic & Linear & Quadratic\\
			\hline 
			3 TeV& $[-153, 153]$&$[-169, 142]$&$[-65.8, 65.8]$&$[-68.2, 63.6]$ \\
			\hline 
			10 TeV& $[-12.8, 12.8]$&$[-13.9, 11.9]$&$[-6.14, 6.14]$&$[-6.37, 5.93]$ \\
			\hline
			14 TeV& $[-6.40, 6.40]$&$[-6.95, 5.99]$&$[-3.17, 3.17]$&$[-3.29, 3.06]$ \\
			\hline 
			30 TeV& $[-1.34, 1.34]$&$[-1.44, 1.25]$&$[-0.71, 0.71]$&$[-0.737, 0.686]$ \\
			\hline
		\end{tabular}
		
		\vspace{1cm}
		
		\begin{tabular}{||c|| c | c || c |  c ||} 
			\cline{2-5}
			\multicolumn{1}{c|}{}&\multicolumn{4}{|c|}{Marginalized (Exclusive-only)  [$95\%$~CL]}\\
			\cline{2-5}
			\multicolumn{1}{c|}{}&\multicolumn{2}{|c||}{$C_B[ 10^{-4}\text{ TeV}^{-2}]$}&\multicolumn{2}{|c||}{$C_W [ 10^{-4}\text{ TeV}^{-2}]$} \\ 
			\cline{2-5}
			\multicolumn{1}{c|}{}& Linear & Quadratic & Linear & Quadratic\\
			\hline 
			3 TeV& $[-478, 478]$&$[-352, 596]$&$[-217, 217]$&$[-583, 125]$ \\
			\hline 
			10 TeV& $[-53.2, 53.2]$&$[-35.2, 50]$&$[-28.4, 28.4]$&$[-53.5, 14.2]$ \\
			\hline
			14 TeV& $[-29.4, 29.4]$&$[-18.6, 25]$&$[-16.5, 16.5]$&$[-27.5, 7.82]$ \\
			\hline 
			30 TeV& $[-7.98, 7.98]$&$[-4.45, 5.19]$&$[-5.04, 5.04]$&$[-6.16, 2.05]$ \\
			\hline
		\end{tabular}
		
		\vspace{0.2cm}
		
		\begin{tabular}{||c|| c | c || c |  c ||} 
			\cline{2-5}
			\multicolumn{1}{c|}{}&\multicolumn{4}{|c|}{Marginalized (Combined)  [$95\%$~CL]}\\
			\cline{2-5}
			\multicolumn{1}{c|}{}&\multicolumn{2}{|c||}{$C_B[ 10^{-4}\text{ TeV}^{-2}]$}&\multicolumn{2}{|c||}{$C_W [ 10^{-4}\text{ TeV}^{-2}]$} \\ 
			\cline{2-5}
			\multicolumn{1}{c|}{}& Linear & Quadratic & Linear & Quadratic\\
			\hline 
			3 TeV& $[-442, 442]$&$[-341, 535]$&$[-189, 189]$&$[-426, 115]$ \\
			\hline 
			10 TeV& $[-44, 44]$&$[-33.4, 43.4]$&$[-21.1, 21.1]$&$[-35.1, 12.3]$ \\
			\hline
			14 TeV& $[-23.1, 23.1]$&$[-17.6, 21.6]$&$[-11.4, 11.4]$&$[-17.6, 6.6]$ \\
			\hline 
			30 TeV& $[-5.24, 5.24]$&$[-4.12, 4.43]$&$[-2.79, 2.79]$&$[-3.70, 1.62]$ \\
			\hline
		\end{tabular}	
		\caption{Single operator and marginalized $95\%$ reach on $C_B$ and $C_W$, at different muon collider energies. The sensitivity of exclusive cross-section measurements alone is shown separately from the combination of all the measurements. The significant degradation of the marginalized bounds relative to the single-operators ones, and the strong sensitivity to the quadratic terms at the marginalized level, is due to the approximately flat direction displayed in Figure~\ref{Fig:ExclusionReachDiboson} \label{Tab:dibosonBounds}}
	\end{center}
\end{table}

The results of our analysis are reported in Table~\ref{Tab:dibosonBounds} and on the left panel of Figure~\ref{Fig:ExclusionReachDiboson}, in terms of the dimensionful coefficients ($C_B$ and $C_W$) of the $O_B'$ and $O_W'$ operators of Table~\ref{Tab:OpWarsaw}. Our finding are quantitatively similar to the ones of Ref.~\cite{Buttazzo:2020uzc}. We can thus refer to that article for the (very favorable) assessment of the muon collider sensitivity to $C_B$ and $C_W$ in comparison with current knowledge and with other future colliders. We devote the rest of this section to discuss the approximate flat direction of the likelihood in the $(C_B,C_W)$ plane, which we observe in Figure~\ref{Fig:ExclusionReachDiboson} (left panel).

The flat direction entails a strong degradation of the marginalized sensitivity, as in Table~\ref{Tab:dibosonBounds}. Furthermore this degradation brings the marginalized $C_B$ and $C_W$ limits to large values, in a region where the likelihood is considerably affected by the contributions to the cross-sections of the terms that are quadratic in the new physics parameters. In theories like Composite Higgs where $C_{B,W}\sim1/m_*^2$, this fact implies that the marginalized limits correspond to a new physics scale $m_*$  not much above the collider energy. In fact, looking at Table~\ref{Tab:dibosonBounds} we notice that the $30$~TeV $C_B$ reach corresponds to $m_*=43$~TeV. Thus, if new physics happened  to sit along the flat direction in Figure~\ref{Fig:ExclusionReachDiboson}, diboson processes would  fail to extend the muon collider sensitivity well above the direct mass-reach. We do not have reasons to expect new physics to lie in that direction. Actually in certain Composite Higgs models one expects it to lie in the nearly orthogonal direction $C_B=C_W$~\cite{Buttazzo:2020uzc}. However the presence of the flat direction is an obstruction to the broad exploration of new physics and to the characterization of a putative discovery. It is thus worth explaining its origin and discussing strategies to eliminate it.

\begin{figure}
	\begin{minipage}{.6\linewidth}
		\renewcommand*{\arraystretch}{1.5}
		\begin{tabular}{c|c}
			Process & $\sigma_B/\sigma_0$  \\\hline
			$(\mu \mu)_L\to Z h$& $\left[\tw^2(1+E_{\rm{cm}}^2 C_B)-(1+E_{\rm{cm}}^2 C_W)\right]^2$\\
			\hline
			$(\mu \mu)_L\to WW$ & $\left[\tw^2(1+E_{\rm{cm}}^2 C_B) +(1+E_{\rm{cm}}^2 C_W)\right]^2$\\
			\hline
			$(\mu \mu)_R\to WW/Zh$ & $\left[2\tw^2(1+E_{\rm{cm}}^2 C_B)\right]^2$\\
			\hline
			$ (\mu\nu)_L\to WZ/Wh$ & $\left[\sqrt2(1+E_{\rm{cm}}^2 C_W)\right]^2$\\
		\end{tabular}
	\end{minipage}\hfill
	\begin{minipage}{.4\linewidth}
		\centering
		\includegraphics[scale=.4]{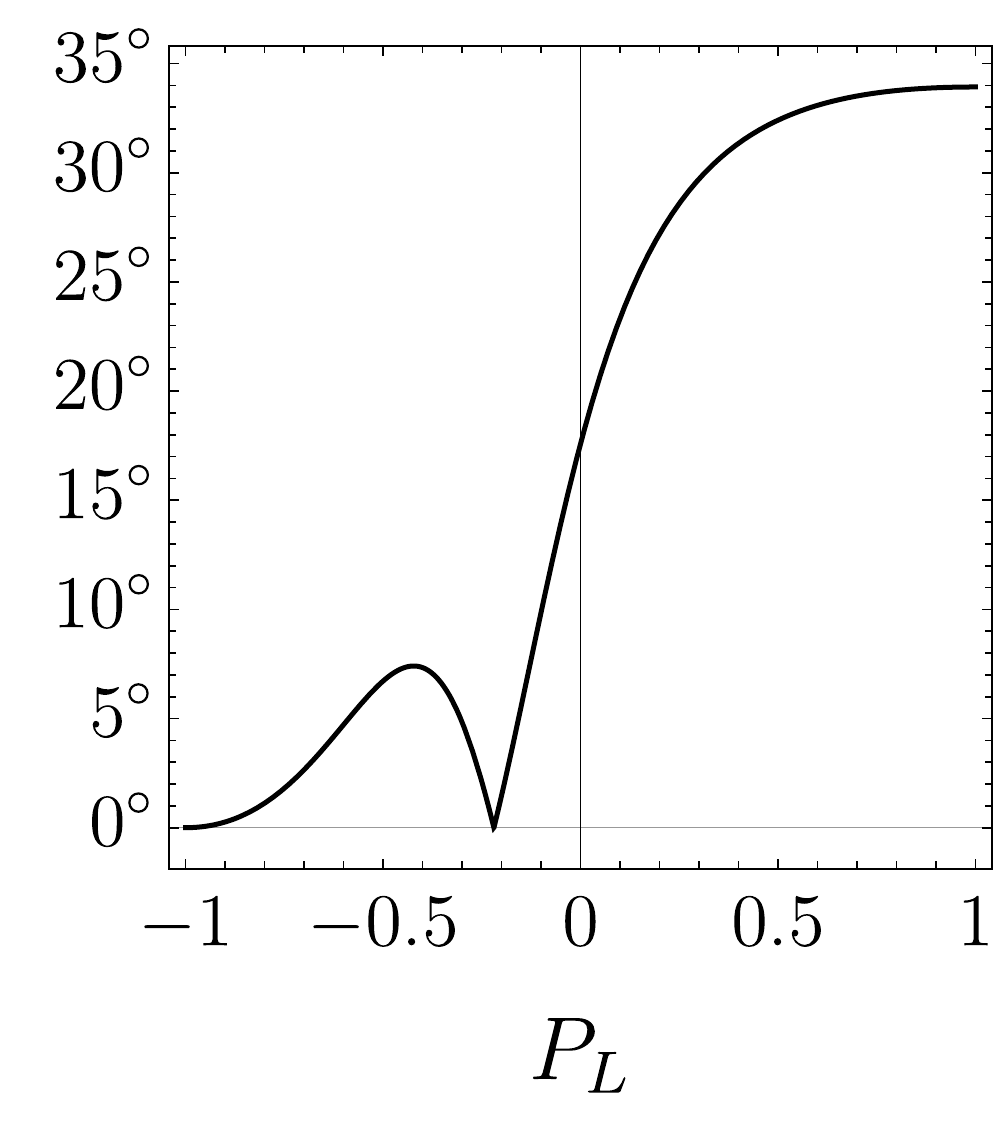}
	\end{minipage}
	\caption{Left: Born-level cross-sections, with $\tw$ the tangent of the Weinberg angle, normalized to a common $\sigma_0$ (whose expression is irrelevant). Right: the angle between the $ZH$ and $WW$ cross-section gradients as a function of the beam polarization fraction.\label{Tab:DibosonAmplitudes}}
\end{figure}

The origin of the flat direction in the tree-level sensitivity contour (showed dashed, on the left panel of Figure~\ref{Fig:ExclusionReachDiboson}) is readily understood analytically, by considering the gradients  ``$\nabla$'' of  the Born-level cross-sections  in the $(C_B,C_W)$ plane, around the SM point  $(0,0)$.  Using the results for $WW$ and $Zh$ shown in Figure~\ref{Tab:DibosonAmplitudes}
and rescaling the gradients by the common factor  $2\,E_{\rm{cm}}^2\sigma_0$ one finds
\beq\label{LRgrad}
\nabla_L^{Zh}=(1-\tw^2)\{-\tw^2,+1\} \,,\;\;\;\;\;
\nabla_L^{WW}= (1+\tw^2)\{+\tw^2,+1\} \,,\;\;\;\;\;
\nabla_R^{Zh}=\nabla_R^{WW}= 4\,\tw^4\{1,0\} \,,
\eeq
where sup- and sub-scripts refer respectively to the final states and to the chirality of the incoming fermions.
Notice that the $Zh$ and  $WW$ gradients for right-handed initial states are perfectly aligned, so that this contribution to the cross sections has a flat direction  (orthogonal to the gradient). The degeneracy can only be lifted by the left-handed contribution to the cross sections. However, given the small value of $\tw^2\simeq0.3$,  the gradients $\nabla_L^{Zh}$ and $\nabla_L^{WW}$ also form a relatively small angle, $\sim 30^\circ$. They are thus not very effective in lifting the flat direction when considering the total ($L+R$) contribution to the $WW$ and $Zh$ cross-section. Indeed, the angle between $\nabla_L^{Zh}+\nabla_R^{Zh}$ and $\nabla_L^{WW}+\nabla_R^{WW}$ is in the end only $\sim 17^\circ$ and thus the flat directions of the two cross-section measurements essentially coincide, as the right panel of Figure~\ref{Fig:ExclusionReachDiboson} shows. The combined likelihood is consequently also flat, in the same direction.

\begin{figure}
	\begin{minipage}{.5\linewidth}
		\centering
		\includegraphics[scale=.5]{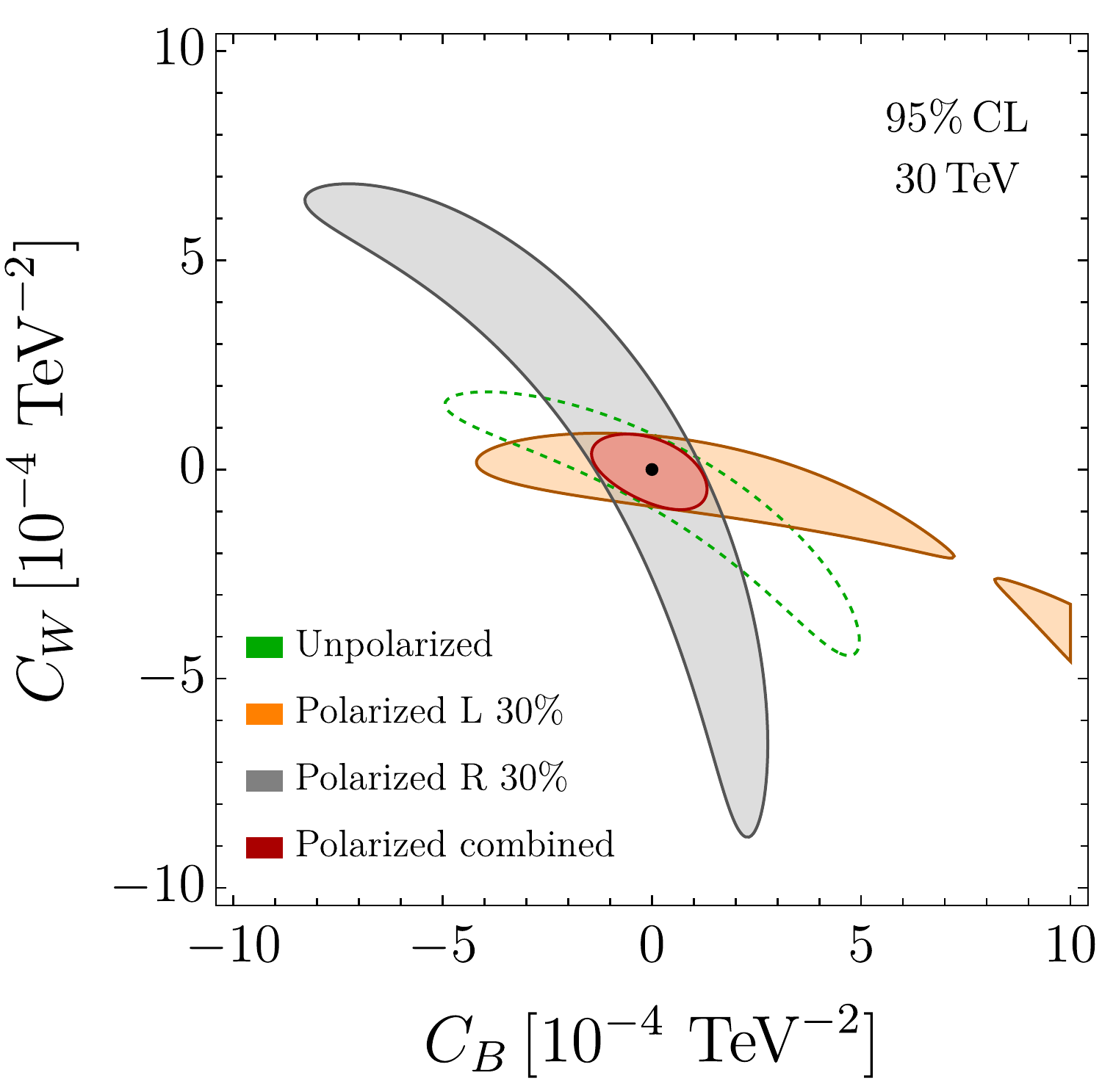}
	\end{minipage}%
	\begin{minipage}{.5\linewidth}
		\centering
		\includegraphics[scale=.5]{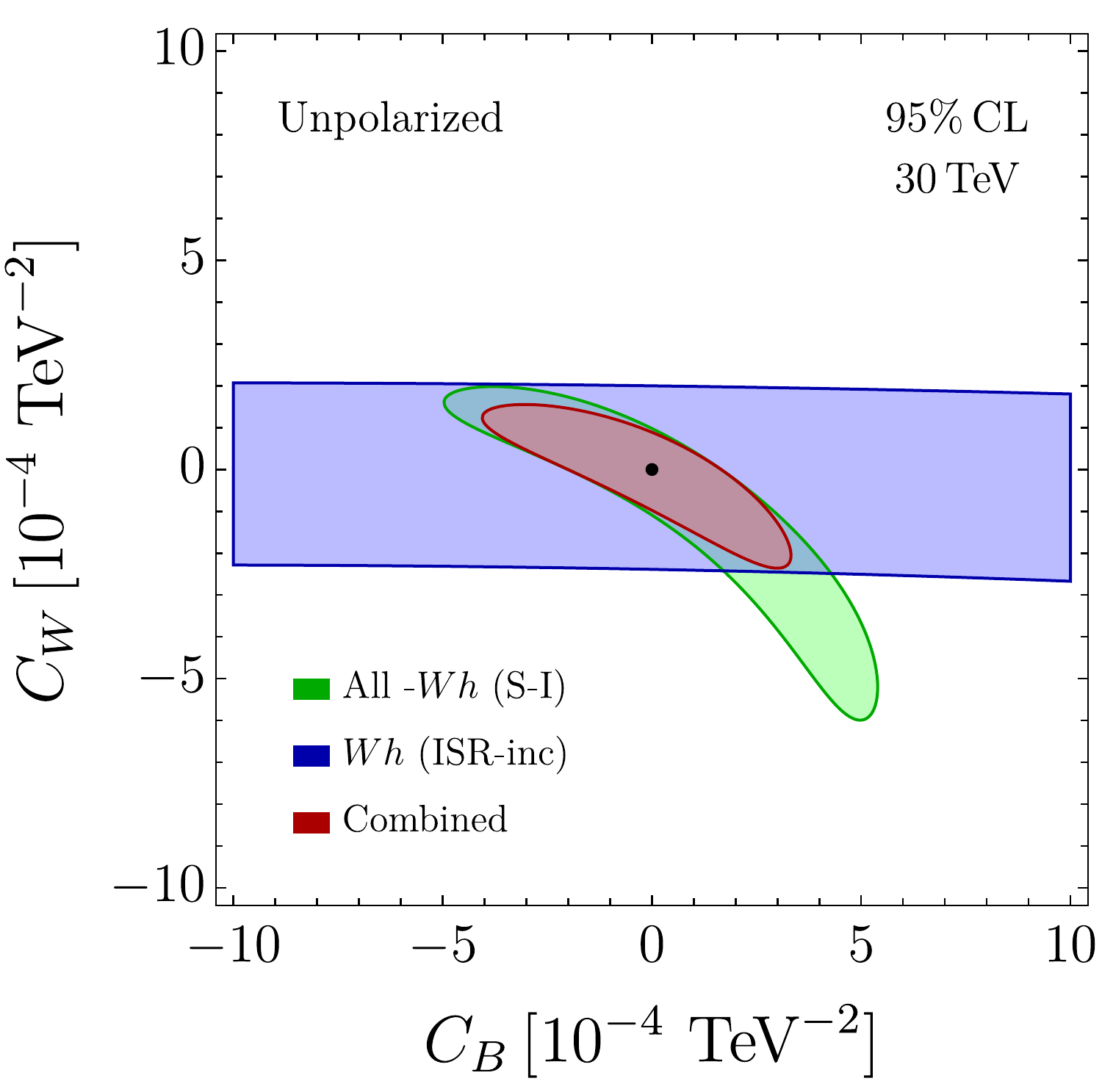}
	\end{minipage}
	\caption{Left Panel: $95\%$~CL contours in for $P_L=\pm30\%$ beam polarization. Right Panel: the impact of the ISR-inclusive $Wh$ cross-section measurement.}
	\label{Fig:DibosonPolarized}
\end{figure}

As evident in eq.~(\ref{LRgrad}), the $L$-gradients form  a large angle with the $R$-gradient. Therefore, if one could use polarized beams, the degeneracy would be eliminated by measuring the contribution of each chirality.
Considering  a polarization fraction $-P_L$ for the muon, and $+P_L$ for the anti-muon beam, the cross-section gradients read (we indicate by $\nabla_R$ the identical $\nabla_R^{Zh}$, $\nabla_R^{WW}$)
\beq
\nabla_{P_L}^{Zh}=\frac{(1+P_L)^2}4\nabla_L^{Zh}+\frac{(1-P_L)^2}4\nabla_R\,,\;\;\;\;\;
\nabla_{P_L}^{WW}=\frac{(1+P_L)^2}4\nabla_L^{WW}+\frac{(1-P_L)^2}4\nabla_R\,.
\eeq
The angle between the two gradients steeply increases for positive $P_L$, as indicated by the plot in the right panel of Figure~\ref{Tab:DibosonAmplitudes}. Correspondingly, even a modest amount of polarization has a considerable impact on the sensitivity. The left panel of Figure~\ref{Fig:DibosonPolarized} displays our sensitivity projections in a scheme where the VHEL integrated luminosity is equally split between positive and negative $P_L=\pm30\%$. The likelihood contour (in green) corresponding to  $P_L=+30\%$  is significantly smaller than that (in blue) for  $P_L=-30\%$, owing to the lifting  of the flat direction achieved for positive $P_L$. On the other hand, the   measurements  at $P_L=-30\%$ probe a direction complementary to that probed at $P_L=+30\%$. The combination of the two measurements  thus benefits the sensitivity. The impact of beam polarization was emphasized already in Ref.~\cite{Buttazzo:2020uzc}. Here we confirm that result, using more accurate predictions and including the entire set of exclusive and semi-inclusive cross-section measurements previously described.

Up to this point, we discussed the flat direction in the un-polarized likelihood (left panel of Figure~\ref{Fig:ExclusionReachDiboson}) by employing the tree-level cross-sections. When considering also EW radiation, the predictions are significantly affected, but the flat direction is not fully eliminated. For the exclusive $Zh$ and $WW$ cross-sections this is easily understood, since virtual radiation suppresses the $L$-processes more than the $R$ ones, owing to the larger Sudakov for incoming left-handed muons. The exclusive $Zh$ and $WW$ cross-sections gradients are thus even more aligned than the gradients of the corresponding tree-level predictions. The semi-inclusive cross-sections for $Zh$ and $WW$ production are also quite aligned, among them and with the exclusive cross-sections. This was expected because the partial cancellation between real and virtual logarithms make semi-inclusive cross-sections not vastly different from the tree-level ones.

On the contrary, the measurement of the charged processes $WZ$ and $Wh$ could have been expected to eliminate or mitigate the flat direction, because they are strongly  sensitive to the Born cross-section of the charged scattering $(\mu\nu)_L\to WZ/Wh$ (see the left panel of Figure~\ref{Tab:DibosonAmplitudes}). The associated gradient 
\beq
\nabla_L^{\rm{ch}}=2\{0,+1\}\,,
\eeq
points in a different direction than $\nabla_L^{Zh}$, $\nabla_L^{WW}$ and $\nabla_R$. Therefore the gradient of $\sigma^{Wh/WZ}$ could in principle point in a direction that is completely different from that of the (nearly parallel) gradients of the $Zh$ and $WW$ cross-sections. However, by expanding at the first order in ${\cal{L}}=g^2/16\pi^2\log^2(E_{\rm{cm}}^2/\MW^2)$, the unpolarized (longitudinal) $WZ$ and $Wh$ cross-sections are approximately equal and read
\beq\label{xsapp}
\sigma^{WZ}\simeq\sigma^{Wh}\simeq\frac14 {\cal{L}}\cdot (\sigma_{B}^{Zh}+\sigma_{B}^{WW}+\sigma_{B}^{{\rm{ch}}})\,,
\eeq
where $\sigma_{B}^{{\rm{ch}}}$ is the charged Born cross-section reported on the left panel of Figure~\ref{Tab:DibosonAmplitudes} (times $1/4$, from the polarization average) and ${\sigma_{B}}^{\hspace{-7pt}Zh,WW}$ are the Born cross-sections of the neutral processes. Therefore, the charged cross-section gradient $\nabla_L^{\rm{ch}}$ must compete with the (nearly parallel) gradients of $Zh$ and $WW$, and its size happens to be insufficient to produce a large misalignment angle between the $\sigma^{Wh/WZ}$ and $\sigma^{Zh/WW}$ gradients. 

The situation would be improved, if we could tailor an observable in which the  $\sigma_{B}^{Zh}$ and $\sigma_{B}^{WW}$ contributions in eq.~(\ref{xsapp}) are eliminated or reduced. Notice, for that purpose, that the $Zh$ and $WW$ terms in eq.~(\ref{xsapp}) can be interpreted as due to one hard $\mu\mu$ neutral-current scattering, followed by the radiation of one charged $W$ boson from the final legs of the hard process. The $W$ is thus preferentially collinear to the final states. The $\sigma_{B}^{{\rm{ch}}}$ term comes instead from the radiation of a $W$ from the initial state, collinear to the beam axis, followed by a hard $\mu\nu$ scattering.\footnote{This interpretation would straightforwardly correspond to Feynman diagrams in a physical gauge, where  DL's are associated to emissions from individual legs. We already remarked that in covariant gauges instead they arise from the interference between emission from strictly different legs.} This suggests to consider alternative  $Wh$ and $WZ$ cross-sections that exclude final state radiation (FSR) while being inclusive  on initial state radiation (ISR). FSR consists of soft radiation collinear to  the hard particles in the final state, which is precisely the source of the first two terms in eq.~(\ref{xsapp}). Excluding FSR, the resulting ``ISR-inclusive'' cross-sections are expected be roughly
\beq
\sigma^{WZ/Wh}_{\rm{ISR-inc}}\simeq\frac14{\cal{L}}\,\sigma_{B}^{{\rm{ch}}}\,.
\eeq
This observable  should thus be mostly sensitive to $C_W$, and its measurement should produce a nearly horizontal band in the $(C_B,C_W)$ plane, thus eliminating the flat direction.

Unfortunately we are unable to produce resummed predictions for the ISR-inclusive cross-sections with the IREE methodology. We can however illustrate the impact of such measurements using tree-level {\texttt{MadGraph}}~\cite{Alwall:2014hca} predictions with the SMEFT@NLO model~\cite{Degrande:2020evl}, focusing in particular on the $Wh$ channel. Specifically, we simulate the process
\beq
\mu^+\mu^-\to W^+W^-h\,,
\eeq
at $E_{\rm{cm}}=30$~TeV, with the following selection cuts. First, we identify as ``hard'' the $W$ boson that forms, together with the Higgs, the pair with the highest invariant mass. Secondly  we ask  this mass to be above $0.85\cdot E_{\rm{cm}}=25.5$~TeV and the hard $W$ and $h$  to be within the central region $\theta_* \in [30^\circ, 150^\circ]$. These selections enforce the occurrence of a hard scattering, and correspond to our definition of a ``semi-inclusive'' process. We further restrict to the ``ISR-inclusive'' region by asking the other (``soft'') $W$ to be parallel to the beam, in a cone of $20^\circ$. Since the emission of at least one soft $W$ is required for $Wh$ production, the latter cut effectively corresponds to a veto on central EW radiation.\footnote{The attempt made in Ref.~\cite{Buttazzo:2020uzc} to exploit the $WWh$ process  did not impose the crucial angular cut that defines the ISR-inclusive region.}

The above estimate of the ISR-inclusive cross-section produces the blue band on the right panel of Figure~\ref{Fig:DibosonPolarized}.  As expected, the band is nearly horizontal. In the figure we also display, in green, the $95\%$~CL contour of the likelihood including all the measurements discussed in the present section, apart from the measurement of the semi-inclusive $Wh$ cross-section which is correlated with the ISR-inclusive measurement. The combination of the two contours, shown in red, strongly mitigates the flat direction issue. Notice however that our tree-level estimate of the ISR-inclusive cross-section could be subject to large errors, and resummed predictions should be employed for a conclusive assessment of the sensitivity gain.

\subsection{BSM sensitivity}\label{Sec:4}

\begin{figure}
	\begin{minipage}{.5\linewidth}
		\centering
		\includegraphics[scale=.5]{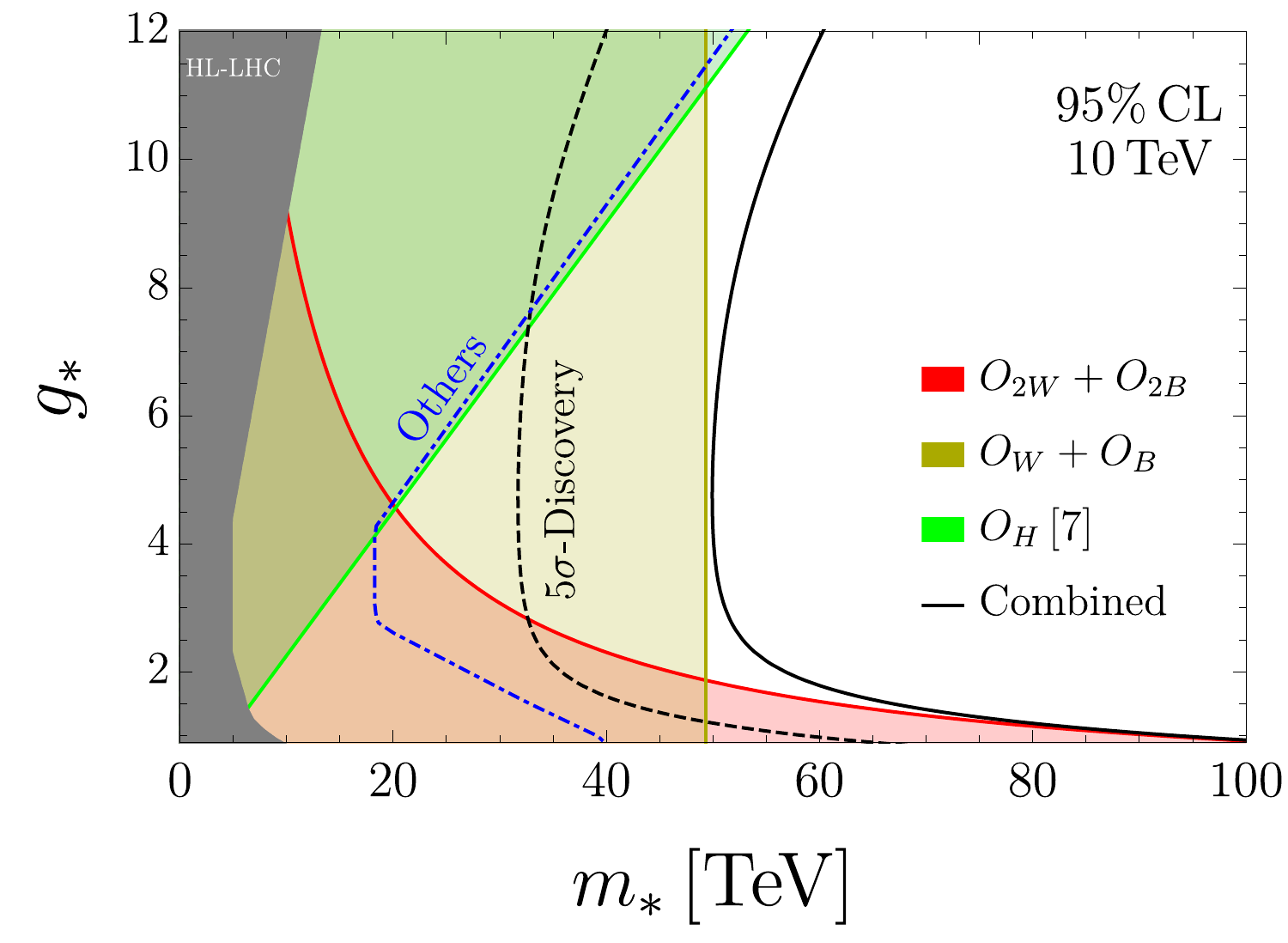}
	\end{minipage}%
	\begin{minipage}{.5\linewidth}
		\centering
		\includegraphics[scale=.5]{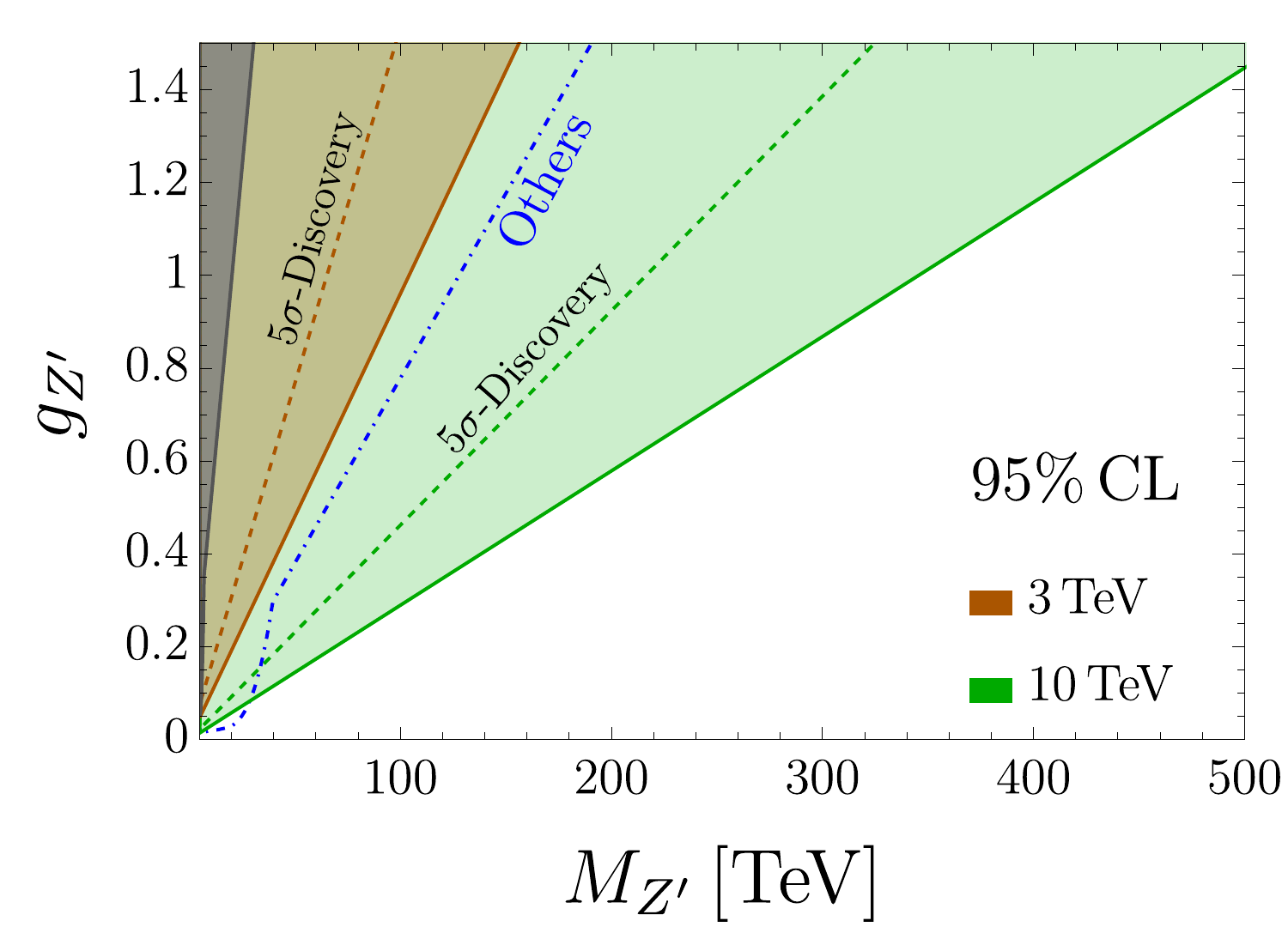}
	\end{minipage}
	
	\caption{Left Panel: $95{\%}$ exclusion reach on the Composite Higgs coupling-mass parameter space. The reach for $O_H$ is taken from~\cite{Buttazzo:2020uzc}. Right Panel: sensitivity projections for a $Y$-universal $Z'$ model.  The gray band and the blue dash-dot line represent respectively the region probed by the HL-LHC program and the sensitivity projections for all other future collider projects~\cite{EuropeanStrategyforParticlePhysicsPreparatoryGroup:2019qin}.}
	\label{Fig:CHandZprime}
\end{figure}

\paragraph{Composite Higgs.}

As a first concrete scenario of new physics we consider Composite Higgs~\cite{Kaplan:1983fs,Agashe:2004rs,Panico:2015jxa}. In this scenario, the Higgs is a composite Pseudo-Nambu-Goldstone boson emerging from some strong dynamics at a scale $m_*$. In principle the underlying dynamics could arise from gauge interactions, like in QCD. However the only concrete realistic constructions, accounting for the origin of both the fermion masses and the scale $m_*$ itself, have been obtained in the context of warped compactifications. In these constructions, compositeness occurs in a holographic sense. Within the Composite Higgs scenario, the size of the Wilson coefficients in the resulting low energy EFT, can be estimated, under simple but robust dynamical assumptions, in terms of the mass scale $m_*$ and overall coupling strength $g_*$ of the underlying strong dynamics~\cite{Giudice:2007fh}. Furthermore, simple considerations suggest $g_*\lsim4\pi$, while the existence of $\mathcal{O}(1)$ couplings within the SM implies $g_*\gsim 1$. The Composite Higgs power-counting rules predict the Wilson coefficients of the operators in the left column of Table~\ref{Tab:OpWarsaw} to scale as
\begin{equation}
C_{2W} = -\frac{c_{2W}}{2} \frac{g^2}{g_*^2m_*^2},\quad C_{2B} = -\frac{c_{2B}}{2}\frac{g^{\prime 2}}{g_*^2m_*^2},\quad C_W = c_W\frac{1}{m_*^2},\quad C_B = c_B\frac{1}{m_*^2}\,,
\end{equation}
where the dimensionless coefficients $c_{2W}$, $c_{2B}$, $c_W$, $c_B$ are expected to be of order~$1$. Even though it does not affect the processes studied in this paper, an important role is also played by
\begin{equation}
O_H = \partial_\mu(H^\dagger H)\partial^\mu(H^\dagger H),\qquad C_H=\frac{c_H}{2}\frac{g_*^2}{m_*^2}\,.
\end{equation}
In our sensitivity projections we will report  the corresponding bounds, as obtained in~\cite{Buttazzo:2020uzc} by studying the process $\mu^+\mu^- \to hh\nu\nu$ at tree level. Other probes of $C_H$ at the muon collider, from Higgs coupling measurements, are superior or competitive at the lower energy muon colliders~\cite{Buttazzo:2020uzc}, but they are not considered in the sensitivity plots.

Using the above scalings, and setting all the $c$ coefficients to $1$, we can translate the bounds of Section~\ref{Sec:3} for a $10\text{ ~TeV}$ muon collider into sensitivity estimates in the plane ($m_*$, $g_*$),  as in Figure~\ref{Fig:CHandZprime}. In the same plot we also report the HL-LHC sensitivity projections, the envelope of the $95\%$~CL sensitivity contours of all the future collider projects that have been considered for the~2020 update of the European Strategy for Particle Physics~\cite{EuropeanStrategyforParticlePhysicsPreparatoryGroup:2019qin}. The advantage of the muon collider is evident. Results at muon colliders with different energies, with an integrated luminosity scaling as in eq.~(\ref{intum}), are reported in Appendix~\ref{App:Results}.

\begin{figure}
	\begin{minipage}{.5\linewidth}
		\centering
		\includegraphics[scale=.5]{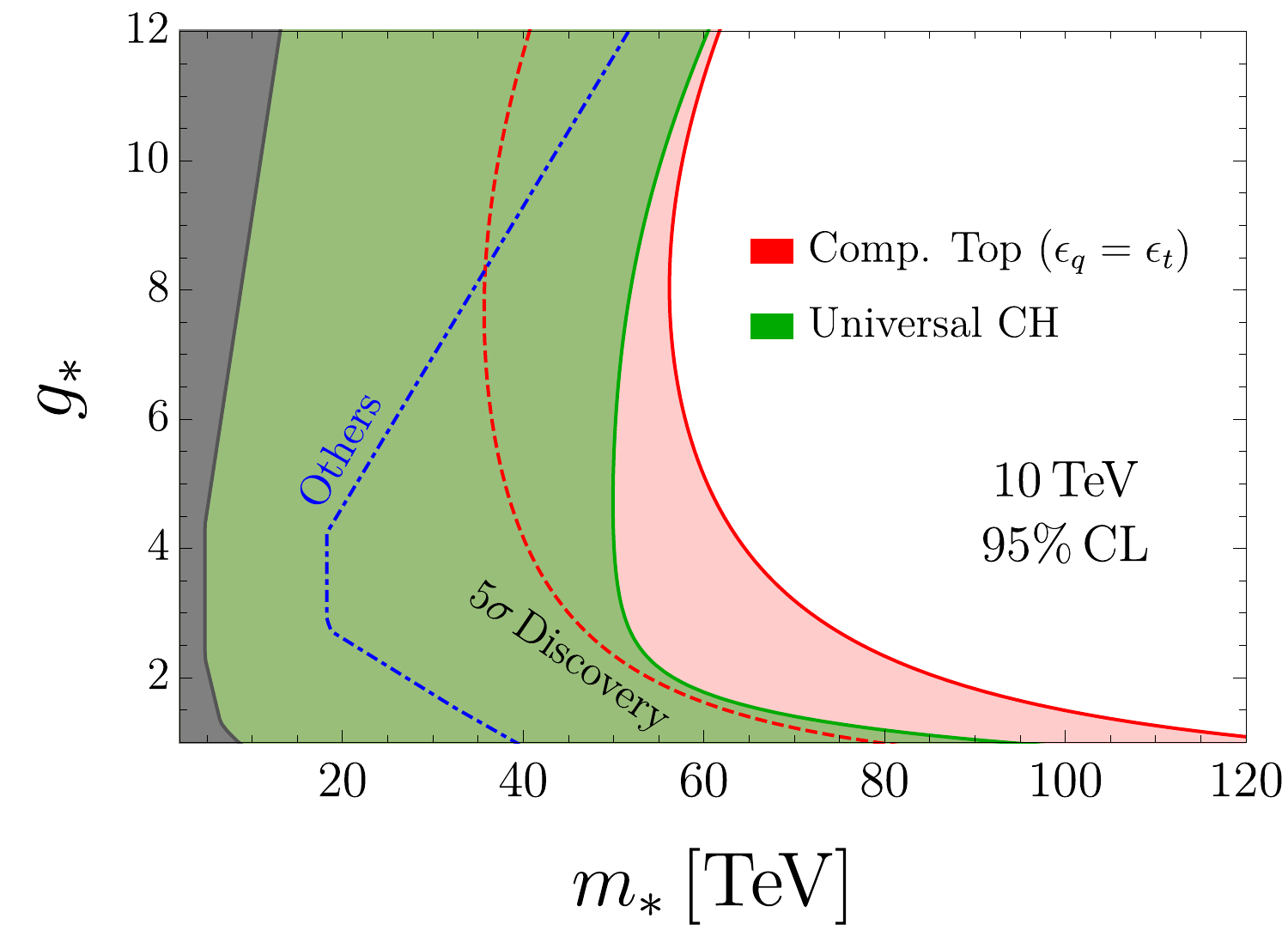}
	\end{minipage}%
	\begin{minipage}{.5\linewidth}
		\centering
		\includegraphics[scale=.5]{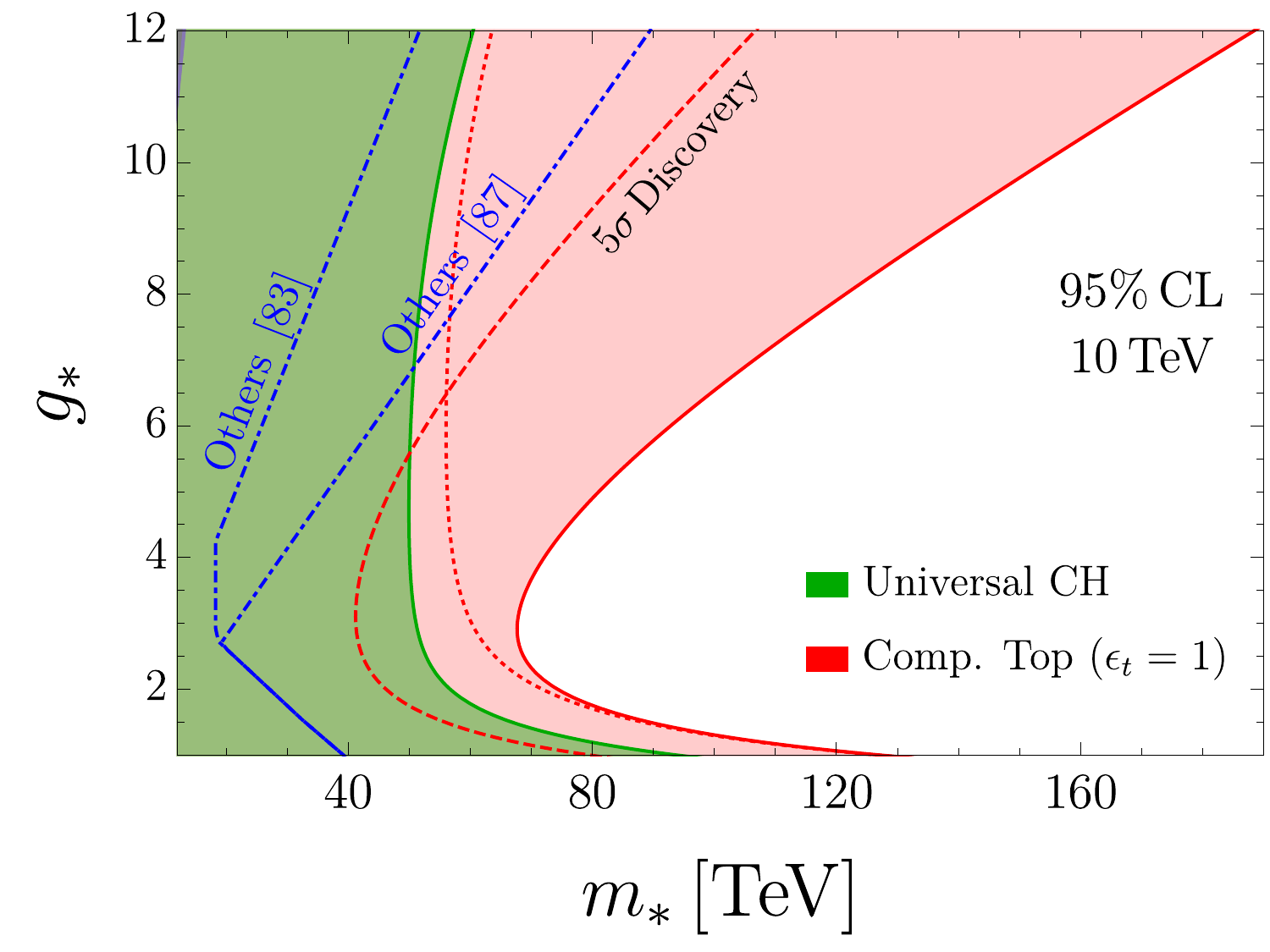}
	\end{minipage}
	\caption{$95{\%}$ exclusion reach for the two partial compositeness scenarios under consideration. The green shapes represent the combined bound from the flavor universal measurements, while the red contours also includes the di-top and di-bottom constraints. The projected sensitivity of other future collider projects and the gray band of HL-LHC are taken from Ref.~\cite{EuropeanStrategyforParticlePhysicsPreparatoryGroup:2019qin}. The right panel ($\epsilon_t=1$) also includes the stronger CLIC sensitivity estimated in Ref.~\cite{Banelli:2020iau}. }
	\label{Fig:CT}
\end{figure}

\paragraph{Composite top.} The results for purely bosonic operators we just discussed  apply robustly to basically all composite Higgs scenarios. Operators involving fermions are more sensitive to the assumptions on the flavor dynamics, but one convenient option is offered by the mechanism of partial compositeness \cite{Kaplan:1991dc}, under which the elementary fermions mix linearly with heavy partners in the strong sector. Due to its large Yukawa coupling, the top quark is expected to have a large mixing with its partners and therefore precise measurements involving the third family represent an appealing opportunity to probe new physics.  

At the muon collider the most relevant effects are expected in $t\bar t$  and $b\bar b$ production.\footnote{See~\cite{Durieux:2018tev} for a similar analysis for CLIC.} Indeed  the dimension-6 operators in the last block of Table~\ref{Tab:OpWarsaw} gives rise to contributions that grow with $E_{\rm{cm}}$ and which can be exploited at the large energy of the muon collider. In a model-independent approach one can  parametrize the  ``amount of compositeness'' of  respectively the $3^{\text{rd}}$ quark family left-handed  doublet and right-handed up-type singlet by $\epsilon_q$ and $\epsilon_t$. These quantities range from $0$ to $1$. Given the universal coupling strength $g_*$ of the strong sector the resulting top Yukawa coupling scale as~\cite{Giudice:2007fh}
\begin{equation}
y_t \sim  \epsilon_q \epsilon_t g_* \,.
\end{equation}
The relevant Wilson coefficients are then expected to scale as (see~\cite{Durieux:2018ekg} for a short review) 
\begin{align}
C_{qD}^{(3)} = c_{qD}^{(3)} \frac{g \, \epsilon_q^2}{m_{*}^2}\,, &&C_{qD}^{(1)} = c_{qD}^{(1)} \frac{g' \epsilon_q^2}{m_{*}^2}\,, && C_{tD} = c_{tD} \frac{g' \epsilon_t^2}{m_{*}^2}\,,
\label{Eq:PCTop}
\end{align}
where the $c_i$ are, as usual, expected to be order $1$ coefficient.  For concreteness we focus on two benchmark scenarios, where we fix $\epsilon_t$ and   $\epsilon_q$ and leave  $g_*$ and $m_*$ free. In the first scenario, the right-handed top quark is assumed to be fully composite, corresponding to  $\epsilon_t=1$ and $\epsilon_q=y_t/g_*$. In the second, the two top chiralities are assumed equally composite, that is $\epsilon_q=\epsilon_t = \sqrt{y_t/g_*}$.  

Notice that the contribution of the operator 
\beq
O_{tt}\equiv \frac{1}{2}( \bar{t}_R \gamma^{\mu} t_R)( \bar{t}_R \gamma_{\mu} t_R)\,,
\eeq
to the Wilson coefficients of the $O_{tD}$, through Renormalization Group (RG) evolution, is not negligible in the scenario of total right-handed top quark compositeness~\cite{Banelli:2020iau}. Using the power-counting estimate
\beq
C_{tt}= \epsilon_t^4 \frac{g_*^2}{m_*^2}c_{tt}\,,
\eeq
we obtain a correction~\cite{Banelli:2020iau} to the $C_{tD}$ coefficient at a scale $\mu=E_{\rm{cm}}$
\begin{align}
\displaystyle
C_{tD} (\mu) = C_{tD} (m_{*}) + C_{tt}(m_{*}) \frac{32}{9} \frac{g'}{16 \pi^2} \log \left( \frac{m_*^2}{\mu^2} \right)= \epsilon_t^2 \frac{g'}{m_*^2} \left(c_{tD} +c_{tt} \frac{32}{9} \frac{\epsilon_t^2g_*^2}{16 \pi^2}\log \left( \frac{m_*^2}{\mu^2} \right) \right)\,.
\label{Eq:RunnCtt}
\end{align}
This correction is sizable if $\epsilon_t\sim 1$, especially for large $g_*$, because the sensitivity of the muon collider extends to a scale $m_*$ that is significantly larger than $E_{\rm{cm}}$.

There are in principle three more operators \parbox[c][0pt]{92pt}{$O_{Ht}$,  $O_{Hq}^{(1)}$ and $O_{Hq}^{(3)}$} (defined as in the ``Warsaw'' basis \cite{Grzadkowski:2010es}) that mix significantly with those in eq.~(\ref{Eq:PCTop}) through  RG evolution. However, the their effects can only be important in the case where $\epsilon_q\sim 1$, which we do not contemplate in our analysis. We will therefore neglect the RG effects of the latter three operators and consider only those of $O_{tt}$.

Our results are summarized in Figure~\ref{Fig:CT}, where we report the projected exclusion reach in the $g_{*}$ and $m_{*}$ plane in the two scenarios under consideration for $E_{\rm{cm}}=10$ TeV. Additional results can be found in Appendix~\ref{App:Results}. Starting from the  scenario of equal compositeness (left panel) we  notice that at $g_{*}$ the additional hypothesis of top compositeness extends the muon collider potential to probe the scale of Higgs compositeness $m_*$.  The effect is even stronger for fully composite $t_R$ (right panel), which shows that di-top measurements can cover up to $m_*\sim 150$ TeV for $g_{*}\gtrsim 8$. We should point out, however, that this result depends on the exact $\mathcal{O}(1)$ value of the $c_{tt}$, $c_{tD}$ coefficients in eq.~\eqref{Eq:RunnCtt}. This dependence is illustrated in Figure~\ref{Fig:CTAppctt}, where we set $c_{tD}=1$ and we vary the value of $c_{tt}$.

Finally we remark that a detailed analysis of the composite Higgs scenario with partial compositeness would require specific hypotheses on the flavor dynamics and a detailed inspection of the flavor observables. Depending on those hypotheses, principally the maximality or minimality of the underling flavor symmetry, the resulting flavor constraints on the new physics scale $m_*$ can vary dramatically.  While a comprehensive analysis clearly exceeds the purposes of this work, a perspective can be gained by considering available studies. As shown in Ref.~\cite{Barbieri:2019kfa}, under the strongest assumptions, that is for a symmetry structure offering the best protection from unwanted effects, flavour and CP observables could start exploring the range $m_{*}= O(10)$~TeV in the next decade or so, given the availability of better measurements and assuming better theoretical predictions. This is roughly the same range explored by a $3$~TeV muon collider. Moreover the $m_* \sim 50$ TeV reach of a $10$ TeV muon collider vastly surpasses any conceivable improvement of flavour constraints, and competes with the more stringent flavour bounds obtained by making more generic assumptions on the flavor dynamics. Notice also that the present lepton flavor universality anomalies in B-decays, at least the seemingly more prominent ones in neutral currents, suggest a new physics scale in the $\sim 30$ TeV range, which could be explored both directly and indirectly by the muon collider.

\paragraph{$Y$-Universal $Z'$ model}
The $Y$-universal $Z'$ model represents a simple extension of the SM, featuring an additional heavy gauge boson, of mass $M_{Z'}$, on top of the SM particles.\footnote{See \cite{EuropeanStrategyforParticlePhysicsPreparatoryGroup:2019qin} for details.  Concrete BSM scenarios featuring additional $Z'$s can be found, for instance, in Ref.~\cite{Appelquist:2002mw}.} In this benchmark scenario the new vector charges are aligned with the SM hypercharge with coupling $g_{Z'}$. Requiring the width of the $Z'$ to not exceed $0.3 M_{Z'}$ sets the perturbative limit on the coupling to be $g_{Z'} \lesssim 1.5$. At energies below $M_{Z'}$, integrating out the $Z'$ only generates the $O_{2B}'$ operator of Table~\ref{Tab:OpWarsaw}. The Wilson coefficient of the operators corresponds, by eq.~(\ref{WYW}), to 
\beq
{\rm{Y}}=\left( \frac{g_{Z'} \,m_W}{g^{\prime }\, M_{Z'}}\right)^2\,.
\eeq

The sensitivity projections are reported in the right panel of Figure~\ref{Fig:CHandZprime}.  The orange and green regions are the ones probed by muon colliders at $3$ and $10$~TeV energy, respectively.  The gray band represents the expected exclusion reach from HL-LHC, while the blue line indicates the combined sensitivity from other future collider projects (dominantly FCC, and the $3$~TeV stage of CLIC). The $3$~TeV muon collider sensitivity is obviously similar to the one of CLIC. A $10$ TeV machine would greatly improve this result probing up to $500$~TeV for large (but still perturbative) coupling.  The dashed lines represent the discovery reach, showing that already at $3$~TeV there are vast opportunities for indirect discovery, well above the region that the HL-LHC might exclude.  Results at higher muon collider energies are reported in Figure~\ref{Fig:ZPrime}.

\begin{figure}[t]
	\centering
	\includegraphics[scale=.5]{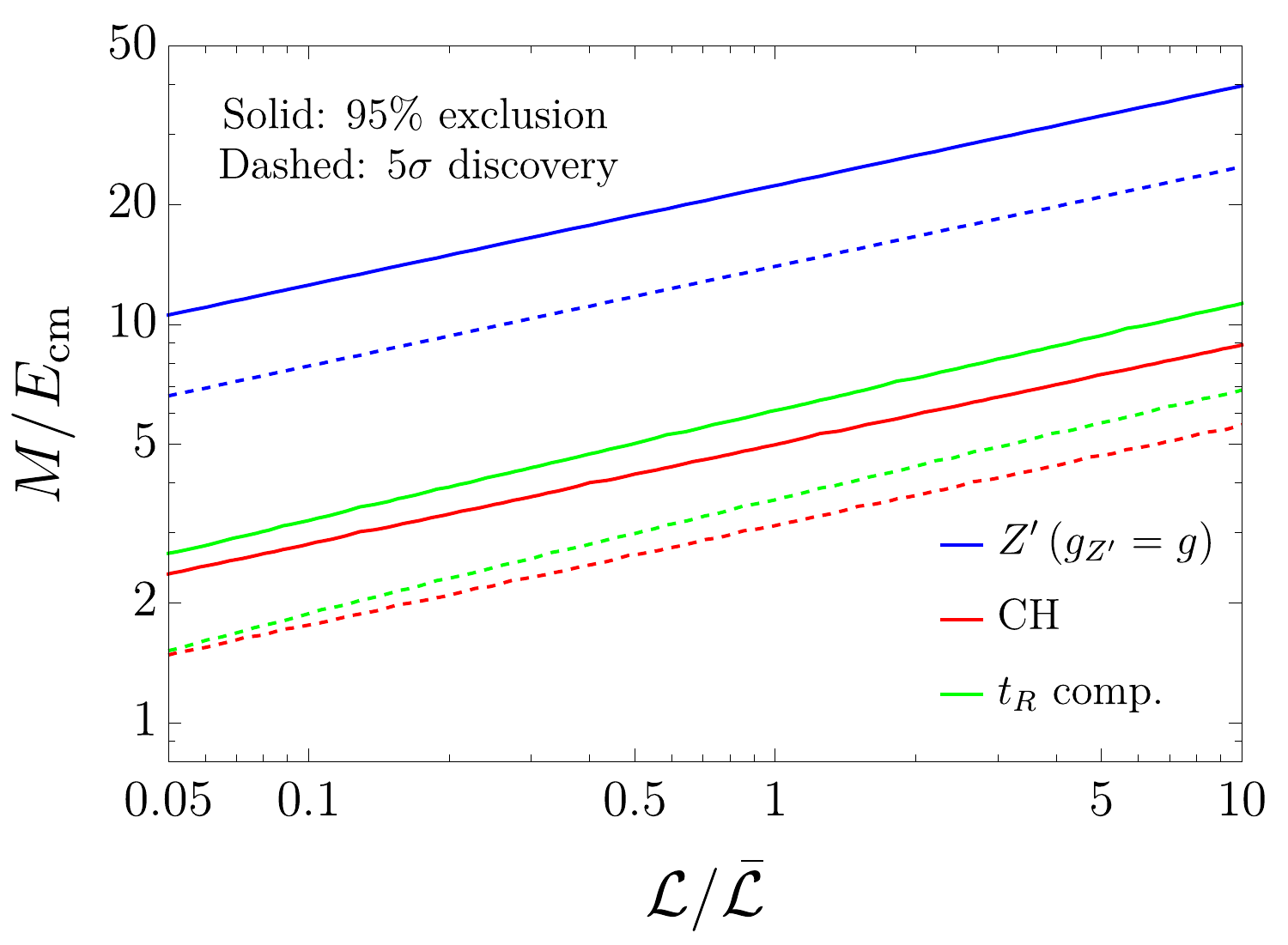}
	\caption{Reach on the new physics scale at $2\sigma$ (continuous) and at $5\sigma$ (dashed), relative to the collider energy, as a function of the integrated luminosity normalized to eq.~(\ref{intum}). The red lines are for Universal manifestations of Higgs compositeness, while the green ones include the effect of Top compositeness in the $\epsilon_t=1,\,\epsilon_q=y_t/g_*$ scenario. The blue lines are for the $Y$-universal $Z'$ for a fixed coupling $g_{Z'} = g$.}
	\label{Fig:lumivsbound}
\end{figure}

\section{Conclusions and outlook}\label{Sec:5}

We have studied the interplay between two classes of novel phenomena, which can be observed at future lepton colliders with very large center of mass energy. The first class consists of hard scattering processes induced by new physics at around $100$~TeV. The second class consists of the long-distance effects of EW radiation. Both phenomena play a relevant role at  lepton colliders with $E_{\rm{cm}}\sim10$~TeV. In particular they are relevant  at muon colliders, which are the main target of the present work.

The interplay manifests itself in two ways. The first one is simply that EW radiation effects on the SM predictions are large (see Sections~\ref{Sec:Diferm} and \ref{Sec:Dibos}) and require to be included and resummed with high accuracy in order to isolate the putative BSM contribution to the measurements. EW radiation thus plays for muon colliders a similar role as QCD radiation for the LHC, with the difference (discussed in the Introduction) that its effects can not be mitigated by the choice of suitable (inclusive) observables. Therefore they are actually even more important for muon colliders than QCD is for the LHC.

The second and possibly more interesting aspect of the interplay is given by the influence on the pattern of EW radiation operated by the presence of new physics in the hard scattering amplitude. This makes observables that require or that exclude the presence of radiation display a different dependence on the new physics parameters, and the sensitivity profits from their combined measurements as illustrated in Sections~\ref{Sec:3.1} and~\ref{Sec:3.2}, and in Appendix~\ref{3rd}. 

Our sensitivity projections rely on putative measurements of exclusive and semi-inclusive cross-sections. Both classes of processes are characterized by the occurrence of a hard scattering, with two particles in the final state carrying almost all the available energy. The emission of additional EW bosons and hard photons is vetoed in the exclusive case and allowed in the semi-inclusive one. We computed the resummed  semi-inclusive cross-sections in double logarithm (DL) approximation by extending the IREE methodology \cite{Fadin:1999bq}, as described in Section~\ref{Sec:IREE}. The exclusive cross-sections were  computed at DL, but also including single logarithms  at 1-loop, which we found to be sizable.

The studies performed in this paper should be improved and extended in many directions. Better predictions will be definitely needed in order to approach the percent-level accuracy target that is needed to fully exploit the statistical precision potential of a muon collider. Moreover, given the magnitude of the radiation effects we observed, it is possible that more accurate predictions will  considerably  affect some of our sensitivity projections. A first step in that direction, which we leave for future work, is the inclusion of single logarithms at fixed leading order in the semi-inclusive predictions. That could be achieved by combining one loop virtual logarithms with a factorized treatment of  real emission. That calculation,  would possibly  also help clarify the connection between  soft-collinear effects (studied in this paper) and the PDF/Fragmentation Function treatment of EW radiation. It is not unconceivable that the same approach could be extended at the two loop order. In parallel, the impact of resummation beyond DL  should be assessed. The SCET methodology currently offers the most promising approaches.

Another priority is to investigate further  classes of cross-sections, sensitive to different EW radiation patterns. Our results indicate that investigation should be done on the basis of the structure of short distance new physics. At the end of Section~\ref{Sec:3.2} we took one step in that direction, showing that  the approximately flat direction in the $(C_B,C_W)$ likelihood contours is mitigated, in the absence of polarized beams, by considering an ``ISR-inclusive'' cross-section. This third cross-section type is inclusive on  radiation collinear to the beam, but vetoes centrally emitted radiation. We could not compute the ISR-inclusive cross-section at  DL  with our IREE methodology and limited ourself to a tree-level estimate. A straightforward direction for progress would be to perform that calculation and verify if and how it impacts our findings.

The definition and study of cross-sections should be also based on experimental considerations. We already pointed out that exclusive cross-sections are problematic in that regard. Indeed imposing the radiation veto requires experimental sensitivity to EW radiation that is emitted in all directions, including the forward and backward regions along the beam line. The angular coverage of the muon collider detector is still to be quantified, however we expect that it will be insufficient for the measurement of exclusive cross-sections.

In view of the above, it is important to emphasize that our sensitivity projections have been verified  to not change radically when exclusive cross-section measurements are not available. This conclusion is not in contradiction with (and cannot be inferred from) our sensitivity plots, where (see e.g. Figure~\ref{Fig:ExclusionReachWY}) we observe a strong complementarity between ``exclusive'' observables   and observables ``with radiation''. Indeed for neutral processes the latter observables consists of  the difference between the semi-inclusive and the exclusive cross-sections. Therefore the impact of eliminating the exclusive measurements can not be estimated by suppressing the ``exclusive'' cross section measurement  in the computation of the likelihood. The proper estimate is obtained by employing the semi-inclusive neutral processes without subtraction, combined with the charged measurements, produces a combined reach that is not much inferior to the one that exploits the exclusive measurements. In essence, the main sensitivity gain due to radiation stems from the emergence of the charged processes and from their complementarity with the neutral ones. The complementarity between neutral measurements with different degrees of radiation inclusiveness (e.g., exclusive versus semi-inclusive) plays instead a less relevant role in our results. This same qualitative behavior can be observed in the comparison between the neutral and charged lepton production processes on the right panel of Figure~\ref{Fig:ExclusionReachWY0}.

On the other hand, the complementarity between exclusive and semi-inclusive measurements exists and can benefit  the sensitivity as we illustrated on the left panel of Figure~\ref{Fig:ExclusionReachWY0}. It plays a marginal role in the combined fit to the limited number of EFT operators we considered in this paper. It could however be relevant in a more global exploration and characterization of putative new physics. One way to recover sensitivity, if exclusive measurements were  indeed unavailable, could be to exploit the ISR-inclusive cross-sections. These are easier to measure because they do not require sensitivity to radiation in the forward and backward regions. This aspect should be investigated.

Our phenomenological results strengthen and extend previous estimates of the muon collider sensitivity to heavy new physics. We have a considered a variety of BSM scenarios with new physics coupled to the SM with electroweak  strength. We found that a $E_{\rm{cm}}=10$~TeV  muon collider can probe up to a scale ranging  from $50$ to $200$~TeV. The reach improves linearly with $E_{\rm{cm}}$. These figures are significantly above the  potential (direct and indirect) sensitivity of other future collider projects, and above the direct sensitivity reach of the muon collider itself, which is obviously bounded by the collider energy. 

The indirect sensitivity to scales that are well above the direct reach is a great addition to the physics case of a muon collider, whose relevance would not be diminished, but on the contrary augmented, by the occurrence of a direct discovery. Indeed, direct hints for new particles observed at the muon collider will turn into a full-fledged discovery of new physics only after unveiling the underlying theoretical description of their dynamics. The possibility of probing this dynamics well above the particle's mass will play a decisive role in this context. Furthermore, the direct manifestation of new physics might be hard to detect. Perhaps, indirect probes will provide the first hint of its existence, to be eventually confirmed by targeted direct searches. Finally, indirect searches for BSM phenomena based on precise measurements guarantee  a sound output of the project. The connection with the phenomenon of EW radiation, which is interesting per se, adds scientific value to the program. 

Before concluding, we discuss the impact of the integrated luminosity on our results. We employed the baseline luminosity in eq.~(\ref{intum}), which corresponds to $90$~ab$^{-1}$ for a $30$~TeV muon collider. Since the possibility of reducing the $90$~ab$^{-1}$ target is under discussion, it is worth assessing the impact of a lower integrated luminosity on our conclusions. An important aspect is associated with the actual experimental feasibility of the relevant measurements. While a conclusive assessment will require dedicated studies, the expected number of events in Table~\ref{Tab:Processes}~\footnote{The table is for $E_{\rm{cm}}=10$~TeV, however with the scaling in eq.~(\ref{intum}) the results depend weakly on the energy.} and the corresponding efficiencies show that, with a factor $10$ reduction in luminosity, some of the processes we employed would be left with a handful of observed events, possibly preventing the corresponding measurements. If the reduction in luminosity were less extreme, the sensitivity to  the scale of new physics would simply deteriorate as the fourth root of the luminosity, as shown in Figure~\ref{Fig:lumivsbound} for some of the BSM scenarios we studied in Section~\ref{Sec:4}. The figure displays the exclusion and discovery reach on the new physics scale normalized to the collider energy $E_{\rm{cm}}=30$~TeV. This is the right figure of merit, since the goal is to extend the muon collider sensitivity above the direct reach. The reduction by a factor of $10^{1/4}=1.8$ due to a factor $10$ luminosity reduction partially undermines this goal, especially for what concerns the generic manifestations of the Composite Higgs scenario.

\section*{Acknowledgements}
The Swiss National
Science Foundation supported the work S.C, L.R, A.W. under contract 200021-178999 and the work of A.G. and R.R. under contract 200020-188671.   The work of A.W is also supported by the Italian Ministry of Research  under  PRIN contract   2017FMJFMW. 

\newpage
\appendix

\section{Radiation integrals}\label{AppDL}

The contribution of virtual radiation to the amplitude variation in eq.~(\ref{Eq:VarAmp}) is proportional to the integral 
\beq
I\equiv\frac{-i}{(2\pi)^4}\int_{\ds\delta\sigma}
\hspace{-2pt} d^4 q\,
\frac{1}{q^2 -m_V^2+ i \epsilon} \frac{(k_i \cdot k_j)}{(q \cdot k_i)(q \cdot k_j)}\,,
\label{Eq:IntVirt}
\eeq
where we included a mass $m_V\sim\MW$ for the virtual vector in order to verify explicitly that the integral is log-enhanced only in the $\lambda\gg\MW^2$ regime, where the IR cutoff is much above the EW scale. We now proceed (following~\cite{Landau}) to the evaluation of $I$ assuming, for simplicity, exactly massless hard $4$-momenta $k_i^2=k_j^2=0$. The integral is Lorentz-invariant, therefore it can only depend on the scalar product $(k_i\cdot k_j)$, that we set to
\beq
(k_i\cdot k_j)=\frac12(k_i+k_j)^2=\frac12 E^2\,,
\eeq
in what follows.

The calculation is conveniently performed in Sudakov coordinates~\cite{Sudakov:478063}. Namely we parametrize the loop momentum $q$ as
\beq\label{sud}
q=u k_i +v k_j + q_{\perp}^1\zeta_1+q_{\perp}^2\zeta_2\,,
\eeq
where $(\zeta_1)^2=(\zeta_2)^2=-1$ and $\zeta_{1,2}\cdot k_i=\zeta_{1,2}\cdot k_j=0$. In these coordinates
\beq
q^2=u\,v\, E^2-|q_\perp|^2\,,
\eeq
and the infinitesimal strip $\delta\sigma$~\eqref{strip} that defines the integration region is expressed as
\beq\label{uvbcstrip}
|u\, v| E^2 \in \left[ \lambda, \lambda + \delta \lambda \right]\,.
\eeq
After the change of variables, the integral reads
\beq\label{inttemp}
I = \frac{i }{(2 \pi)^4}\int_{\delta \sigma} \hspace{-2pt} \frac{du dv }{uv}\int \frac{d^2q_{\perp}}{|q_{\perp}|^2 - u v E^2 + m_V^2 - i \epsilon}\,.
\eeq

The $d^2q_\perp$ integral must be performed up to an upper cutoff that justifies the usage of the Eikonal approximation formula in eq.~\eqref{Eq:EikAppr} for the gauge boson vertices. In particular we notice that the actual denominators of the virtual legs in the diagram are not $(k_{i,j}\cdot q)$ as in the Eikonal formula, but rather $(k_{i,j}\cdot q) \pm q^2/2$, with plus or minus for incoming and outgoing particles, respectively. Neglecting $q^2/2$ is justified only if $|q_\perp|^2/2$ is not as large as to compete with the minimum among $|k_{i}\cdot q|=|u|\,E^2/2$ and $|k_{j}\cdot q|=|v|\,E^2/2$. Therefore the $q_\perp$  integral should be cutoff at
\beq\label{qtib}
|q_\perp|^2<\Lambda^2\,,\;\;\;\;\;{\rm{with}}\;\;\;\;\;\Lambda^2= E^2\min\big[|u|,|v|\big]\,.
\eeq
Similarly, the term proportional to $uv$ in $q^2$ should not be large compared to $(k_{i,j}\cdot q)$, therefore the $u$ and $v$ integrals are also bounded, in the region
\beq\label{uvbc}
|u|<1\,,\;\;\;\;\;|v|<1\,,
\eeq
supplemented by eq.~(\ref{uvbcstrip}).

The integration boundaries of $u$ and $v$ are invariant under $u\to-u$ and under $v\to-v$ reflections. We can thus perform the integral~(\ref{inttemp}) in the first quadrant of the $(u,v)$ plane, provided we duly symmetrize the integrand. Furthermore we notice that first-quadrant integration region (restricted to $u<1$ and $v<1$, owing to eq.~(\ref{uvbc})) is conveniently described by the coordinates $\tau$ and $y$, defined by 
\beq\label{tauy}
u = \sqrt{\tau} e^y\,, \;\;\;\;\; v= \sqrt{\tau} e^{-y}\,.
\eeq
Indeed in these coordinates the condition~(\ref{uvbcstrip}) merely sets the value of $\tau$ to
\beq
\tau=\frac{\lambda}{E^2}\,,\;\;\;\;\;{\rm{with}}\;\;\;\;\;\delta\tau=\frac{\delta\lambda}{E^2}\,,
\eeq
while the upper bound~(\ref{qtib}) on the $|q_\perp|$ integral reads
\beq
\Lambda^2=E^2 \sqrt\tau e^{-|y|}=E\sqrt{\lambda}\, e^{-|y|}\,.
\eeq
We can thus express the integral as
\beq\label{int2}
I= \frac{2\, i}{(2 \pi)^3} \frac{\delta\lambda}{\lambda} \int_0^{y_{\rm{M}}}\hspace{-10pt} dy\int_0^{E\sqrt\lambda e^{-y}} \hspace{-20pt} d |q_\perp|^2 \left[
\frac1{|q_{\perp}|^2 - \lambda + m_V^2 - i \epsilon}
-\frac1{|q_{\perp}|^2 + \lambda + m_V^2 - i \epsilon}
\right]\,.
\eeq
where, since $\Lambda^2$ is symmetric under $y\to-y$, we could take the $y$ integral from $0$ to
\beq
y_{\rm{M}} = -\frac{1}{2} \log \tau= \log\frac{E}{\sqrt\lambda}\,,
\eeq
and multiply by a factor $2$.

The $d|q_\perp|^2$ integral in eq.~(\ref{int2}) is readily performed. It is convenient to separate the terms that emerge from the upper $|q_\perp|^2$ integration extreme from the one of the lower extreme  $|q_\perp|^2=0$, obtaining two contributions to $I$, $I_{\rm{U}}$ and $I_{\rm{L}}$. We will readily see that fhe former contribution is suppressed, therefore
\beq
I\simeq I_{\rm{L}}=\frac1{8\pi^2}\frac{\delta\lambda}{\lambda}\log\frac{E^2}\lambda\frac{i}{\pi}
\log\left[
\frac{\lambda+m_V^2-i\,\epsilon}{-\lambda+m_V^2-i\,\epsilon}
\right]\,.
\eeq
The logarithm gives quite different results in the two regimes $\lambda\gg\MW^2\sim m_V^2$ and $\lambda\ll\MW^2\sim m_V^2$. In the second one, the argument of the logarithm has positive real part, almost equal to to $1$ up to $m_V^2/\lambda$ power-corrections. In the first regime, the argument has negative real part and the log equals $+i\,\pi$, plus $\lambda/m_V^2$ corrections. Namely
\beq
I\overset{\; {\lambda\gg\MW^2}}{{=}}\;-\frac1{8\pi^2}\frac{\delta\lambda}{\lambda}\log\frac{E^2}\lambda
\big[1+{\mathcal{O}}(m_V^2/\lambda)
\big]\,,\;\;\;\;\;
I\overset{\; {\lambda\ll\MW^2}}{{=}}\;-\frac1{8\pi^2}\frac{\delta\lambda}{\lambda}\log\frac{E^2}\lambda\cdot{\mathcal{O}}(\lambda/m_V^2)\,,
\eeq
from which we recover eq.~(\ref{Eq:VarAmp1}). More precisely, notice that the integral for each pair of external legs $ij$ is controlled by the specific scale $E^2=2(k_i\cdot k_j)$. In eq.~(\ref{Eq:VarAmp1}) we set all these scales equal up to corrections that are not log enhanced, but of order one. This in turn corresponds to order $1/\log$ corrections to the evolution kernel and to single-logarithm corrections to the solutions of the IREE.

The contribution to $I$ from the upper $|q_\perp|^2$ integration extreme is suppressed. To see this it is convenient to employ the integration variable $\rho=\exp(y-y_{\rm{M}})$, obtaining 
\beq
I_{\rm{U}}=\frac{2\, i}{(2 \pi)^3} \frac{\delta\lambda}{\lambda} \int_{\sqrt{\lambda}/E}^{1}\hspace{-10pt} d\rho\,
\frac1\rho
\log\left[
\frac
{1-\rho (1-m_V^2/\lambda)}
{1+\rho(1+m_V^2/\lambda)}
\right]\,,
\eeq
where we could drop the $-i\epsilon$ because the argument of the logarithm has positive real part in the entire range of integration. The $\rho$ integral is finite for $\sqrt\lambda/E\to0$, therefore it does not produce log-enhanced contributions. In particular the integral gives $-\pi^2/4$ for $\lambda\gg\MW^2\sim m_V^2$ and it is power-suppressed in the opposite regime.

We now compute the contribution of real radiation to the density matrix variation, which we employ in eq.~\eqref{Eq:VarDM}. The relevant integral that controls the contribution from a real radiation diagram like those in Figure~\ref{Fig:DMEvDiag}, reads
\beq
I_{\rm{R}}\equiv -\frac1{(2\pi)^3}\int_{\delta\sigma}
\hspace{-2pt}
\frac{d^3q}{2q^0} \frac{(k_i \cdot k_j)}{(k_i \cdot q)(k_j \cdot q)}
= -\frac1{(2\pi)^3}\int_{\ds\delta\sigma}
\hspace{-8pt} d^4q\,\delta(q^2-m_V^2)\,\theta(q^0)\frac{(k_i \cdot k_j)}{(k_i \cdot q)(k_j \cdot q)}
\,,
\eeq
where we employed the eikonal formula in eq.~(\ref{Eq:EikAppr}), but ignored the ``$G$'' factors that are included separately in eq.~\eqref{Eq:VarDM}. Notice the presence of a minus sign, which is due to three factors of $-1$. The first minus is due to the fact that applying the eikonal formula to the conjugate amplitude gives the complex conjugate of the generator matrix ``$G_i$'' of the corresponding leg, while eq.~\eqref{Eq:VarDM} is expressed in terms of the generators $G_{i^c}=-G_i^*$ of the conjugate representation. The second minus sign emerges from the sum over the polarizations of the real vector boson, which gives $-\eta_{\mu\nu}$. The third minus is because the contribution to the variation is minus the integral over the strip $\delta\sigma$, since the $\lambda$ cutoff is a lower bound on the hardness.

In Sudakov coordinates~(\ref{sud}), and setting $(k_i\cdot k_j)=E^2/2$, the integral becomes
\beq\label{realint}
I_{\rm{R}}= -\frac1{(2\pi)^2}
\int_{\delta \sigma} \hspace{-2pt} \frac{du dv }{uv}\,\theta(u+ v) \int d|q_{\perp}|^2
\delta \left(uv E^2 - |q_{\perp}|^2 -m_V^2 \right)
\,.
\eeq
The integration extremes of all the variables are like those of the virtual integral, including the condition~(\ref{uvbcstrip}) that defines the infinitesimal integration strip $\delta\sigma$. The delta function in eq.~(\ref{realint}) has support only if $u\,v>0$, and the theta function further restricts the integral to the first quadrant of the $(u,v)$ plane. We can thus employ the $\tau$ and $y$ coordinates in eq.~(\ref{tauy}) and readily obtain
\beq
I_{\rm{R}}= -\frac1{8\pi^2} \frac{\delta \lambda}{\lambda} \log\frac{E^2}\lambda\cdot\theta(\lambda-m_V^2)\,.
\eeq

Evidently, the theta function condition is not satisfied for $\lambda\ll\MW^2\sim m_V^2$, therefore the integral vanishes in this regime. In the other regime
\beq
I_{\rm{R}}\overset{\; {\lambda\gg\MW^2}}{{=}}\;= -\frac1{8\pi^2}  \frac{\delta \lambda}{\lambda} \log\frac{E^2}\lambda\,,
\eeq
which is equal to the virtual radiation integral, as anticipated in the main text. Notice that the pre-factor of eq.~(\ref{Eq:VarDM}) contains an additional $1/2$, due to the fact that the real radiation contribution is effectively counted twice in the equation by the two terms proportional to  \parbox[c][0pt]{52pt}{$(G^{A}_i)(G_{j^c}^A)$} and to  \parbox[c][0pt]{52pt}{$(G^{A}_{i^c})(G_{j}^A)$}, which are equal after summing over $i$ and $j$.

\section{High-energy EW multiplets}\label{AppEWBasis}

\begin{table}
	\begin{center}
		\renewcommand{\arraystretch}{1.2}
		\begin{tabular}{| c | c | c |} 
			\hline
			Bosons &  \mbox{SU$(2)_L$}  & ${\cal{Y}}$\\
			\hline
		$W$ &  t  & $0$\\
		\hline
		$B$& s  & $0$\\
		\hline
		$ H $&  d  & $+1/2$\\
		\hline
		$\bar{H}$&  ${\bar{\rm{d}}}$  & $-1/2$\\
			\hline
		\end{tabular}
		\hspace{10pt}
		\begin{tabular}{| c | c | c |} 
			\hline
			Leptons &  \mbox{SU$(2)_L$}  & ${\cal{Y}}$\\
			\hline
			$(\nu_{\ell,-1/2}\,,\ell^-_{-1/2}) $ &  d  & $-1/2$\\
			\hline
			$({\bar{\nu}}_{\ell,+1/2},\,\ell^+_{+1/2})^t$&  ${\bar{\rm{d}}}$  & $-1/2$\\
			\hline
			$\ell^-_{+1/2} $&  s  & $-1$\\
			\hline
			$\ell^+_{-1/2} $&  s  & $+1$\\
			\hline
		\end{tabular}
	\end{center}
	\begin{center}
		\renewcommand{\arraystretch}{1.2}
		\begin{tabular}{| c | c | c |} 
				\hline
			Quarks &  \mbox{SU$(2)_L$}  & ${\cal{Y}}$\\
			\hline
			$(u_{-1/2},\,d_{-1/2})^t$ &  d  & $+1/6$\\
			\hline
			$({\bar{u}}_{+1/2},\,{\bar{d}}_{+1/2})^t$&  ${\bar{\rm{d}}}$  & $-1/6$\\
			\hline
			$u_{+1/2} $&  s  & $+2/3$\\
			\hline
			${\bar{u}}_{-1/2} $&  s  & $-2/3$\\
			\hline
			$d_{+1/2} $ &  s  & $-1/3$\\
			\hline
			${\bar{d}}_{-1/2}$ &  s  & $+1/3$\\
			\hline
		\end{tabular}
	\end{center}
	\caption{\mbox{SU$(2)_L\times$U$(1)_Y$} quantum numbers of the SM particles in the high-energy regime.\label{tabrep}}
\end{table}

The EW symmetry is effectively unbroken at energies much above the EW scale. Therefore in this regime it is convenient to describe the SM particles in terms of representations of the (linearly-realized) SM group \mbox{SU$(2)_L\times$U$(1)_Y$}, with generators
\beq
{\cal{T}}^A=\{{\cal{T}}^{a},{\cal{Y}}\}\,,\;\;\;\;\;a=1,2,3\,.
\eeq
The generators act on the single particle states as
\beq
{\cal{T}}^A |{\rm{p}}(k,\alpha)\rangle=|{\rm{p}}(k,\beta)\rangle (T^A_{\rm{r}})^\beta_{\;\alpha}\,,
\eeq
with generator matrices $T^A_{\rm{r}}$ that define the representation ``r'' of the particle multiplet. The field $\Phi^\alpha_{\rm{r}}$ that interpolates the multiplet from the vacuum, namely
\beq
\langle0|\Phi_{\rm{r}}^\alpha(0)|{\rm{p}}(k,\beta)\rangle\propto\delta^\alpha_\beta\,,
\eeq
transforms with the same generator matrix, i.e.
\beq
\left[\Phi_{\rm{r}}^\alpha(x),{\cal{T}}^A\right]=(T^A_{\rm{r}})^\alpha_{\;\beta}\Phi_{\rm{r}}^\alpha(x)\,.
\eeq
The \mbox{SU$(2)_L$} representations of the SM particles and the corresponding \mbox{U$(1)_Y$} charges are listed in Table~\ref{tabrep}. 

All the fermionic particles with helicity $-1/2$ transforms as doublets (i.e., ${\rm{r}}={\rm{d}}$), the anti-particles with helicity $+1/2$ transform in the conjugate (${\rm{r}}={\bar{\rm{d}}}$) of the doublet representation, while all the others are singlets. Obviously this is true only in the high energy limit where the fermions are effectively massless and the helicity corresponds the chirality of the corresponding interpolating fields. The doublet representation matrices are the standard 
\beq\label{doub}
T^{a}_{\rm{d}}=\left\{
\left(
\begin{matrix}
	0 & +1/2\\
	+1/2 & 0
\end{matrix}
\right)\,,
\left(
\begin{matrix}
	0 & -i/2\\
	+i/2 & 0
\end{matrix}
\right)\,,
\left(
\begin{matrix}
	+1/2 & 0\\
	0 & -1/2
\end{matrix}
\right)
\right\}\,,\;\;\;\;\; T^{a}_{\bar{\rm{d}}}=-(T^{a}_{\rm{d}})^*=-(T^{a}_{\rm{d}})^t\,.
\eeq

The EW multiplets of bosonic particles are less well-known, but equally straightforward to work out employing the standard Goldstone Boson Equivalence Theorem, or better its stronger formulation in Ref.s~\cite{Cuomo:2019siu,Wulzer:2013mza}. The point is that for massive $W^\pm$ and $Z$ vector bosons with $0$ (longitudinal) helicity one can employ interpolating fields that are a specific combination of the regular gauge fields $W^\pm_\mu$ and $Z_\mu$ and of the Goldstone boson scalar fields $\pi^\pm$ and $\pi_Z$. The longitudinal states are thus a linear combination of the quantum fluctuation modes associated to this two different type of fields, and the scattering amplitudes with external longitudinal states are a linear combination of Feynman diagrams where the external states are represented either as gauge fields or as Goldstone fields. This is a convenient formalism in the high energy limit because the polarization vector associated with gauge external lines vanishes as $\MW/E$ (unlike the regular longitudinal polarization, that diverges as $E/\MW$), and only the Goldstone diagrams survive. In essence this means that the Goldstones, and not the gauge, are the adequate interpolating fields for the longitudinal particles at high energy. Therefore the EW quantum numbers of the longitudinal particles are the ones of the Goldstones, and not of the gauge fields. The Goldstone bosons, together with the Higgs, form a doublet with $+1/2$ hypercharge, and the corresponding conjugate doublet
\beq\label{Lgauge}
H = \left( \pi^+ \,,\; \frac{h + i \pi_Z}{\sqrt 2}\right)^t \,,\;\;\;\;\;  {\bar{H}} = \left( \pi^- \,,\; \frac{h - i \pi_Z}{\sqrt 2}\right)^t \,.
\eeq

Vector bosons with transverse helicity ${\rm{T}}=\pm1$ are instead well-described by gauge fields even at high energy.  Therefore they decompose into a triplet plus a singlet EW multiplet, which are readily obtained by undoing the Weinberg rotation
\beq\label{Tgauge}
W^{a} = \left(\frac{W^+_{\rm{T}} + W^-_{\rm{T}}}{\sqrt 2}\,,\; i\frac{W^+_{\rm{T}} - W^-_{\rm{T}}}{\sqrt 2 } \,,\; \CW Z_{\rm{T}} + \SW \gamma_{\rm{T}}  \right)^t\,, \;\;\;\;\; B = -\SW Z _{\rm{T}}+ \CW \gamma_{\rm{T}}\,.
\eeq
Explicitly, the triplet generator matrices are
\beq\label{trip}
T^{a}_{\rm{t}}=\left\{
\left(
\begin{matrix}
	0 & 0 & 0\\
	0 & 0 & -i\\
	0 & i & 0
\end{matrix}
\right)\,,
\left(
\begin{matrix}
	0 & 0 & i\\
	0 & 0 & 0\\
	-i & 0 & 0
\end{matrix}
\right)\,,
\left(
\begin{matrix}
	0 & -i & 0\\
	i & 0 & 0\\
	0 & 0 & 0
\end{matrix}
\right)
\right\}\,.
\eeq

We now proceed to the evaluation of the $K_i$ exponentials in the explicit formula for the semi-inclusive density matrix~(\ref{sidm}), for external legs in the doublet (or anti-doublet) or in the triplet canonical representations defined by eq.~(\ref{doub}) and~(\ref{trip}). The $K$ tensors in eq.~(\ref{KernelDMHIGH}) are
\begin{eqnarray}
&\ds
\big[{{K}}_{\rm{{d}}}\big]^{\alpha\bar\alpha}_{\beta\bar\beta}=
\big[{{K}}_{\bar{\rm{d}}}\big]^{\alpha\bar\alpha}_{\beta\bar\beta}=
c_{\rm{d}}\,\delta^{\alpha}_{\beta}\delta^{\bar\alpha}_{\bar\beta}+
\sum_a(T^a_{\rm{d}})^\alpha_{\;\;\beta} (T^a_{\bar{\rm{d}}})_{\;\;\bar\beta}^{\bar\alpha}  =
\delta^\alpha_{\beta} \delta_{\bar\beta}^{\bar\alpha}-\frac12\delta^{\alpha\bar\alpha} \delta_{\beta\bar\beta}\,,&\nonumber\\
&\ds
\big[{{K}}_{\rm{{t}}}\big]^{\alpha\bar\alpha}_{\beta\bar\beta}=
c_{\rm{t}}\,\delta^{\alpha}_{\beta}\delta^{\bar\alpha}_{\bar\beta}+
\sum_a(T^a_{\rm{t}})^\alpha_{\;\;\beta} (T^a_{\rm{t}})_{\;\;\bar\beta}^{\bar\alpha} =
2\,\delta^{\alpha}_{\beta}\delta^{\bar\alpha}_{\bar\beta}+
\delta^\alpha_{\bar\beta} \delta_{\beta}^{\bar\alpha}-\delta^{\alpha\bar\alpha} \delta_{\beta\bar\beta}\,.
&
\end{eqnarray}
By exponentiating them, we readily obtain
\begin{eqnarray}
\label{Eq:ExpK}
&\ds\label{dfact}
\bigg[\exp\big(
{-{\mathcal{L}}\,K_{\rm{d}}}
\big)\bigg]^{\alpha\bar\alpha}_{\beta\bar\beta}=
e^{-\mathcal{L}}\delta^{\alpha}_{\beta}\delta^{\bar\alpha}_{\bar\beta}
+{e^{-{\mathcal{L}}/2}}\sinh({\mathcal{L}}/2)\delta^{\alpha\bar\alpha} \delta_{\beta\bar\beta}\,,&\\
&
\bigg[\exp\big(
{-{\mathcal{L}}\,K_{\rm{t}}}
\big)\bigg]^{\alpha\bar\alpha}_{\beta\bar\beta}=
{e^{-2{\mathcal{L}}}} \left( \cosh({\mathcal{L}})\delta^{\alpha}_{\beta}\delta^{\bar\alpha}_{\bar\beta}-\sinh({\mathcal{L}})\delta^\alpha_{\bar\beta} \delta_{\beta}^{\bar\alpha}\right) +
\frac23
{e^{-\frac{3}{2}{\mathcal{L}}}}\sinh \left(3{\mathcal{L}/2} \right)
\delta^{\alpha\bar\alpha} \delta_{\beta\bar\beta}\,, 
&
\label{tfact}
\end{eqnarray}
where we defined
\beq
{\mathcal{L}}=\frac{g^2}{16\pi^2} \log^2(E^2/\MW^2)\,.
\eeq

\section{${\boldsymbol{3}}^{\text{\bf{rd}}}$ family operators}\label{3rd}
\begin{figure}
	\begin{minipage}{.5\linewidth}
		\centering
		\includegraphics[scale=.5]{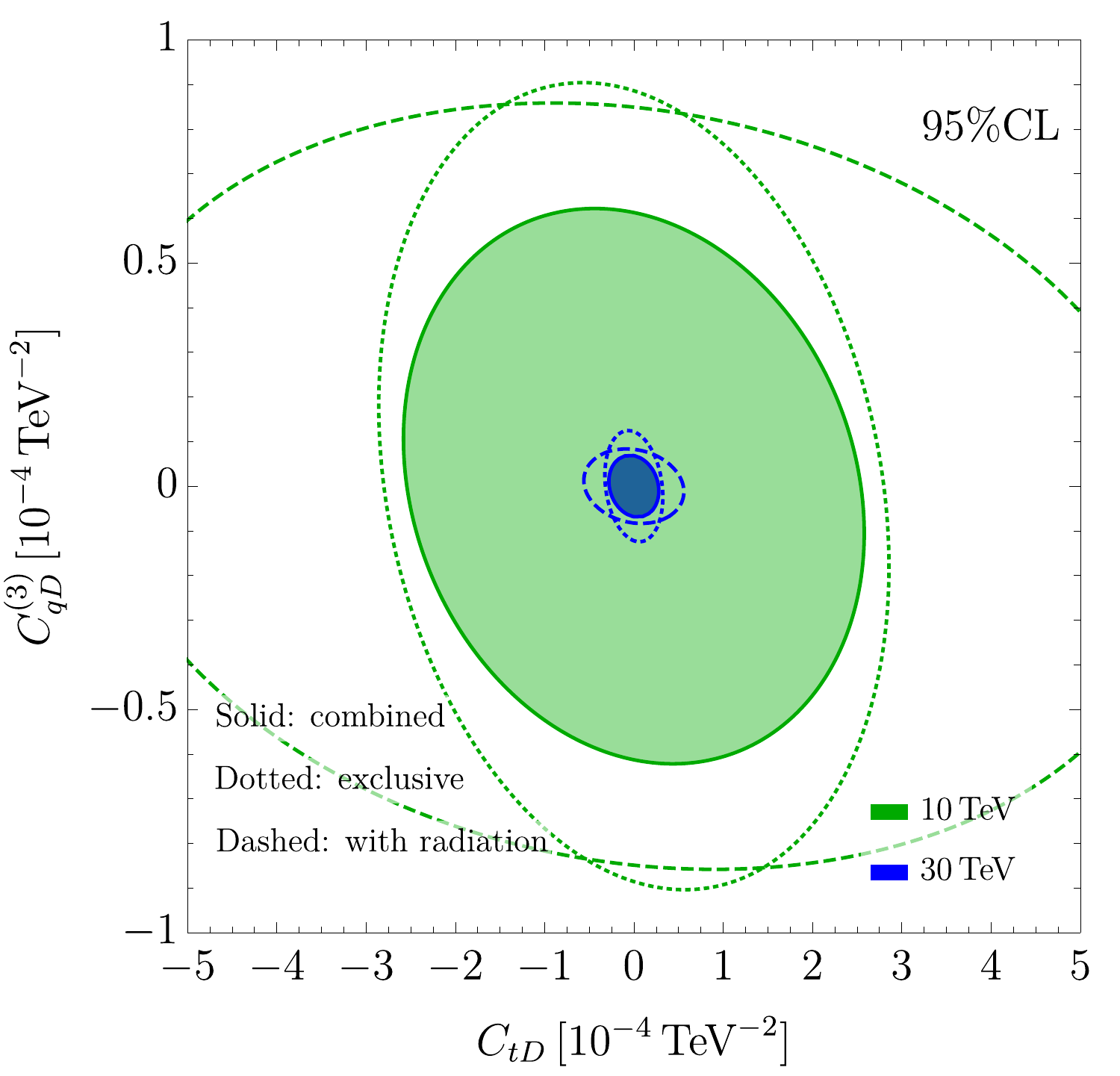}
	\end{minipage}%
	\begin{minipage}{.5\linewidth}
		\centering
		\includegraphics[scale=.5]{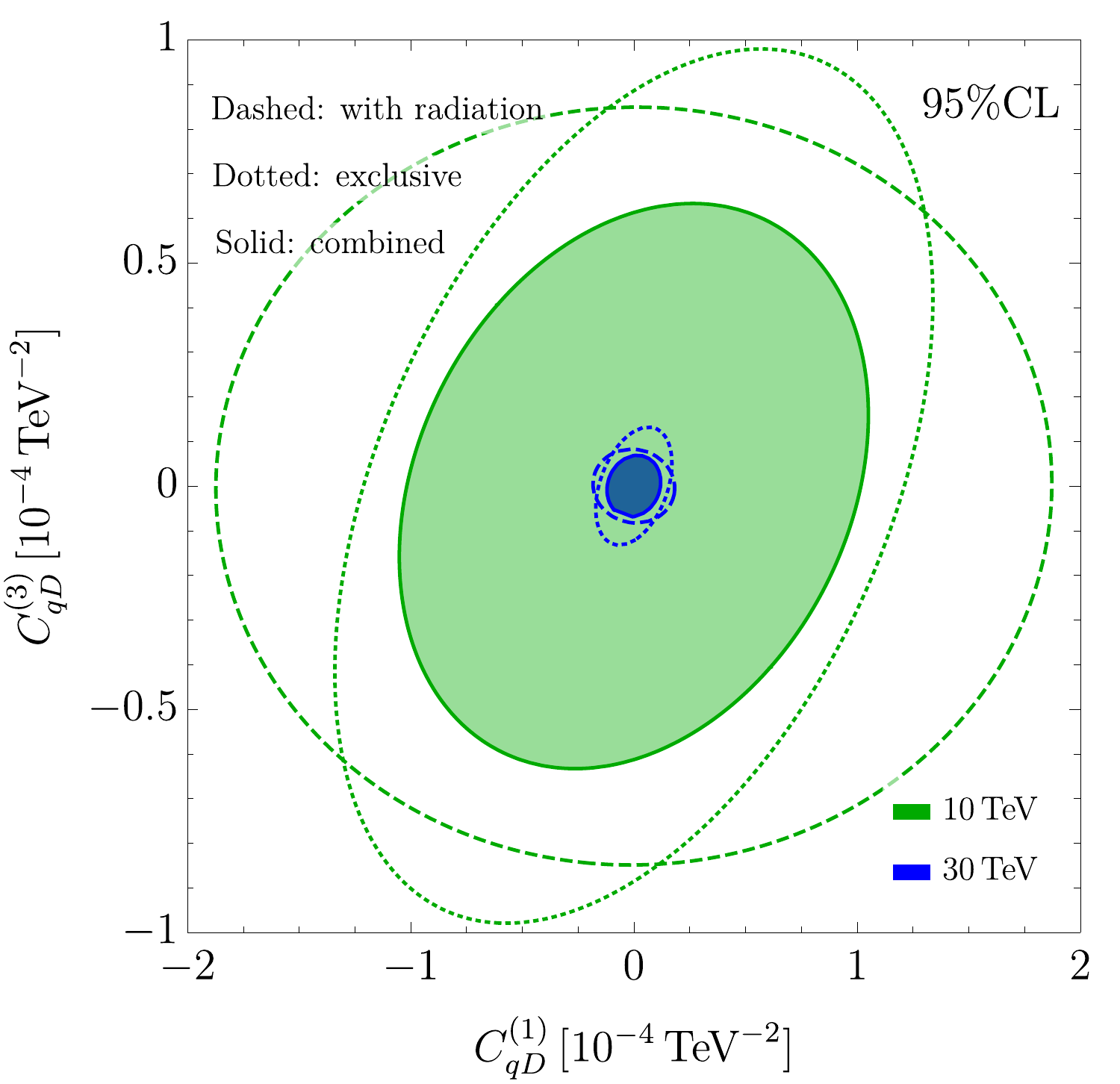}
	\end{minipage}
	\caption{$95\%$~CL contours in the ($C_{tD},\,C_{qD}^{(3)}$) (left) and ($C_{qd}^{(1)},\,C_{qD}^{(3)}$) (right) planes at the $10$ and $30$~TeV muon colliders.  \label{Fig:2D3rd}}
\end{figure}

The sensitivity to the $3^{\rm{rd}}$ family operators in Table~\ref{Tab:OpWarsaw} are summarized in this section.  In Figure~\ref{Fig:2D3rd} we report the two-dimensional contours in the  \parbox[c][0pt]{52pt}{($C_{tD},\,C_{qD}^{(3)}$)} and  \parbox[c][0pt]{52pt}{($C_{qd}^{(1)},\,C_{qD}^{(3)}$)} planes, with the third operator set to zero. We notice that the ``with radiation'' cross-section measurements (see the main text) is mostly effective to probe $C_{qD}^{(3)}$ producing a significant sensitivity improvement on the combined bound in this direction. The effect is milder in the orthogonal directions. The likelihood is dominated by the linear term in the new physics parameters so all our result can be expressed in terms of the single operator reaches (at $95\%$~CL) of and the correlations matrices in Table~\ref{Tab:SORTopOp}. In the table we report the sensitivity of exclusive cross-section measurements alone, and the combination of all the measurements. 

\newpage

\begin{table}[h]
\begin{subfigure}{.5\textwidth}
		\centering
		\includegraphics[width=0.9\textwidth]{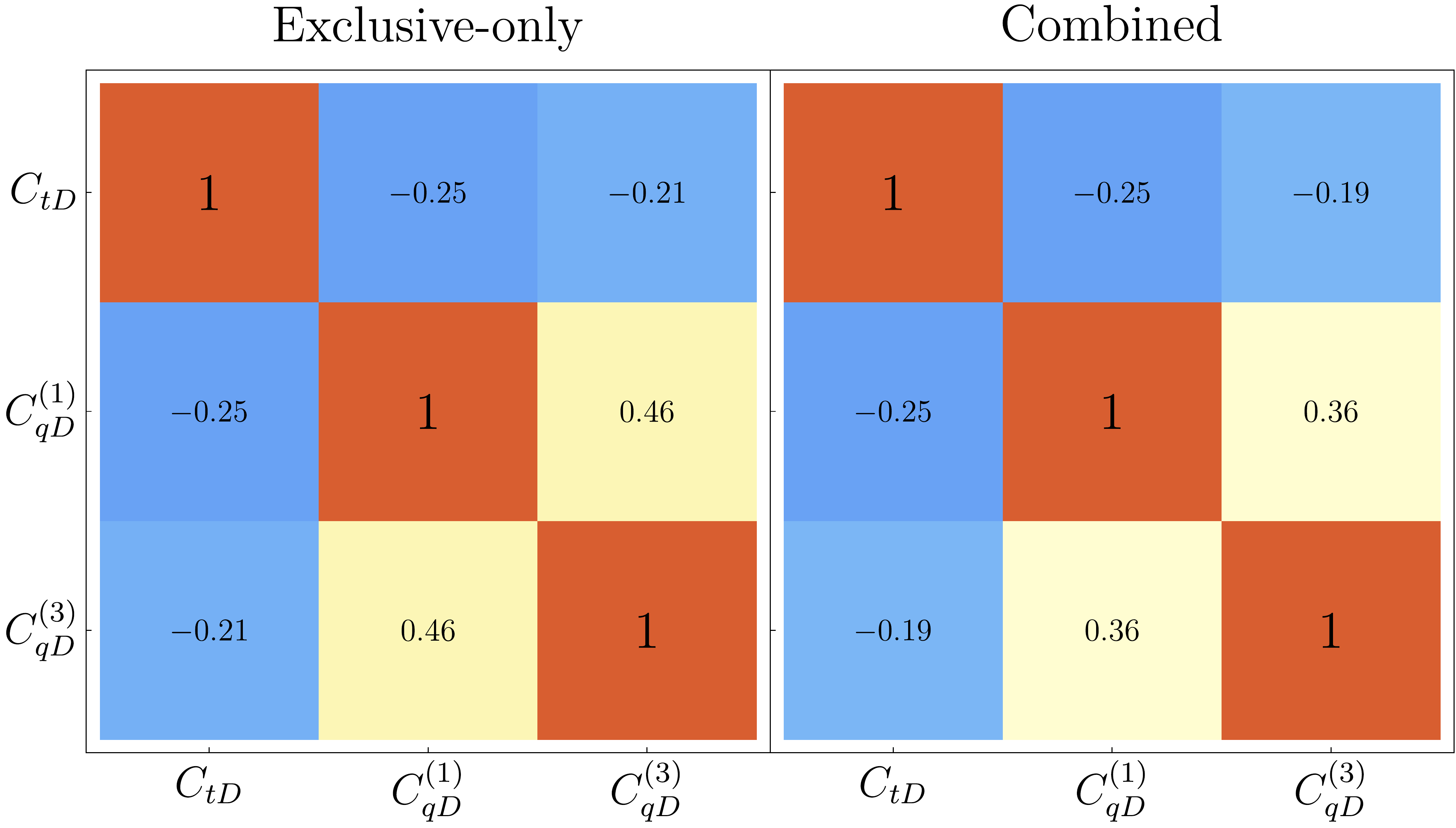}  
		\caption*{$E_{\rm{cm}}=3$ TeV \label{Fig:CorrA}}
	\end{subfigure}
	\begin{subfigure}{.5\textwidth}
		\centering
		\includegraphics[width=0.9\textwidth]{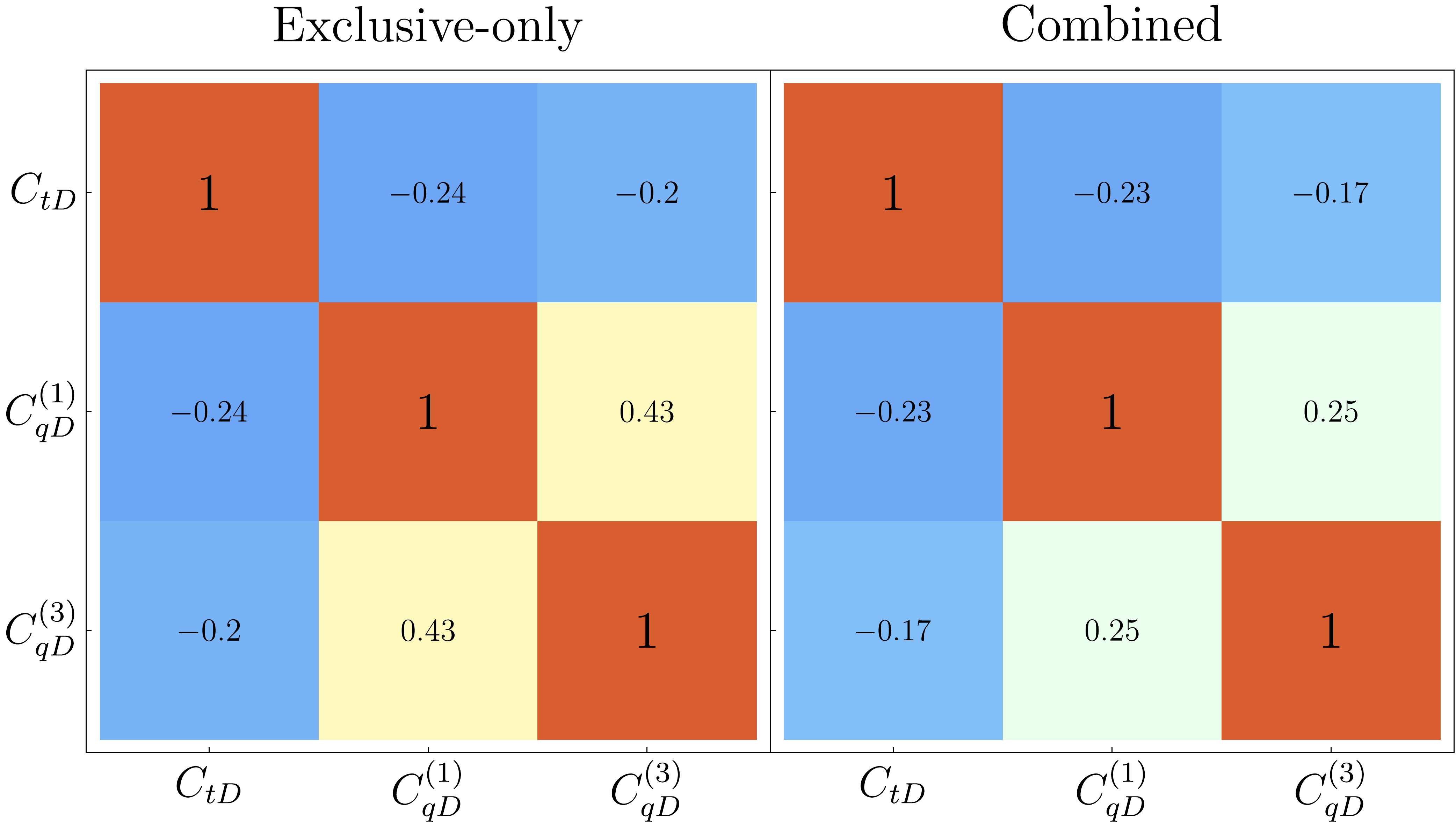}  
		\caption*{$E_{\rm{cm}}=10$ TeV}
	\end{subfigure}
	\begin{subfigure}{.5\textwidth}
		\centering
		\includegraphics[width=0.9\textwidth]{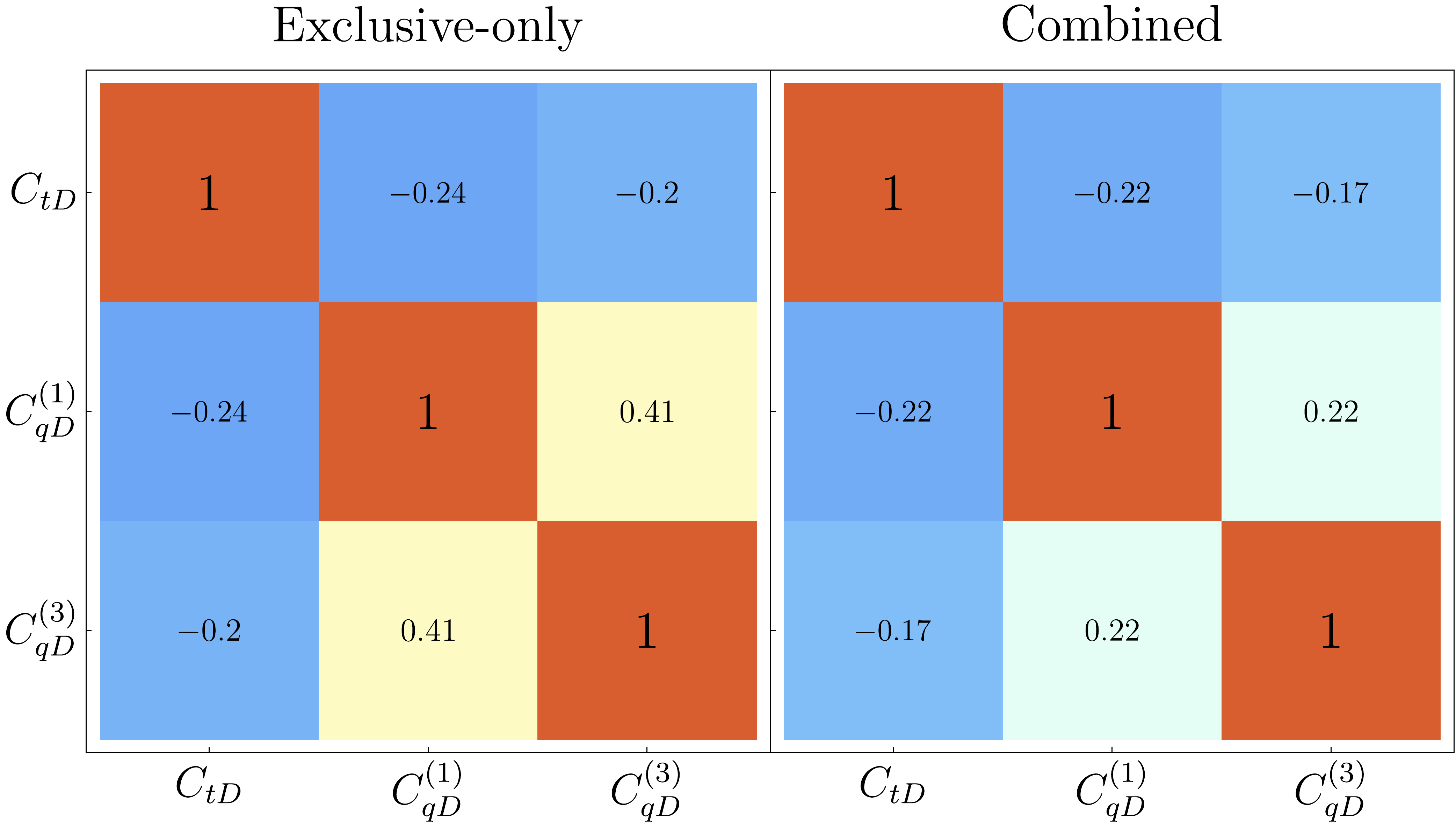}  
		\caption*{$E_{\rm{cm}}=14$ TeV}
	\end{subfigure}
	\begin{subfigure}{.5\textwidth}
		\centering
		\includegraphics[width=0.9\textwidth]{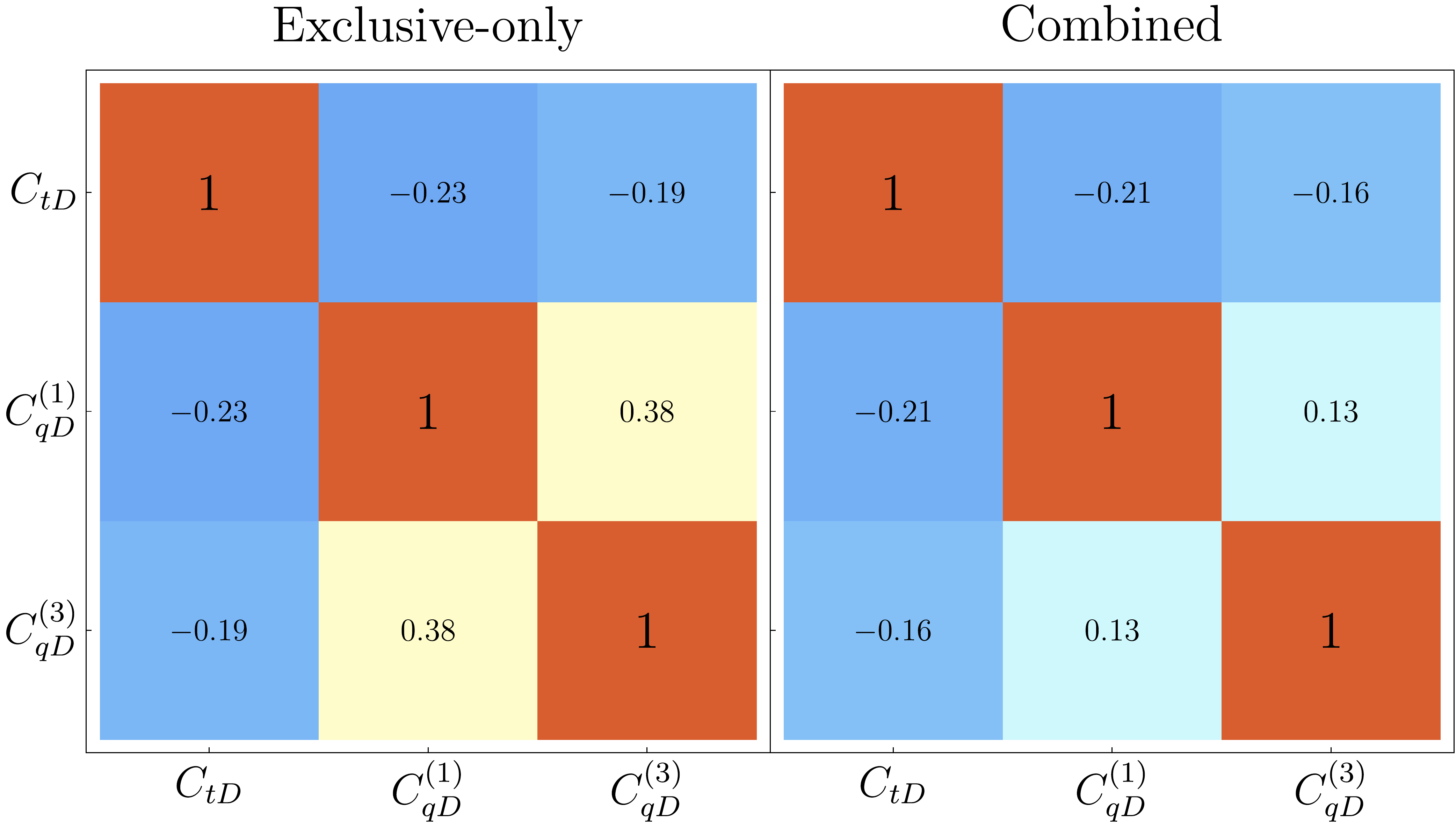}  
		\caption*{$E_{\rm{cm}}=30$ TeV}
	\end{subfigure}
	{\small
		\begin{center}
			\begingroup
			\renewcommand*{\arraystretch}{1.4}
			\begin{tabular}{||c|| c | c | c || c | c | c ||} 
				\cline{2-7}
				\multicolumn{1}{c|}{}&\multicolumn{3}{|c||}{Exclusive-only  [$95\%$~CL]}&\multicolumn{3}{c|}{Combined  [$95\%$~CL]}\\
				\cline{2-7}
				\multicolumn{1}{c|}{}&$C_{tD}$&$C_{qD}^{(1)}$&$C_{qD}^{(3)}$&$C_{tD}$&$C_{qD}^{(1)}$&$C_{qD}^{(3)}$\\ \hline 3 TeV&$[-24.4	,24.4	$]&[$-9.47,	9.47$]&[$-6.68,	6.68$]&[$	-23.1	,23.1$]&[$	-8.59,	8.59$]&[$	-5.45,	5.45 $]\\ \hline 10 TeV&$[-2.24	,2.24	]$&$[-0.97,	0.97]$&$[	-0.71	,0.71]$&$[	-2.04,	2.04]$&$[	-0.81	,0.81	]$&$[-0.49,	0.49]$ \\ \hline 14 TeV& $[-1.15	,1.15$]&[$	-0.52,	0.52$]&[$	-0.38,	0.38	$]&[$-1.03	,1.03$]&[$	-0.42	,0.42$]&[$	-0.25,	0.25$]\\ \hline 30 TeV& $[-0.26,	0.26$]&[$	-0.13	,0.13$]&[$	-0.10,	0.10$]&[$	-0.22	,0.22$]&[$	-0.10,	0.10$]&[$	-0.05,	0.05$]\\ \hline
			\end{tabular}
			\endgroup
			\caption{Single-operator $95\%$~CL reach and correlation matrices for the Wlison coefficient $C_{qD}^{(3)}$, $C_{qD}^{(1)}$ and $C_{qD}^{(3)}$ of the operators of Table~\ref{Tab:OpWarsaw} at different collider energies. All results include exclusive cross-sections or combined measurements.  The Wilson coefficient are expressed in $10^{-4} \,\text{TeV}^{-2}$.  Since the likelihood is dominated by the linear terms in the new physics parameters, the single parameter reach and the correlation characterize our results completely. \label{Tab:SORTopOp}}
	\end{center}}
\end{table}

\section{Summary plots}\label{App:Results}
\begin{figure}
	\begin{minipage}{.5\linewidth}
		\centering
		\includegraphics[scale=.55]{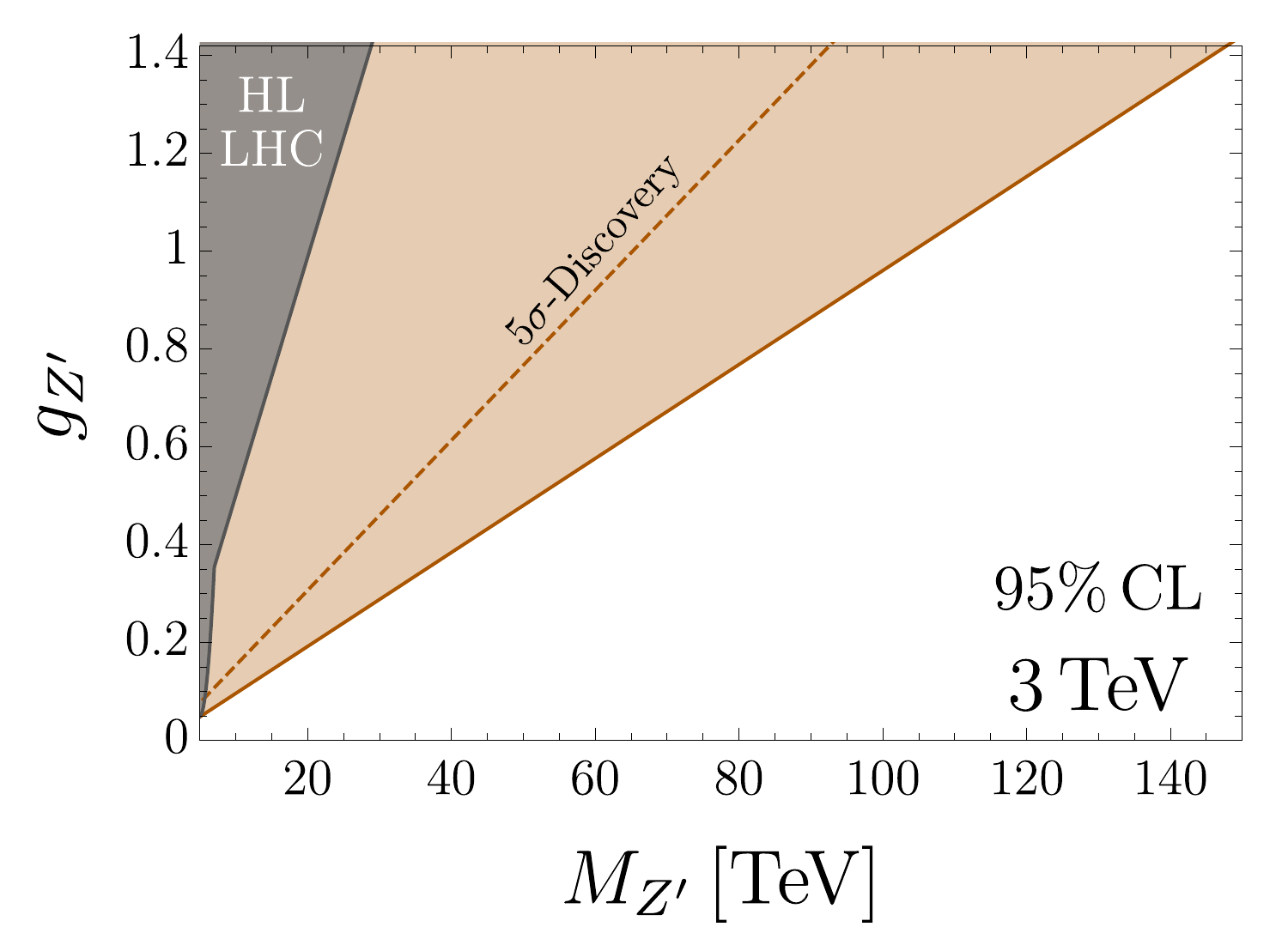}
	\end{minipage}%
	\begin{minipage}{.5\linewidth}
		\centering
		\includegraphics[scale=.55]{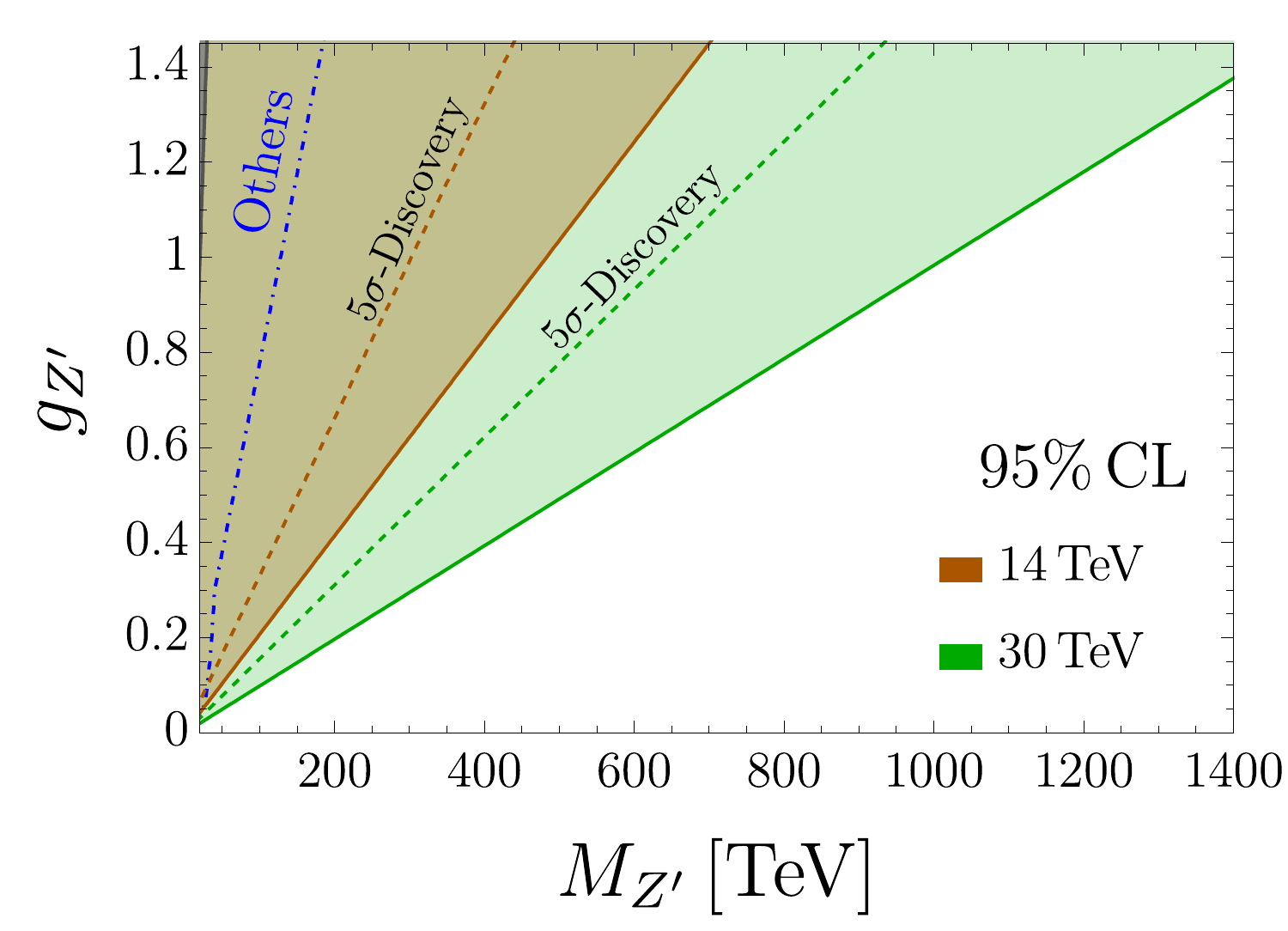}
	\end{minipage}
	\caption{The same as the right panel of Figure~\ref{Fig:CHandZprime} for various collider energies.\label{Fig:ZPrime}}
\end{figure}

\begin{figure}
	\begin{subfigure}{.5\linewidth}
		\centering
		\includegraphics[scale=.5]{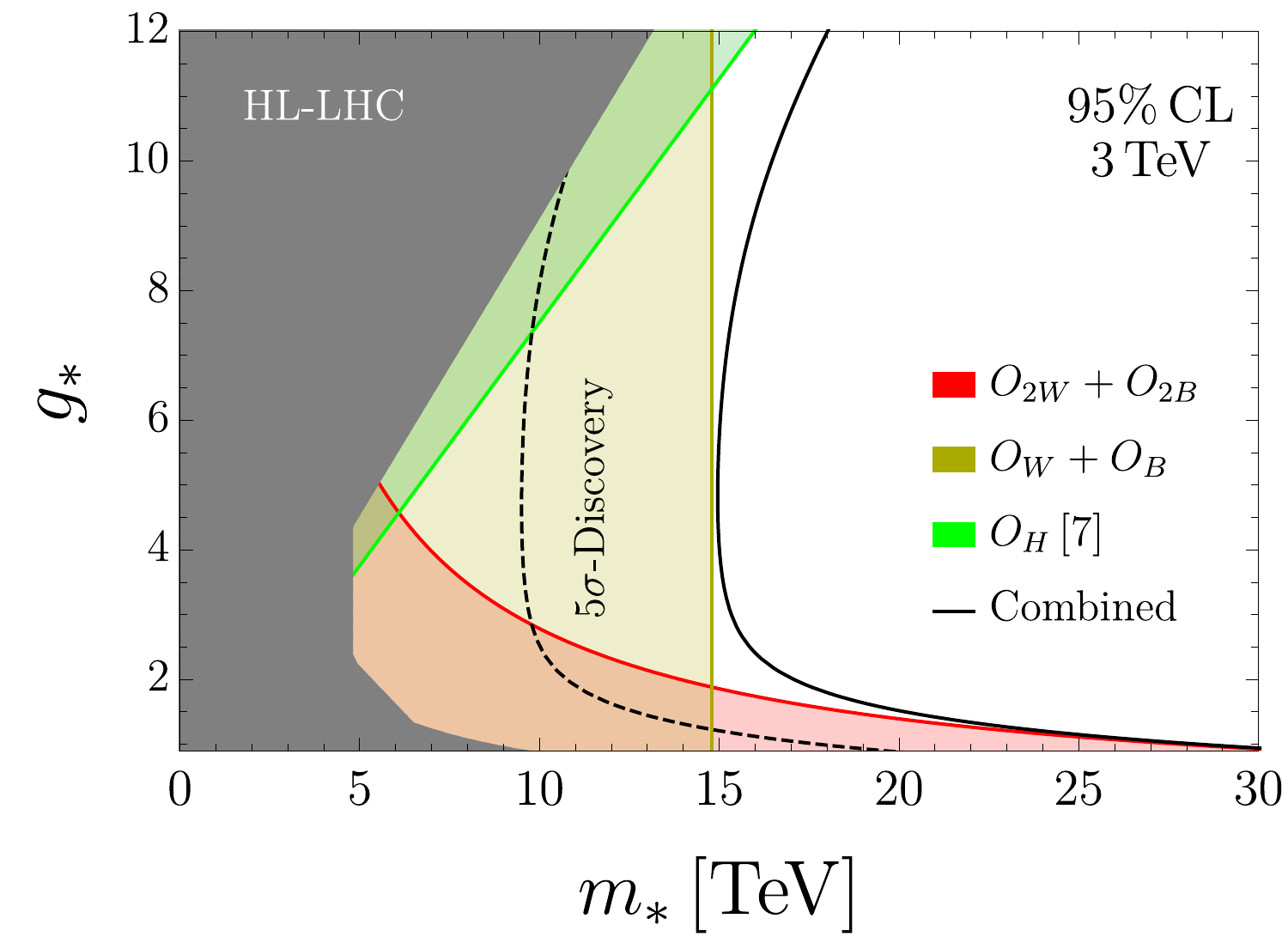}
	\end{subfigure}%
	\begin{subfigure}{.5\linewidth}
		\centering
		\includegraphics[scale=.5]{Figures/10TeVcombined.pdf}
	\end{subfigure}
	\begin{subfigure}{.5\linewidth}
	\centering
	\includegraphics[scale=.5]{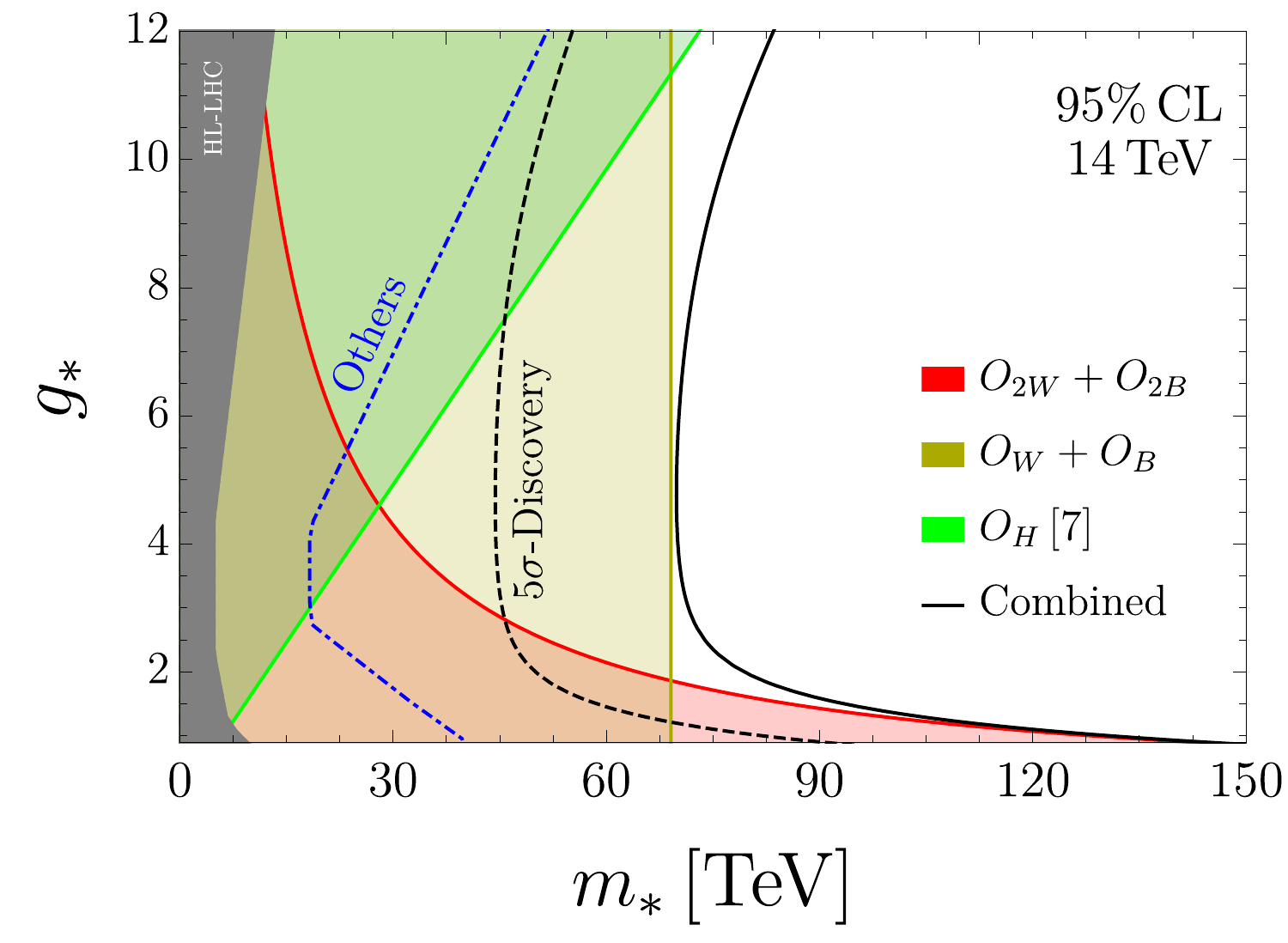}
	\end{subfigure}%
	\begin{subfigure}{.5\linewidth}
	\centering
	\includegraphics[scale=.5]{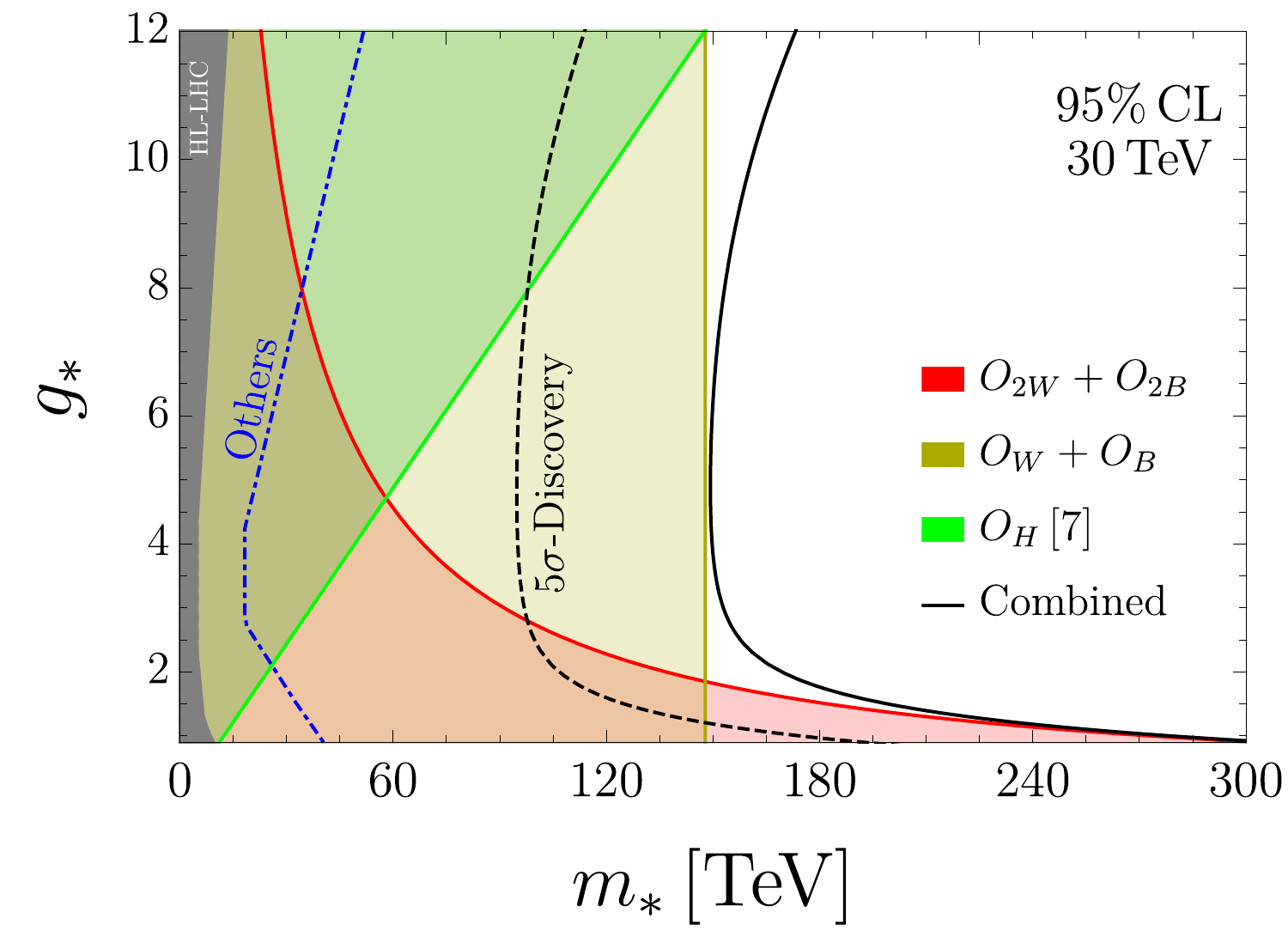}
	\end{subfigure}
	\caption{The same as the left panel of Figure~\ref{Fig:CHandZprime} \label{Fig:ch} for various collider energies.	\label{Fig:chApp}}
\end{figure}

\begin{figure}
	\centering
	\includegraphics[scale=.7]{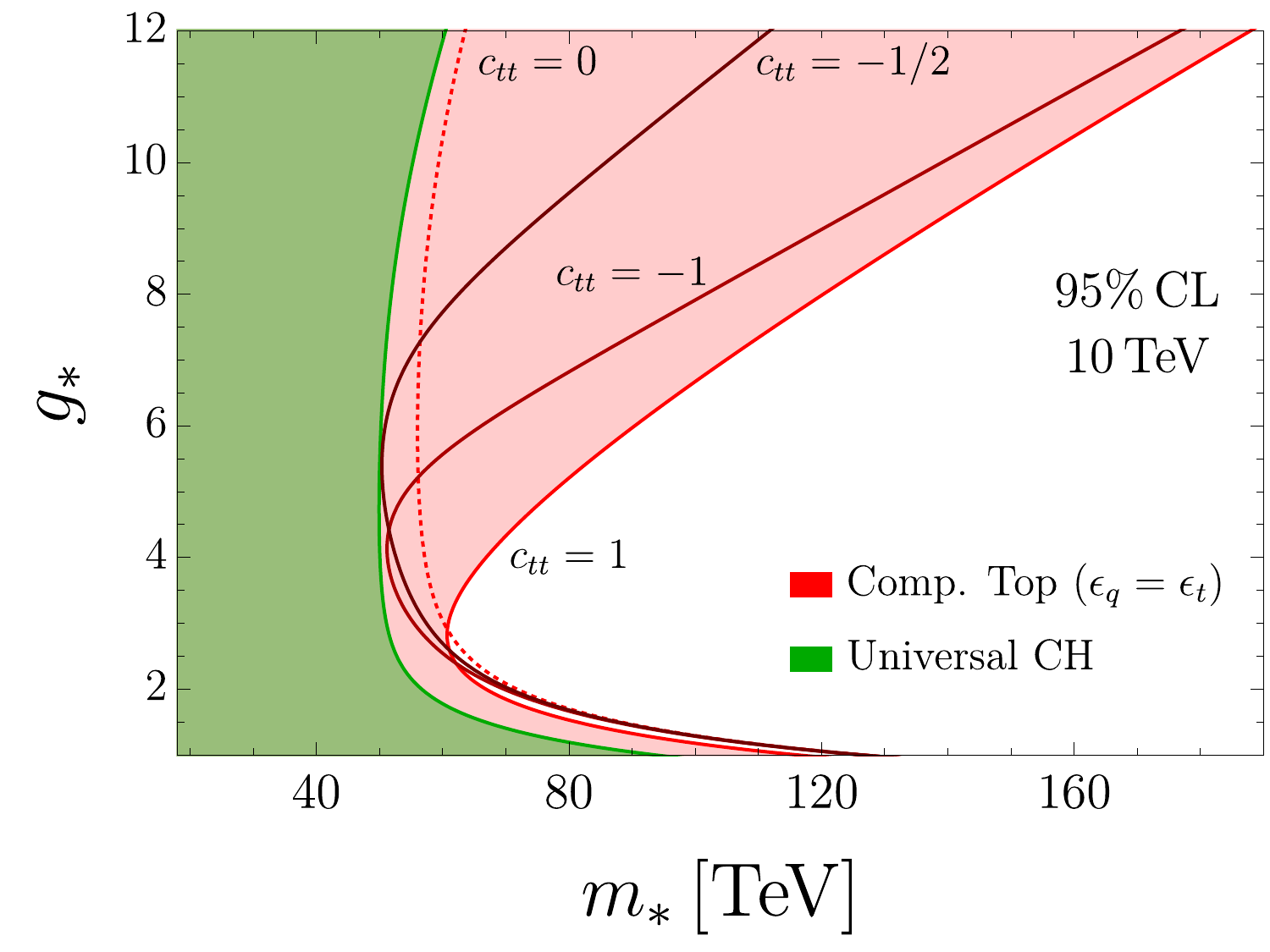}
	\caption{The same as the right panel of Figure~\ref{Fig:CT} for different values of $c_{tt}$ to show the model dependence of the result.\label{Fig:CTAppctt}}
\end{figure}

\begin{figure}
	\begin{subfigure}{.5\linewidth}
		\centering
		\includegraphics[scale=.5]{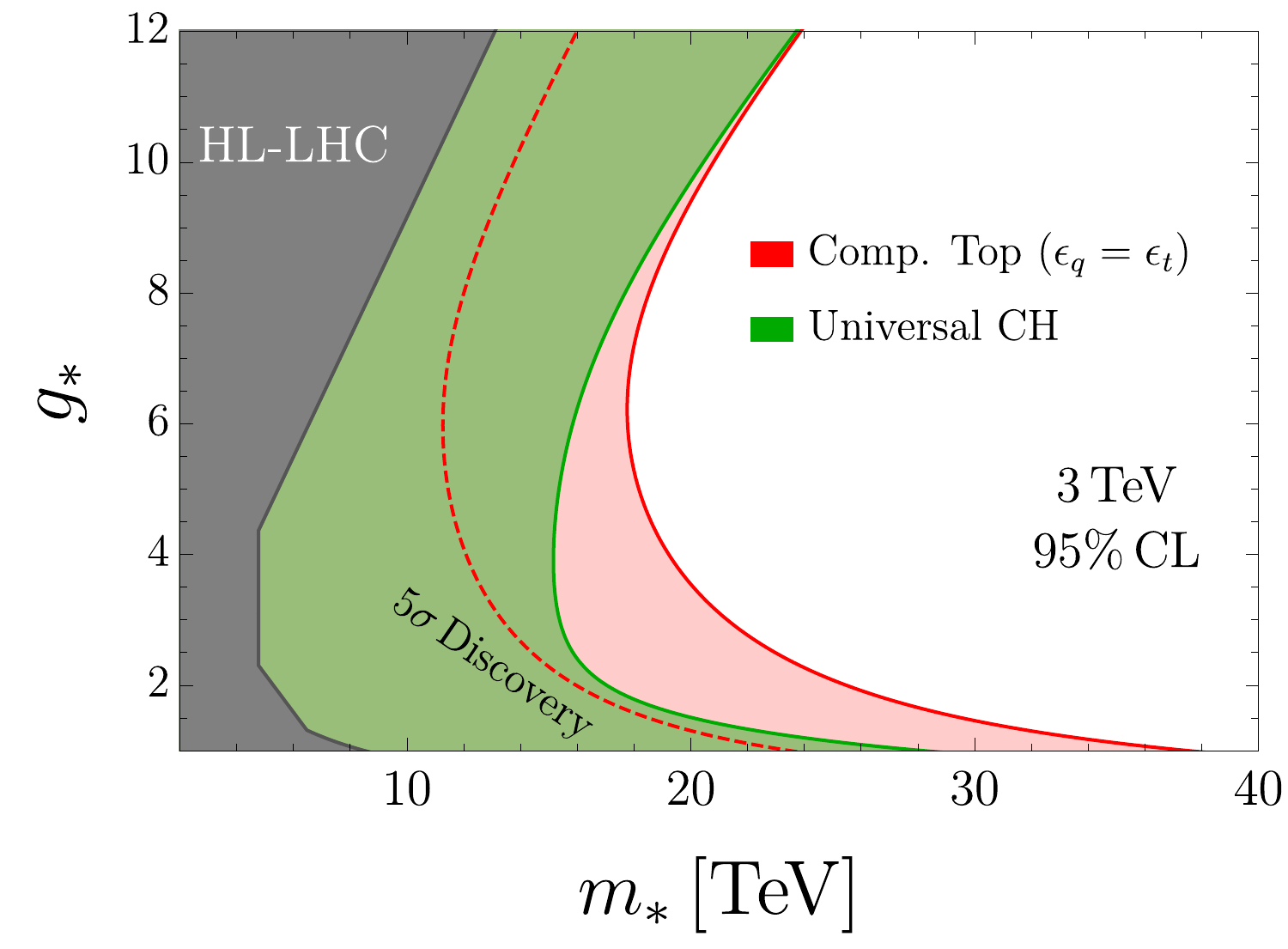}
	\end{subfigure}%
	\begin{subfigure}{.5\linewidth}
		\centering
		\includegraphics[scale=.5]{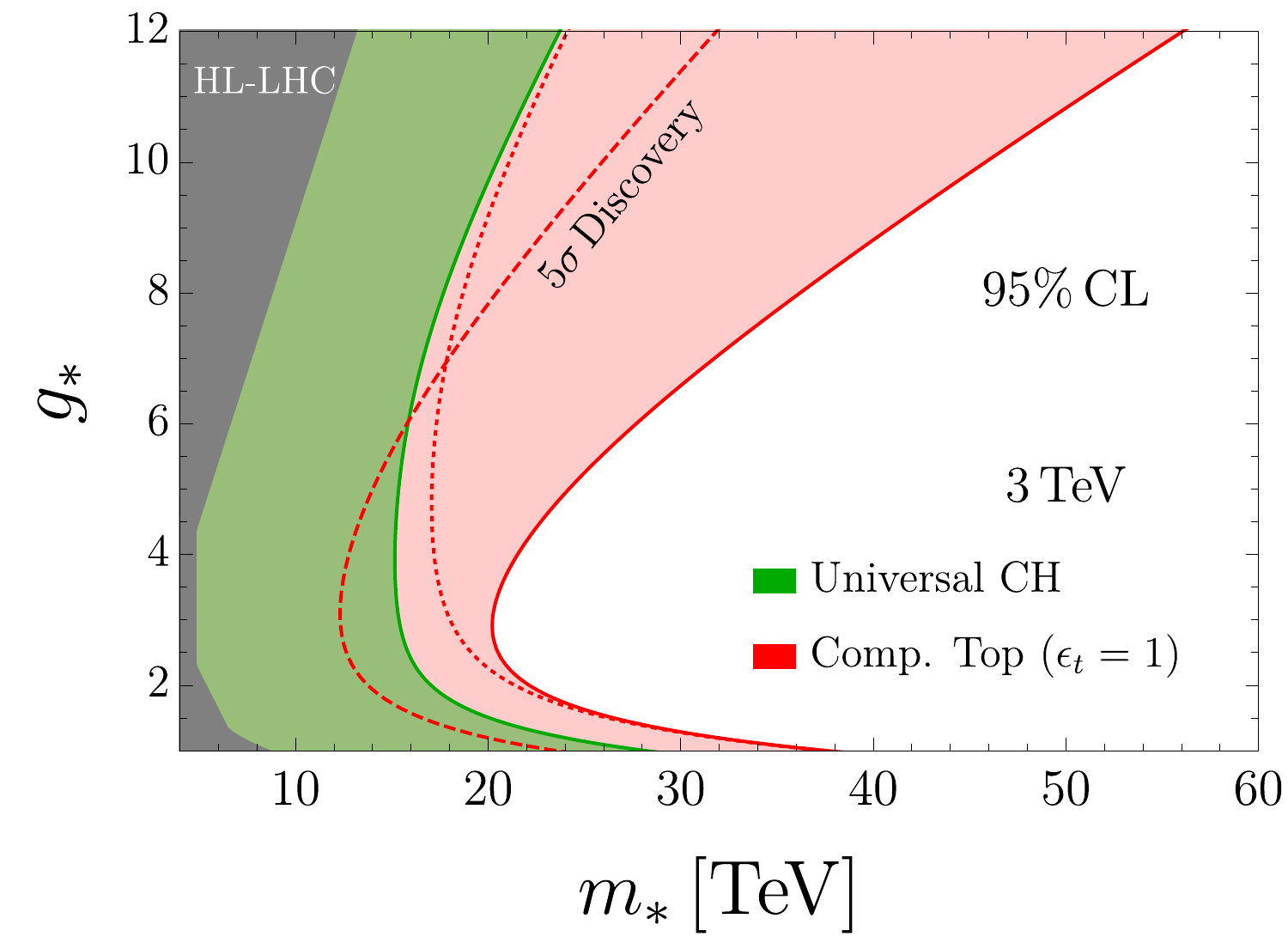}
	\end{subfigure}
	\begin{subfigure}{.5\linewidth}
		\centering
		\includegraphics[scale=.5]{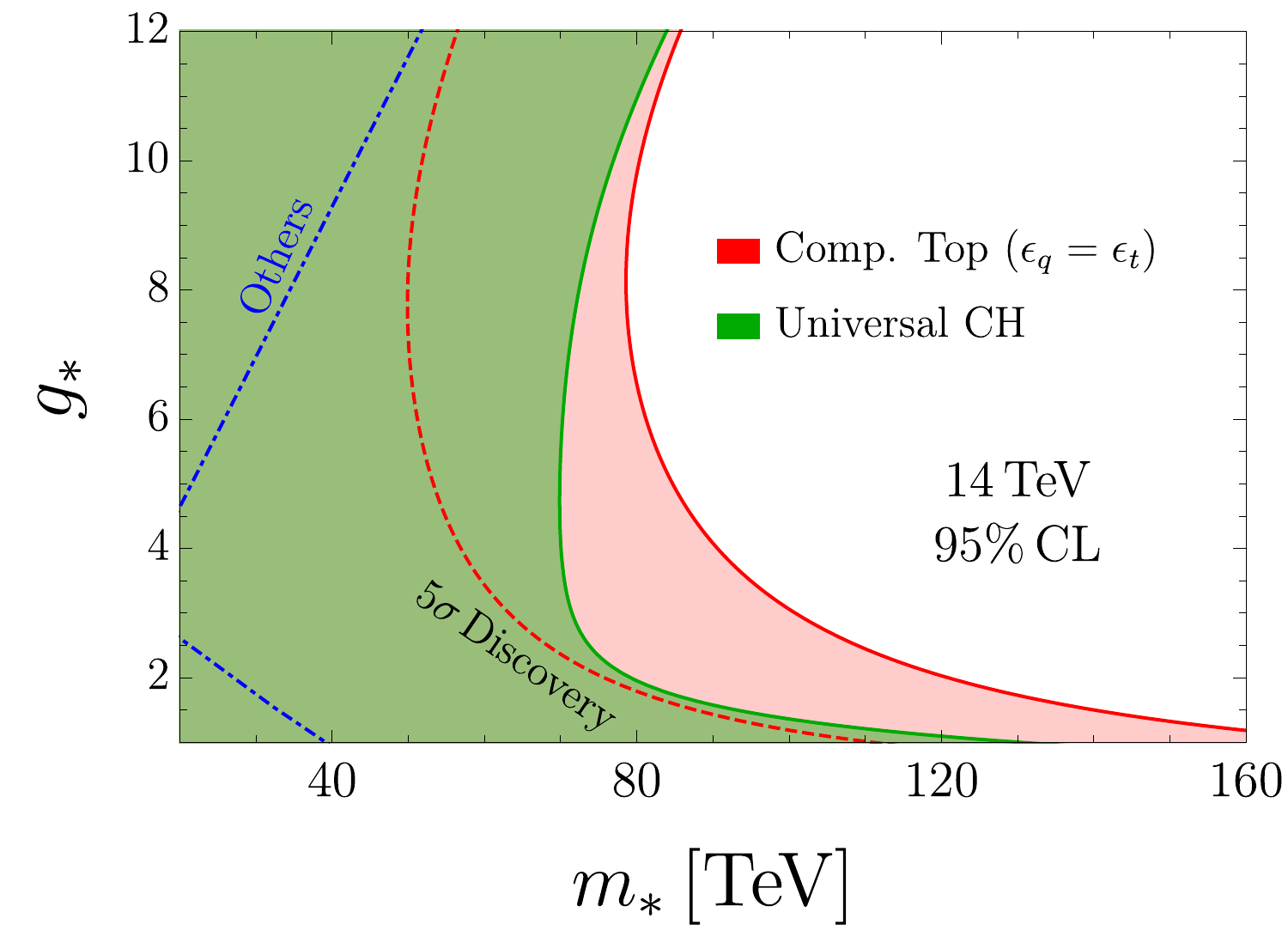}
	\end{subfigure}%
	\begin{subfigure}{.5\linewidth}
		\centering
		\includegraphics[scale=.5]{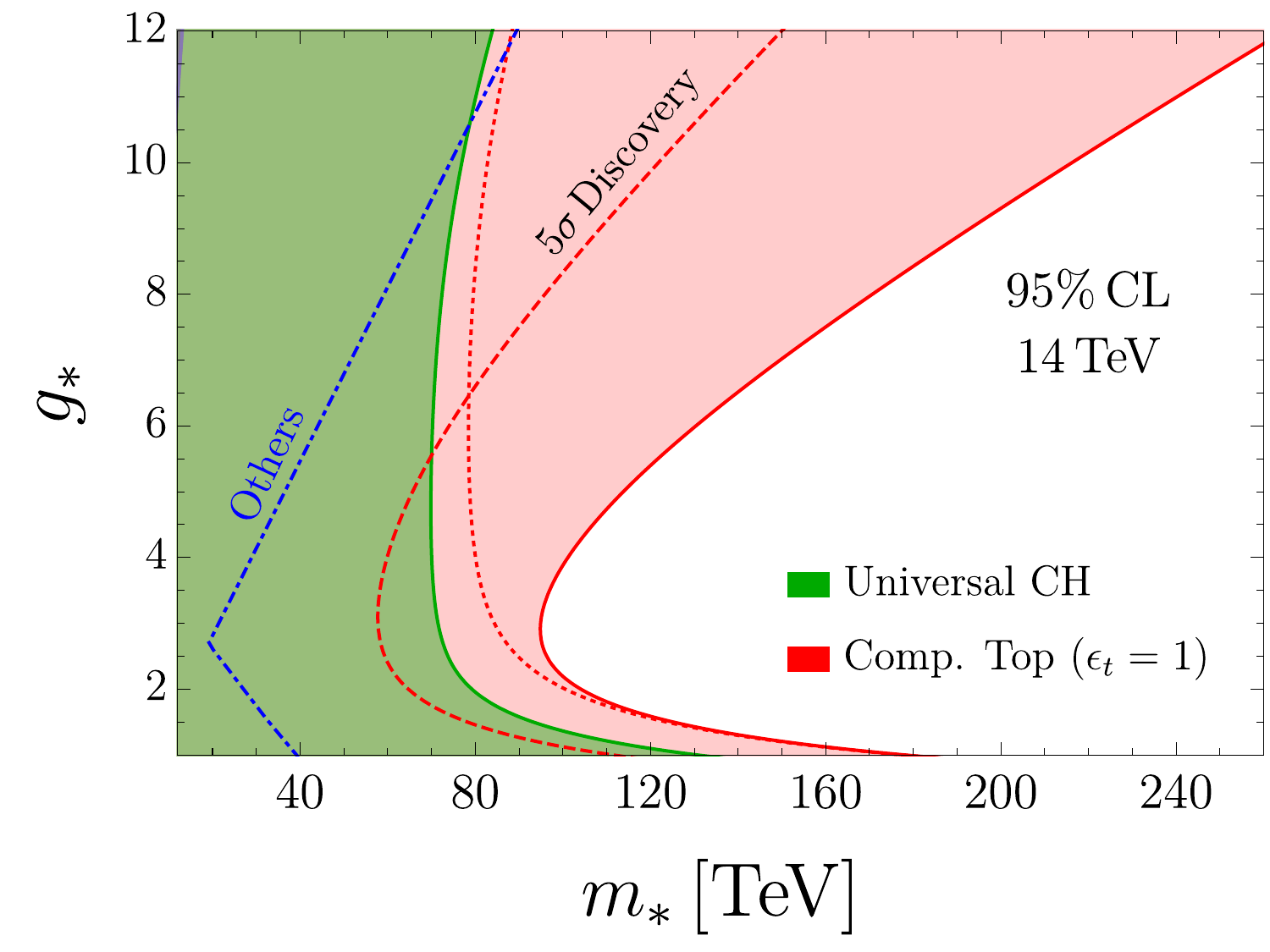}
	\end{subfigure}
	\begin{subfigure}{.5\linewidth}
		\centering
		\includegraphics[scale=.5]{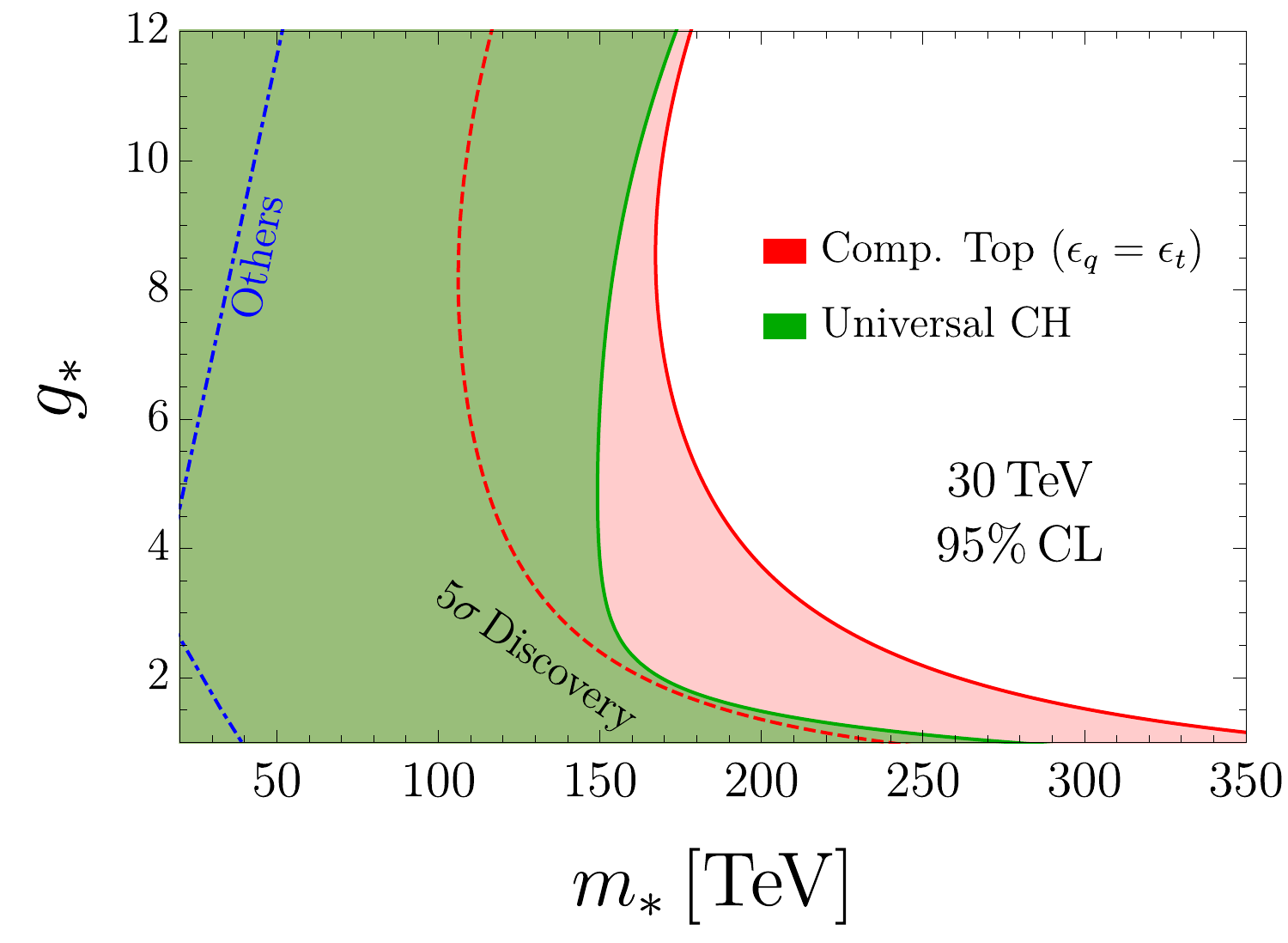}
	\end{subfigure}%
	\begin{subfigure}{.5\linewidth}
		\centering
		\includegraphics[scale=.5]{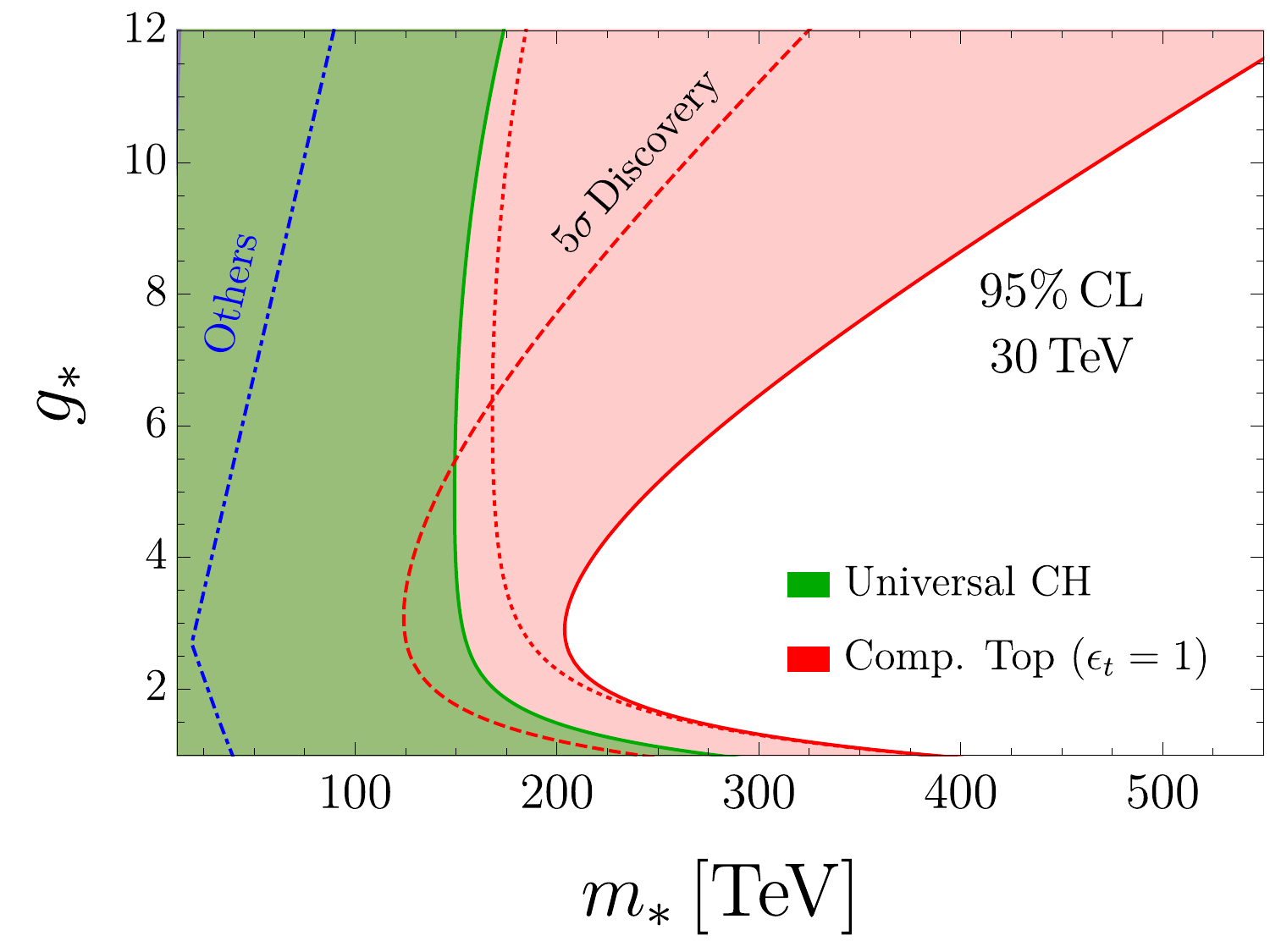}
	\end{subfigure}
	\caption{The same as Figure~\ref{Fig:CT}\label{Fig:CTApp} for various collider energies. The blue line on the equally-composite (left panels) projections are taken from \cite{EuropeanStrategyforParticlePhysicsPreparatoryGroup:2019qin}, while on the right-handed composite top scenario are take from \cite{Banelli:2020iau}.}
\end{figure}

In this appendix we collect additional results skipped in the main text. In particular in Figure~\ref{Fig:ZPrime} we report the sensitivity projections for the $Y$-universal $Z'$ model, in the $(g_{Z'}, M_{Z'})$ plane for the different collider energies. In Figure~\ref{Fig:chApp} we collect the sensitivity projections for the composite Higgs model in the $(m_*, g_*)$ plane for $E_{\rm{cm}}=3,\,14,\,30$ TeV. Projections including composite top measurements can be found in Figure~\ref{Fig:CTApp}. Finally, Figure~\ref{Fig:CTAppctt} shows the dependence of the bound on the value of the $c_{tt}$ coefficients, as explained in the main text. 

\newpage 


\providecommand{\href}[2]{#2}\begingroup\raggedright\endgroup

\end{document}